\documentclass[reprint,aps,sort&compress,amsmath,amssymb]{revtex4-2}
\usepackage{bm}
\usepackage{graphicx}
\usepackage{units}
\usepackage[english]{babel}
\usepackage[utf8]{inputenc}
\usepackage[colorlinks, citecolor=blue, urlcolor=blue, linkcolor=red,breaklinks]{hyperref}

\begin{document}

\title{Diagrammatic perturbation approach to moiré bands in twisted bilayer graphene}

\author{Federico Escudero}
\email{federico.escudero@uns.edu.ar}

\affiliation{IFISUR, Departamento de F\'isica, CONICET, Universidad Nacional del Sur, B8000 Bah\'ia Blanca, Argentina}

\begin{abstract}
	
	We develop a diagrammatic perturbation theory to account for the emergence of moiré bands in the continuum model of twisted bilayer graphene. Our framework is build upon treating the moiré potential as a perturbation that transfers electrons from one layer to another through the exchange of the three wave vectors that define the moiré Brillouin zone. By working in the two-band basis of each monolayer, we analyze the one-particle Green's function and introduce a diagrammatic representation for the scattering processes. We then identify the moiré-induced self-energy, relate it to the quasiparticle weight and velocity of the moiré bands, and show how it can be obtained by summing irreducible diagrams. We also connect the emergence of flat bands to the behavior of the static self-energy at the magic angle. In particular, we show that a vanishing Dirac velocity is a direct consequence of the relative orientation of the momentum transfer vectors, suggesting that the origin of magic angles in twisted bilayer graphene is intrinsically connected to its geometrical properties. Our approach provides a diagrammatic framework that highlights the physical properties of the moiré bands.	
	
\end{abstract}

\maketitle  

\section{Introduction}\label{sec:Introduction}

In recent years, moiré heterostructures have emerged as a promising platform to study strongly correlated phases of matter \cite{cao_unconventional_2018,Cao2018,Kerelsky2019,Oh2021,Jiang2019}. The most prominent example is twisted bilayer graphene (TBG) \cite{Andrei2020,Nimbalkar2020,Aggarwal2023}, where superconductivity and metal-insulator transitions have been experimentally observed around the \textit{magic} angle ($\theta\sim1^{\circ}$) \cite{Lopes dos Santos2007,Andrei2021}. These correlated effects are thought to be induced by a sharp increase in the density of states as the lowest moiré bands become flat at the magic angle \cite{Lopes dos Santos2007,Bistritzer2011}. The ability to easily manipulate the parameters of the system, such as the twist angle or the carrier density, provides a rich experimental platform to test and understand the nature of the correlated phases \cite{Yankowitz2019,Nimbalkar2020,Andrei2020}. Such understanding may prove to be not only relevant for the phenomena observed in moiré heterostructures, but also for elucidating the mechanisms responsible for unconventional superconductivity in more complex systems, such as in cuprates \cite{Taillefer2010,Singh2021}. 

A first step towards modeling the electronic properties in moiré heterostructures is describing its band structure within the independent electron approximation, from which correlations can then be, when possible, treated perturbatively. There are, in general, two methods to obtain the moiré bands: first principles calculations \cite{Reich2002,SuarezMorell2010,Sboychakov2015,Lin2018,Carr2019}, such as DFT or tight binding computations, and effective continuum models \cite{Lopes dos Santos2007,Bistritzer2011,LopesdosSantos2012,Moon2013}. First principles calculations, although naturally more exact, require commensurate superlattice structures, which for twist-induced heterostructures it only occur at a discrete set of twist angles \cite{Mele2010,Shallcross2010}. In addition, the number of atoms within the supercells increases as the twist angle decreases. Around the magic angle the supercell can typically contain up to ten thousands atoms \cite{Carr2020}, thus requiring a high computational capacity and time to perform first-principles calculations \cite{Leconte2022,Haddadi2020}. This has naturally led to the need of having effective continuum models that can capture the low-twist angle behavior in the system. One of the main advantages of the continuum models is that they can be constructed even if the moiré pattern is incommensurate \cite{Bistritzer2011,Koshino2015}. With only a  few set of parameters, such as the hopping energies between different sublattices, the continuum models can yield results in excellent agreement with those obtained by first-principles calculations \cite{Moon2013,Carr2019a}.

There are different continuum models in the literature \cite{Lopes dos Santos2007,Bistritzer2011,Koshino2015,Guinea2019,Bernevig2021,Kang2023}. The most sophisticated and realistic ones take into account lattice relaxation effects \cite{Carr2019a,Guinea2019,Leconte2022,Kang2023,Miao2023,Ceferino2024}, usually through the inclusion of deformation-induced gauge fields. The rich physical properties arising from the continuum models, especially around the magic angles, have further sparked numerous studies on its mathematical properties \cite{Becker2022,Becker2023,Watson2023}. For twisted bilayer graphene, all of these continuum models share the essence of coupling both layers through a moiré-induced potential, whose strength is modulated by the twist angle \cite{Bistritzer2011}. In momentum space, the leading order Fourier expansion of the moiré potential couples the Dirac points in each layer through the exchange of three twist-dependent wave vectors \cite{Koshino2015}, which are related by a hexagonal symmetry that determines the moiré Brillouin zone \cite{Bistritzer2011}. This leads to one the main features of the continuum models, namely that its Hamiltonian has no inherent momentum cutoff, since any Dirac point can be always coupled to another one. In practice, a sufficiently high momentum cutoff is determined by when the resulting moiré bands of interest (usually the low-energy ones) converge to a steady value. As the twist angle decreases the strength of the moiré coupling, relative to the energy difference between the uncoupled Dirac points, increases. Consequently, around the magic angle one generally needs a relatively large momentum cutoff in order to archive a diagonalization convergence for the low-energy moiré bands, which translates to a continuum model Hamiltonian of high matrix dimensions. Besides being less computational appealing, this increase in the need of a higher matrix dimension hinders the understanding of how the flat moiré bands emerge and behave. Numerous lower dimension models, which still aim to capture the continuum model physics, have been consequently developed, such as two-band models \cite{Lopes dos Santos2007,Bistritzer2011}, helical network models \cite{SanJose2013,Efimkin2018,DeBeule2021}, ten-band models \cite{Carr2019a}, Kondo-lattice models \cite{Chou2023,Hu2023}, $\left(2+2\right)$ models \cite{Bennett2024}, and topological heavy-fermions models \cite{Song2022,Shi2022}. 

In this work, we propose and develop an alternative diagrammatic perturbation approach to study the moiré bands. Our motivation is twofold: On one hand the perturbative approach provides a natural framework to study the increasing effect of the moiré coupling on the one-particle excitations of the system as the twist angle is decreased; on the other hand, the diagrammatic approach gives a direct physical picture of the moiré-induced scattering events taking place, and how this may lead to flat bands. Our framework is build upon the usual continuum model, but explicitly working in the two-band basis of each uncoupled layer. In this basis the moiré potential acts as a one-body operator that transfers electrons from one layer to another by the exchange of the three wave vectors that determine the moiré Brillouin zone. By treating this potential as a perturbation to the uncoupled layers, we obtain the moiré-induced self-energy that renormalizes the linear massless dispersions onto the moiré bands, and discuss its physical interpretation by identifying the contributions of irreducible diagrams. From the self-energy we further obtain the one-particle spectral density, the quasiparticle weight and the velocity of the moiré bands. We also relate the properties of the lowest moiré bands to the behavior of the static self-energy at the magic angle, and show that the emergence of flat bands can be directly connected to the particular geometrical properties of the moiré potential in twisted bilayer graphene.

The developed approach opens a path to deploy diagrammatic techniques to the study of moiré bands. Compared to traditional approaches based on truncated continuum model Hamiltonians, the diagrammatic formalism is entirely based on the perturbation expansion of the Dyson equation. This is in contrast to the usual perturbation theory applied to a truncated TBG Hamiltonian, since in the diagrammatic approach such truncation is already accounted by the perturbation order in the Dyson series. The power of the developed approach is highlighted by showing that properties such as the moiré-induced self-energy, and other related quantities, can be diagrammatically obtained using a simple set of Feynman rules, at any perturbation order. The developed approach further paves a way to obtain effective models for the moiré bands, namely, by identifying those diagrams that make the larger contribution.

The rest of this work is organized as follows: In Sec. \ref{sec:CM} we give a brief recap of the continuum model of twisted bilayer graphene, which serves as a basis for the subsequent perturbation approach.  Next we describe the formulation of the continuum model in the two-band basis of each layer. In Sec. \ref{sec:Perturbation} we develop the proposed diagrammatic perturbation approach to the moiré bands. We introduce the one-particle Green's function and give its exact closed form through a matrix inversion of the continuum model Hamiltonian in the band basis. We identify the moiré-induced self energy, compute it to leading order in perturbation, and give its physical interpretation. We then discuss the one-particle spectral density, the velocity renormalization of the moiré bands, and the emergence and origin of flat bands around the magic angle. We end with the conclusions in Sec. \ref{sec:Conclusions}.

\section{Continuum model of twisted bilayer graphene} \label{sec:CM}

\begin{figure}[t]
	\includegraphics[scale=0.4]{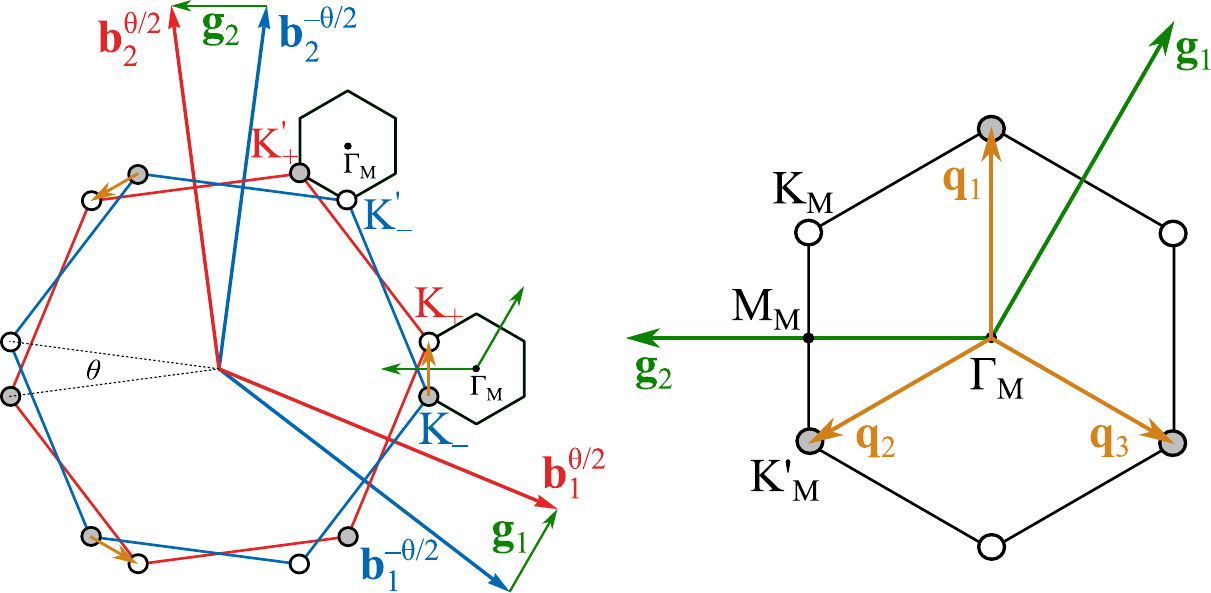}	
	\caption{Reciprocal space representation of two twisted honeycomb Brillouin zones, with corresponding reciprocal vectors $\mathbf{b}_{1,2}^{\pm\theta/2}$ and Dirac points $\mathbf{K}_{\pm},\mathbf{K}_{\pm}^{\prime}$ for the top $\left(+\right)$ and bottom $\left(-\right)$ layer, each rotated by $\pm\theta/2$. The moiré vectors are given by the difference between the rotated reciprocal vectors, $\mathbf{g}_{1,2}=\mathbf{b}_{1,2}^{\theta/2}-\mathbf{b}_{1,2}^{-\theta/2}$. In the continuum model, the twisted Dirac points in different layers are coupled, to leading order, by the exchange of three momentum transfer vectors $\mathbf{q}_{1},\mathbf{q}_{2},\mathbf{q}_{3}$ that determine the borders of the moiré Brillouin zone (right). For momenta originating from the $K$ valley of both layers, the moiré valleys are set by $K_{M}=K_{+}$ and $K'_{M}=K_{-}$, while around the $K'$ valley $K_{M}=K'_{-}$ and $K'_{M}=K'_{+}$. The three momentum transfer vectors relate inequivalent valleys within the moiré Brillouin zone.}\label{fig:TBG}
\end{figure}

Two honeycomb lattices in a bilayer configuration, relatively rotated by an angle $\theta$, give rise to a moiré pattern of characteristic length $L\sim a/2\sin\left(\theta/2\right)$ \cite{Andrei2020}, where $a$ is the lattice constant of the layers ($a\simeq0.246\,\mathrm{nm}$ in graphene). For arbitrary twist angles the moiré pattern is in general incommensurate, i.e., it is not technically a crystal structure \cite{Mele2010,Shallcross2010}. Nevertheless, one can geometrically describe the moiré pattern by defining a set of two moiré vectors $\mathbf{g}_{i}$ ($i=1,2$), constructed by taking the minimum difference between the rotated reciprocal vectors in each layer \cite{Moon2013,Koshino2015}, $\mathbf{g}_{i}=\mathbf{b}_{i}^{\theta/2}-\mathbf{b}_{i}^{-\theta/2}$, where $\mathbf{b}_{i}^{\pm\theta/2}=R\left(\pm\theta/2\right)\mathbf{b}_{i}$ with $R$ the rotation matrix and $\mathbf{b}_{i}$ the reciprocal vectors of a honeycomb lattice. In real space, the moiré vectors so defined connect \emph{all} the AA stacking regimes of the moiré pattern. We choose the primitive lattice vectors of the layers as $\mathbf{a}_{1}=a\left(1,0\right)$ and $\mathbf{a}_{2}=a\left(1/2,\sqrt{3}/2\right)$, which upon rotation lead to the reciprocal space orientation shown in Fig.
\ref{fig:TBG}.

The continuum model captures the low-energy electronic properties of the moiré pattern, in the regime where the moiré length is much larger than the atomic length \cite{Lopes dos Santos2007}. While first-principles models require commensurate superlattice structures, the continuum model holds even if the moiré pattern is incommensurate \cite{Bistritzer2011,Koshino2015}. There are different formulations of the continuum model in the literature; they all share, however, the essence of coupling the Dirac points in each layer through a moiré-induced potential. In this work we will, in particular, consider the Bistritzer--MacDonald model \cite{Bistritzer2011}, for momenta relative to the $K$ valley in both layers (see Fig. \ref{fig:TBG}). The corresponding Hamiltonian reads
\begin{equation}
	H=\left(\begin{array}{cc}
		h_{+}\left(\mathbf{k}\right) & T^{\dagger}\left(\mathbf{r}\right)\\
		T\left(\mathbf{r}\right) & h_{-}\left(\mathbf{k}\right)
	\end{array}\right),\label{eq:H}
\end{equation}
where $h_{\pm}\left(\mathbf{k}\right)=\hbar v\boldsymbol{\sigma}_{\pm\theta/2}\cdot\mathbf{k}_{\pm}$ are the Dirac Hamiltonian in the top $\left(+\right)$ and bottom $\left(-\right)$ layers, with $v=\sqrt{3}ta/2\hbar$ the Fermi velocity in graphene ($t$ being the nearest sublattice hopping energy), $\mathbf{k}_{\pm}$ the momenta relative to the rotated $K$ points, and $\boldsymbol{\sigma}_{\pm\theta/2}=\left(\sigma_{x,\pm\theta/2},\sigma_{y,\pm\theta/2}\right)$ the rotated Pauli matrix vector \cite{CastroNeto2009,DasSarma2011}. The angle dependence in the Pauli matrices leads to a small energy shift in the system; in what follows we neglect this and set $\boldsymbol{\sigma}_{\pm\theta/2}\rightarrow\boldsymbol{\sigma}$ (this restores the particle-hole symmetry in the system) \cite{Bistritzer2011,Song2021}. 

The coupling between both layers is dictated by the moiré potential $T\left(\mathbf{r}\right)$. At small twist angles,  its leading order Fourier expansion  is given by \cite{Bistritzer2011,Moon2013,Koshino2015}
\begin{equation}
	T^{\alpha\beta}\left(\mathbf{r}\right)=w_{\alpha\beta}\sum_{j=1}^{3}e^{i\mathbf{q}_{j}\cdot\mathbf{r}}T_{j}^{\alpha\beta},
\end{equation}
where $\alpha$ and $\beta$ are the sublattice indices in each layer, $w_{\alpha\beta}$ is the corresponding coupling strength, and $\mathbf{q}_{j}$ are the three equal-length wave vectors arising from the difference between the three equivalent rotated Dirac points, cf. Fig. \ref{fig:TBG}. The wave vectors $\mathbf{q}_{j}$ thus determine the moiré Brillouin zone (mBZ), and can be obtained from the moiré vectors as $\mathbf{q}_{1}=\left(2\mathbf{g}_{1}+\mathbf{g}_{2}\right)/3$, $\mathbf{q}_{2}=R\left(2\pi/3\right)\mathbf{q}_{1}$ and $\mathbf{q}_{3}=R\left(4\pi/3\right)\mathbf{q}_{1}$. For an initial AA bilayer configuration, upon rotation the coupling matrices for each wave vector $\mathbf{q}_{j}$ read \cite{Koshino2015}
\begin{equation}
	T_{j}=\left(\begin{array}{cc}
		w_{0} & w_{1}e^{-i\phi_{j}}\\
		w_{1}e^{i\phi_{j}} & w_{0}
	\end{array}\right),\label{eq:Tj}
\end{equation}
where $w_0=w_{AA}$ and $w_1=w_{AB}$ are the AA and AB hopping energies, and $\phi_{j}=\left(j-1\right)2\pi/3$. Due to relaxation effects, the ratio $w_0/w_1$ tends to decrease at low twist angles, effectively reducing the energetically costly AA stacking regimes \cite{Jain2016,Carr2019a,Koshino2020}. In this work, unless otherwise stated, we consider $w_1=0.11$ eV, $w_0=0.8w_1$ and $t=2.8$ eV, which gives the first magic angle at $\theta_{M}\simeq1.06^{\circ}$ \cite{Bistritzer2011,Tarnopolsky2019}. 

\begin{figure}
	\includegraphics[scale=0.72]{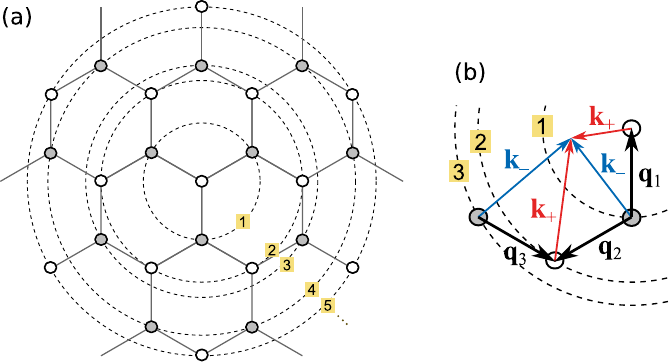}	
	\caption{(a) Tripod model of the TBG Hamiltonian expansion in momentum space, in which the momentum is measured from the Dirac point in one layer. Empty (filled) circles correspond to Dirac points in the top (bottom) layer. Circles indicate the truncation (or cutoffs) at increasing \emph{shells}, corresponding to equidistant Dirac points. (b) Schematic representation of successive momentum exchanges $\mathbf{k}_{-}-\mathbf{k}_{+}=\mathbf{q}_{j}$ between adjacent Dirac points, up to the third shell.}\label{fig:shell}
\end{figure}  

The moiré bands are obtained by diagonalizing the continuum model Hamiltonian in momentum space. As each Dirac point couples to other three by the exchange of the wave vectors $\mathbf{q}_{j}$, the continuum model Hamiltonian has no bound in momentum space. However, only the low-energy moiré bands are relevant, for which the eigenstates effectively converge up to a sufficiently large momentum cutoff. In this work we adopt the original Tripod model expansion of the continuum model Hamiltonian \cite{Bistritzer2011,Bernevig2021}, in which the momentum is measured from a Dirac point and the cutoffs are set by \textit{shells} of equidistant Dirac points from the origin, cf. Fig. \ref{fig:shell}. 

The continuum model Hamiltonian around the $K'$ valley can be obtained from the one in the $K$ valley by replacing $\boldsymbol{\sigma}\rightarrow\boldsymbol{\sigma}'=\left(-\sigma_{x},\sigma_{y}\right)$, $\mathbf{q}_{j}\rightarrow-\mathbf{q}_{j}$ and $\phi_{j}\rightarrow-\phi_{j}$ \cite{Moon2013}. The moiré bands for momenta around the layer valleys $K$ and $K'$ (not to be confused with the moiré valleys $K_M$) only differ slightly around the $\Gamma_{M}$ and $M_{M}$ moiré symmetry points \cite{Moon2013}; their general properties, particularly as a function of the twist angle, are nevertheless the same. As intervalley scattering events within each layer are negligible at low energies, throughout this paper we continue to work with the TBG Hamiltonian for the $K$ valley, keeping in mind that the results for the $K'$ valley can be trivially obtained by the above correspondence. 

\subsection{Band basis}\label{subsec:CMbandbasis}

The continuum model Hamiltonian given by Eq. \eqref{eq:H} is written in the sublattice basis for each layer. To make a connection with perturbation theory, it will prove more convenient to rather express it in the band basis of each layer, in which the Dirac Hamiltonian is diagonalized. In momentum space, the matrix elements of the corresponding TBG Hamiltonian, which we shall denote as $H_{B}$, are given by (cf. Appendix \ref{app:CMband})
\begin{align}
	\left\langle \ell',\mathbf{k}',s'\right\vert \hat{H}_{B}\left\vert \ell,\mathbf{k},s\right\rangle  & = \delta_{\ell,\ell'}s\hbar v\left|\mathbf{k}\right|\delta_{\mathbf{k},\mathbf{k}'}\delta_{s,s'}\nonumber\nonumber \\
	& \quad+\delta_{\ell,-\ell'}\sum_{j=1}^{3}\left(\mathcal{T}_{j,\mathbf{k},s,s'}\delta_{\ell,+}\delta_{\mathbf{k}',\mathbf{k}+\mathbf{q}_{j}}\right.\nonumber\nonumber \\
	& \qquad\;+\left.\mathcal{T}_{j,\mathbf{k}',s',s}^{*}\delta_{\ell,-}\delta_{\mathbf{k}',\mathbf{k}-\mathbf{q}_{j}}\right).\label{eq:HB}
\end{align}
Here $s,s'=\pm1$ are the band indices in the $\ell,\ell'=\pm$ top and bottom layers. The moiré potential reads
\begin{equation}
	\mathcal{T}_{j,\mathbf{k},s,s'}=\psi_{\mathbf{k}+\mathbf{q}_{j},s'}^{\dagger}T_{j}\psi_{\mathbf{k},s},\label{eq:Tjband}
\end{equation}
where $T_{j}$ are the moiré matrices given by Eq. \eqref{eq:Tj}, and 
\begin{equation}
	\psi_{\mathbf{k},s}=\frac{1}{\sqrt{2}}\left(\begin{array}{c}
		1\\
		se^{i\theta_{\mathbf{k}}}
	\end{array}\right)\label{eq:psi}
\end{equation}
are the pseudospinors eigenstates in graphene (accounting for the neglected rotation of the Pauli matrices amounts to replace $\theta_{\mathbf{k}}\rightarrow\theta_{\mathbf{k}}\pm\theta/2$ in  $\psi_{\mathbf{k},s}$). The overlap of these pseudospinors leads to the momentum dependence of the moiré potential in the band basis. Specifically, the AA hoppings lead to a contribution of the form $\mathcal{T}_j\sim w_{0}\psi_{\mathbf{k}+\mathbf{q}_{j},s'}^{\dagger}\psi_{\mathbf{k},s}$, whereas the AB/BA hoppings lead to a contribution of the form $\mathcal{T}_{j}\sim w_{1}\psi_{\mathbf{k}+\mathbf{q}_{j},s'}^{\dagger}\boldsymbol{\sigma}\cdot\boldsymbol{\phi}_{j}\psi_{\mathbf{k},s}$, where $\boldsymbol{\phi}_{j}$ is the versor with an angle $\phi_{j}$ relative to the $x$ axis. The latter overlap is similar to the one obtained for the interaction between two chiral states mediated by a vector potential $\mathbf{A}_{j}\sim w_{1}\boldsymbol{\phi}_{j}$. 

\begin{figure}[t]
	\includegraphics[scale=0.7]{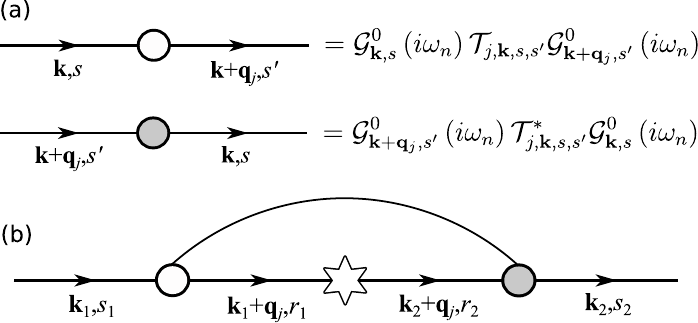}	
	\caption{(a) Diagrammatic representation for the scattering between electrons in different layers induced by the moiré potential. Solid lines represent the \emph{free} Green's function $\mathcal{G}_{\mathbf{k},s}^{0}\left(i\omega\right)=1/\left(i\omega_{n}-\epsilon_{\mathbf{k},s}\right)$ for a Dirac electron with momentum $\mathbf{k}$ in the $s=\pm1$ band and energy $\epsilon_{\mathbf{k},s}=s\hbar v\left|\mathbf{k}\right|$. Empty (filled) circles represent the moiré potential strength $\mathcal{T}$ $\left(\mathcal{T}^{*}\right)$ for the scattering of an electron in the top (bottom) layer to the other layer.  (b) A curved line connecting one empty and one filled circle is used to indicate the same exchanged momentum $\mathbf{q}_{j}$, regardless of possible in-between scattering events that may modify the initial and final momenta.}\label{fig:Frules}
\end{figure}

The use of the band basis provides a natural framework from which one can deploy perturbation methods into the problem. To this end, we  second quantize the continuum model Hamiltonian and separate $\hat{H}=\hat{H}_{0}+\hat{T}$, where 
\begin{align}
	\hat{H}_{0} & =\sum_{\mathbf{k},s}\sum_{\ell=\pm}\epsilon_{\mathbf{k},s}\hat{c}_{\ell,\mathbf{k},s}^{\dagger}\hat{c}_{\ell,\mathbf{k},s},\\
	\hat{T} & =\sum_{\mathbf{k},s,s'}\sum_{j=1}^{3}\mathcal{T}_{j,\mathbf{k},s,s'}\hat{c}_{-,\mathbf{k}+\mathbf{q}_{j},s'}^{\dagger}\hat{c}_{+,\mathbf{k},s}+h.c.\label{eq:Tpot}
\end{align}
Here the operators $\hat{c}_{\ell,\mathbf{k},s}^{\dagger}$ and $\hat{c}_{\ell,\mathbf{k},s}$ create and annihilate an electron in the $\ell$ layer, with momentum $\mathbf{k}$ in the $s$ band and energy $\epsilon_{\mathbf{k},s}=s\hbar v\left|\mathbf{k}\right|$.  We adopt the convention that the momentum is measured relative to a Dirac point in the top layer. The moiré potential $\hat{T}$ can be thought as a one-body operator that exchanges electrons from one layer to another by the transfer of a momentum $\mathbf{q}_{j}$, cf. Fig. \ref{fig:Frules}. 

Since the potential $\hat{T}$ is quadratic in the operators, the total Hamiltonian can be, of course, exactly diagonalized:
\begin{equation}
	\hat{H}=\sum_{\mathbf{k}\in\mathrm{mBZ}}\sum_{\nu}E_{\mathbf{k},\nu}\hat{a}_{\mathbf{k},\nu}^{\dagger}\hat{a}_{\mathbf{k},\nu}.\label{eq:Hsq}
\end{equation}
Here the operators $\hat{a}_{\mathbf{k},\nu}^{\dagger}$ and $\hat{a}_{\mathbf{k},\nu}$ create and annihilate an electron in the $\nu$ moiré band with momentum $\mathbf{k}$ within the moiré BZ, and energy $E_{\mathbf{k},\nu}$. Equation \eqref{eq:Hsq} formally \textit{solves} the problem, provided that the energies $E_{\mathbf{k},\nu}$ are obtained by diagonalizing the continuum model Hamiltonian. However, such matrix diagonalization implicitly requires a truncation of the TBG Hamiltonian, on which the number of moiré bands in the sum over $\nu$, as well as the corresponding expressions for their energies, depend. The same holds for the moiré operators $\hat{a}_{\mathbf{k},\nu}$, whose expansion in terms of the $\hat{c}$ operators generally depends on the momentum cutoff. This sort of perturbation dependence in the continuum model is \textit{hidden} in Eq. \eqref{eq:Hsq}. The effectiveness of the truncation, as measured by diagonalization convergence, further depends on the twist angle: whereas at relatively high angles it may be sufficient to consider a low momentum cutoff to describe the lowest moiré bands, at lower twist angles one generally requires a higher momentum cutoff for convergence. As a result, a physical understanding of the emergent properties of the moiré bands becomes increasingly harder to get as the twist angle decrease, since a larger momentum cutoff translates to a TBG Hamiltonian of high dimensions.

\section{Diagrammatic perturbation approach to moiré bands}\label{sec:Perturbation}

In this work, rather than reducing the problem to a matrix diagonalization, we propose and develop a diagrammatic perturbation approach to study the nature of the moiré bands. We treat the system as an effective many-body problem in which the moiré potential $\hat{T}$ acts as a perturbation to the Hamiltonian $\hat{H}_{0}$. Our interest will be the effect of the perturbation on the one-particle excitations in each layer. In particular, we consider the imaginary time one-particle Green's function \cite{Mahan1990}
\begin{equation}
	\mathcal{G}_{\gamma\gamma'}\left(\tau\right)=-\left\langle T_{\tau}\hat{c}_{\gamma}\left(\tau\right)\hat{c}_{\gamma'}^{\dagger}\left(0\right)\right\rangle,\label{eq:gf}
\end{equation}
where the index $\gamma=\left(\ell,\mathbf{k},s\right)$ accounts for the layer, momentum and band indices of the $\hat{c}$ operators, $T_{\tau}$ is the time-ordering operator, and $\left\langle \cdots\right\rangle $ is the ensemble average over the interacting system. By treating the moiré potential $\hat{T}$ as a perturbation, the ensemble average $\left\langle \cdots\right\rangle $ can be expanded as an ensemble average over the noninteracting system, cf. Appendix \ref{app:PerExpansion}. 

Since the problem is \textit{exactly} solvable, the Green's function can be obtained in closed form. Indeed, in frequency-momentum space the Green's function of the system can be reduced to a matrix inversion \cite{Mahan1990} $\mathcal{G}\left(i\omega_{n}\right)=\left(i\omega_{n}\mathbb{I}-H_{B}\right)^{-1}$, where $\omega_{n}=\left(2n+1\right)\pi/\hbar\beta$ are the fermionic Matsubara frequencies (with $n$ an integer and $\beta=1/k_{B}T$), $H_{B}$ is the band-basis continuum model Hamiltonian in momentum space [cf. Eq. \eqref{eq:HB}], and $\mathbb{I}$ is the identity matrix of the same dimension as $H_{B}$ (which depends on the shell truncation). The Green's function given by Eq. \eqref{eq:gf} can then be obtained as
\begin{equation}
	\mathcal{G}_{\gamma\gamma'}\left(i\omega_{n}\right)=\left(i\omega_{n}\mathbb{I}-H_{B}\right)_{\gamma\gamma'}^{-1},\label{eq:gf2}
\end{equation}
with $\gamma$ and $\gamma'$ being matrix components such that $\left(H_{B}\right)_{\gamma\gamma}=\epsilon_{\gamma}$ and $\left(H_{B}\right)_{\gamma'\gamma'}=\epsilon_{\gamma'}$. The retarded Green's function follows by analytical continuation $i\omega_{n}\rightarrow\omega+i0^{+}$. 

From a numerical point of view, Eq. \eqref{eq:gf2} gives a straightforward method to obtain the Green's function of the system. However, as it stands, it does not provide a direct physical picture of the moiré-induced scattering processes taking place. To this end, in what follows we make a connection between Eq. \eqref{eq:gf2} and the usual many-body diagrammatic techniques. We identify the self-energy, discuss its diagrammatic expansion, and relate it to the spectral density, the quasiparticles velocity, and the emergence of flat bands around the magic angle.

\subsection{Self-energy}\label{subsec:self-energy}

Since the moiré potential, Eq. \eqref{eq:Tpot}, exchanges electrons from one layer to another, with different momenta and arbitrary band indices, in general the Green's function given by Eq. \eqref{eq:gf2} will depend on the indices $\gamma,\gamma'$. Nevertheless, most of the information about the one-particle excitations spectra can be deduced by analyzing the Green's functions with $\gamma=\gamma'$. To simplify the notation we shall write $\mathcal{G}_{\gamma\gamma}\equiv\mathcal{G}_{\gamma}$. The Dyson equation in momentum-frequency space then leads to the exact expression \cite{Jishi2013} (from now on we set $\hbar=1$)
\begin{equation}
	\mathcal{G}_{\gamma}\left(i\omega_{n}\right)=\frac{1}{i\omega_{n}-\bar{\epsilon}_{\gamma}-\Sigma_{\gamma}\left(i\omega_{n}\right)},\label{eq:gfself}
\end{equation}
where $\bar{\epsilon}_{\gamma}=\epsilon_{\gamma}-\mu$ are the unperturbed energies measured relative to the chemical potential $\mu$ (henceforth we set, for simplicity, the energy origin at $\mu=0$, and drop the bar notation), and $\Sigma_{\gamma}\left(i\omega_{n}\right)$ is the proper self-energy (henceforth just the \emph{self-energy}). As with the Green's function, the self-energy can be directly obtained from the continuum model Hamiltonian through Eq. \eqref{eq:gf2}, which implies that
\begin{equation}
	\Sigma_{\gamma}\left(i\omega_{n}\right)=i\omega_{n}-\epsilon_{\gamma}-\frac{1}{\left(i\omega_{n}\mathbb{I}-H_{B}\right)_{\gamma\gamma}^{-1}},\label{eq:self}
\end{equation}
where the component index $\gamma$ should be read as the index at which $\left(H_{B}\right)_{\gamma\gamma}=\epsilon_{\gamma}$.

\begin{figure}[t]
	\includegraphics[scale=0.73]{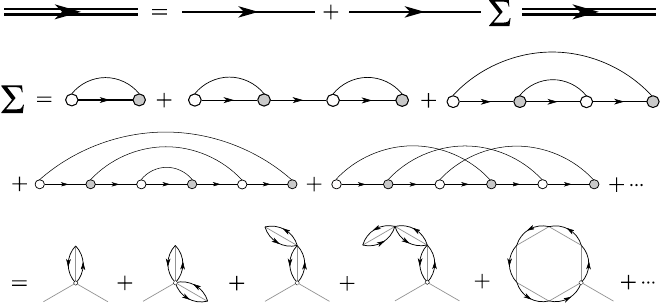}	
	\caption{Top: Diagrammatic representation of the moiré-induced self-energy; double lines indicate the full dressed Green's function. Bottom: leading order examples of irreducible diagrams that contribute to the self-energy. First row is in terms of the Feynman diagrams detailed in Fig. \ref{fig:Frules}. Internal momentum transfers and band indices are to be summed over. Diagrams in which at a midpoint the electron returns to the same initial momentum, as in the second diagram, imply a change of band. Last row is a hexagonal picture representation of the above diagrams, in the same order. Curved lines indicate the path of an electron by coupling to successive Dirac points. For simplicity only one possible equivalent path is shown.}\label{fig:SelfDiag}
\end{figure}

The self-energy describes the renormalization of the uncoupled dispersions $\epsilon_\gamma$ due to the interaction with the moiré potential. One can thus obtain valuable insights into the nature of the moiré bands by solely analyzing the properties of the self-energy. One of the main advantages of the developed approach is that the self-energy can be obtained diagrammatically by summing irreducible diagrams. Some examples of those are shown in Fig. \ref{fig:SelfDiag}. The diagrammatic approach gives not only a straightforward method to compute the self-energy (in principle, to any perturbation order), but also a clear physical picture of the involved moiré-induced scattering processes. This can be very valuable for understanding the emergent properties of the moiré bands, particularly when compared to the numerical diagonalization procedure. 

It is worth noting that the perturbation description put forward here \emph{does not} imply small corrections to the unperturbed energies, since those can actually be strongly renormalized, especially at low twist angles. As long as the dressed Green's function converges (so that Eq. \eqref{eq:gfself} holds), such an increase in renormalization does not, however, invalidate the perturbation approach. In fact, the perturbation expansion of the moiré potential simply accounts, in a diagrammatic approach, the construction of the TBG Hamiltonian in terms of coupled Dirac points. Since the moiré potential given by Eq. \eqref{eq:Tjband} only depends on the angles of the momenta, a coupling to a Dirac point at a general position $-\mathbf{Q}=\sum_{j}\left(\pm\mathbf{q}_{j}\right)$ (see Fig. \ref{fig:shell})  roughly scales, to leading order, as $\sim w_1/\left(\omega-v\left|\mathbf{k}+\mathbf{Q}\right|\right)$ in the total self-energy $\Sigma_{\mathbf{k}}\left(\omega\right)$, cf. Figs. \ref{fig:Frules} and \ref{fig:SelfDiag}.  Therefore, although as the twist angle decreases the contribution of couplings at high Dirac points increases (since $\left|\mathbf{q}_{j}\right|\propto\theta$), their effect can still be effectively neglected at low energies if $w_1/v\left|\mathbf{k+Q}\right|\ll1$, thus justifying the perturbation approach \cite{Bistritzer2011,Bernevig2021}. 

\subsubsection{Physical interpretation} 

\begin{figure}
	\includegraphics[scale=0.77]{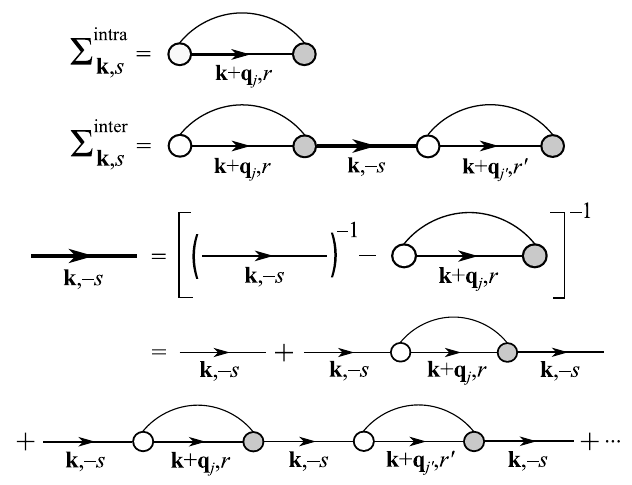}	
	\caption{Diagrammatic representation of the moiré-induced self-energy $\Sigma_{+,\mathbf{k},s}^{\left(1\right)}=\Sigma_{\mathbf{k},s}^{\mathrm{intra}}+\Sigma_{\mathbf{k},s}^{\mathrm{inter}}$ corresponding to the first shell truncation of the continuum model Hamiltonian. The first diagram $\Sigma_{\mathbf{k},s}^{\mathrm{intra}}$ gives the intraband contribution, where an electron scatters back and forth to the opposite layer, returning to the initial state in the same band. The second diagram $\Sigma_{\mathbf{k},s}^{\mathrm{inter}}$ gives the interband contribution, where an electron transitions to the opposite band by scattering back and forth to the opposite layer, before scattering back again to the initial state.}\label{fig:sefl1}
\end{figure}

The self-energy can be directly computed by summing irreducible diagrams, cf. Figs. \ref{fig:Frules} and \ref{fig:SelfDiag}. Due to the strong coupling of the moiré potential at low twist angles, and the nonnegligible contribution of interband transitions, one generally needs to consider many irreducible diagrams in order to achieve a convergence of the self-energy. A valuable guide to visualize the relevant scattering events can be obtained by giving a diagrammatic representation to Eq. \eqref{eq:self}. For conciseness, throughout this section we will focus on the self-energy for the top layer, so that $\gamma=\text{\ensuremath{\left(+,\mathbf{k},s\right)}}$. The results are readily generalized for the bottom layer by replacing $\mathbf{k}\rightarrow\mathbf{k}+\mathbf{q}_{1}$, $\mathbf{q}_{j}\rightarrow-\mathbf{q}_{j}$ and $\mathcal{T}_{j}\rightarrow\mathcal{T}_{j}^{*}$ (see Fig. \ref{fig:TBG}).  

We start by considering the lowest TBG Hamiltonian in the Tripod model, where the momentum is truncated at the first shell (cf. Fig. \ref{fig:shell}). This leads to the well known $8\times8$ TBG Hamiltonian, which in the band basis takes the form
\begin{equation}
	H_{B,\mathbf{k}}^{\left(1\right)}=\left(\begin{array}{cccc}
		h_{\mathbf{k}}^{B} & \mathcal{T}_{1,\mathbf{k}}^{\dagger} & \mathcal{T}_{2,\mathbf{k}}^{\dagger} & \mathcal{T}_{3,\mathbf{k}}^{\dagger}\\
		\mathcal{T}_{1,\mathbf{k}} & h_{\mathbf{k}+\mathbf{q}_{1}}^{B} & 0 & 0\\
		\mathcal{T}_{2,\mathbf{k}} & 0 & h_{\mathbf{k}+\mathbf{q}_{2}}^{B} & 0\\
		\mathcal{T}_{3,\mathbf{k}} & 0 & 0 & h_{\mathbf{k}+\mathbf{q}_{3}}^{B}
	\end{array}\right),\label{eq:H1}
\end{equation}
where $h_{\mathbf{k}}^{B}=-v\left|\mathbf{k}\right|\sigma_{z}$ is the diagonalized Dirac Hamiltonian and [cf. Eq. \eqref{eq:Tjband}]
\begin{equation}
	\mathcal{T}_{j,\mathbf{k}}=\left(\begin{array}{cc}
		\mathcal{T}_{j,\mathbf{k},-,-} & \mathcal{T}_{j,\mathbf{k},+,-}\\
		\mathcal{T}_{j,\mathbf{k},-,+} & \mathcal{T}_{j,\mathbf{k},+,+}
	\end{array}\right).\label{eq:Tjmatrix}
\end{equation}
The self-energy given by Eq. \eqref{eq:self}, for the top layer, can then be recasted as
\begin{equation}
	\Sigma_{+,\mathbf{k},s}^{\left(1\right)}\left(i\omega_{n}\right)=\Sigma_{\mathbf{k},s}^{\mathrm{intra}}\left(i\omega_{n}\right)+\Sigma_{\mathbf{k},s}^{\mathrm{inter}}\left(i\omega_{n}\right),\label{eq:self1}
\end{equation}
where
\begin{align}
	\Sigma_{\mathbf{k},s}^{\mathrm{intra}}\left(i\omega_{n}\right) & =\sum_{j,r}\frac{\left|\mathcal{T}_{j,\mathbf{k},s,r}\right|^{2}}{i\omega_{n}-\epsilon_{\mathbf{k}+\mathbf{q}_{j},r}},\label{eq:self1intra}\\
	\Sigma_{\mathbf{k},s}^{\mathrm{inter}}\left(i\omega_{n}\right) & =\sum_{j,r}\frac{\mathcal{T}_{j,\mathbf{k},s,r}\mathcal{T}_{j,\mathbf{k},-s,r}^{*}}{i\omega_{n}-\epsilon_{\mathbf{k}+\mathbf{q}_{j},r}}\sum_{j,r}\frac{\mathcal{T}_{j,\mathbf{k},-s,r}\mathcal{T}_{j,\mathbf{k},s,r}^{*}}{i\omega_{n}-\epsilon_{\mathbf{k}+\mathbf{q}_{j},r}}\nonumber \\
	& \quad\times\left(i\omega_{n}-\epsilon_{\mathbf{k},-s}-\sum_{j,r}\frac{\left|\mathcal{T}_{j,\mathbf{k},-s,r}\right|^{2}}{i\omega_{n}-\epsilon_{\mathbf{k}+\mathbf{q}_{j},r}}\right)^{-1}.\label{eq:self1inter}
\end{align}
To give a diagrammatic interpretation of these expressions, we first recall the Feynman rules of the system shown in Fig. \ref{fig:Frules}. From it one readily sees that the two irreducible contributions to $\Sigma^{\left(1\right)}$ have a simple physical interpretation:
\begin{itemize}
	\item The term $\Sigma_{\mathbf{k},s}^{\mathrm{intra}}$ accounts for \textit{intraband}
	transitions in which an electron in the state $\left|\mathbf{k},s\right\rangle $ scatters back and forth to the opposite layer only once, returning to the same initial state $\left|\mathbf{k},s\right\rangle $. \item The term $\Sigma_{\mathbf{k},s}^{\mathrm{inter}}$ accounts for all \textit{interband} transitions in which an electron in the state $\left|\mathbf{k},s\right\rangle $ transitions to the opposite band state $\left|\mathbf{k},-s\right\rangle $ by scattering back and forth to the opposite layer, before finally scattering again back to the initial state $\left|\mathbf{k},s\right\rangle $.
\end{itemize}
The corresponding diagrammatic representation is shown in Fig. \ref{fig:sefl1}. The full perturbation calculation that effectively leads to Eq. \eqref{eq:self1} to leading order can be found in the Appendix \ref{app:PerExpansion}. 

The leading order diagrammatic representation of Eq. \eqref{eq:self} suggest that, in general, the self-energy for a shell truncation of the continuum model Hamiltonian is given by the sum of \textit{all} irreducible scattering processes with momentum transfer up to the corresponding cutoff. It is important to note that the perturbation order in the computation of $\Sigma$ (cf. Appendix \ref{app:PerExpansion}) does not generally equal number of shells in the truncated Hamiltonian. This is due to the effect of interband scattering events, which imply higher perturbation orders than intraband ones.

Consider the next truncation of the TBG Hamiltonian at the second shell. The corresponding self-energy $\Sigma{}^{\left(2\right)}$ can be obtained directly from Eq. \eqref{eq:self} by inverting the $20\times20$ truncated Hamiltonian. The resulting expression, however, is highly cumbersome and hard to work with, hindering the inference of physical insight from it. Instead, one can obtain a relatively simple expression for $\Sigma^{\left(2\right)}$ by using the diagrammatic approach, taking into account all the possible irreducible scattering events up to second order in momentum transfer. The full calculation is large and left to the Appendix \ref{app:SEsecond} (see Fig. \ref{fig:A2} for the diagrammatic representation). 

As the number of shells increase, the exact diagrammatic computation of the self-energy gets increasingly harder very rapidly. The situation actually becomes quite complicated beyond the second shell, since then one has to take into account multiple connected paths that an electron can take (see, for instance, the last diagram of $\Sigma$ in Fig. \ref{fig:SelfDiag}). This complexity in the computation of the self-energy, by diagrammatic methods, can be in a way related to the nonlinear matrix dimension increase of the continuum model Hamiltonian as the momentum cutoff increases.

\subsection{Spectral density}\label{subsec:SpectralDensity}

The one-particle spectral density is given by $A_{\gamma}=-\frac{1}{\pi}\mathrm{Im}\mathcal{G}_{\gamma}^{R}$, where $\mathcal{G}_{\gamma}^{R}=\mathcal{G}_{\gamma}\left(i\omega_{n}\rightarrow\omega+i0^+\right)$ is the retarded Green's function (cf. Eq. \eqref{eq:gf2}; we continue to consider the case $\gamma'=\gamma$). The renormalized quasiparticle energies follow from the peaks in the spectral density, which occur when \cite{Coleman2015}
\begin{equation}
	\omega-\epsilon_{\gamma}-\Sigma_{\gamma}^{\prime}\left(\omega\right)=0,\label{eq:Qenergies}
\end{equation}
where $\Sigma^{\prime}\left(\omega\right)=\mathrm{Re}\left[\Sigma\left(i\omega_{n}\rightarrow\omega+i0^{+}\right)\right]$ is the real part of the retarded self-energy. The quasiparticle energies $\omega$ that satisfy the above equation are the same eigenvalues that diagonalize the continuum model Hamiltonian, i.e., the moiré bands. However, each state carries different quasiparticle weight \cite{Coleman2015,Berthod2018}. To see this, we first note that since each quasiparticle state diagonalizes the TBG Hamiltonian, in perturbation terms their lifetime is \emph{infinite}. Consequently, the imaginary part of the self-energy \emph{vanishes} when $\omega=E_{\mathbf{k},\nu}$ [this can be proved rigorously from Eq. \eqref{eq:gf2}]. Thus $A_{\gamma}=\delta\left[\omega-\epsilon_{\gamma}-\Sigma_{\gamma}^{\prime}\left(\omega\right)\right]$, and resolving the delta we obtain 
\begin{equation}
	A_{\gamma}=\sum_{\nu}Z_{\nu,\gamma}\delta\left(\omega-E_{\mathbf{k},\nu}\right),
\end{equation}
where 
\begin{equation}
	Z_{\nu,\gamma}^{-1}=1-\left(\frac{\partial\Sigma_{\gamma}^{\prime}}{\partial\omega}\right)_{\omega=E_{\mathbf{k},\nu}}\label{eq:QW}
\end{equation}
is the quasiparticle weight of the moiré band with energy $E_{\mathbf{k},\nu}$. It can be proved that the normalization condition $\sum_{\nu}Z_{\nu,\gamma}=1$ is satisfied. The quasiparticle weight $Z_{\nu,\gamma}$ \emph{measures} the amplitude of creating a single-particle state with energy $E_{\mathbf{k},\nu}$ from the interaction of the bare electrons with the moiré potential:
\begin{equation}
	Z_{\nu,\gamma}\sim\left|\left\langle \mathbf{k},\nu\right|\hat{c}_{\gamma}^{\dagger}\left|\Omega\right\rangle \right|^{2},\label{eq:QW2}
\end{equation}
where $\left|\Omega\right\rangle $ is the total (\textit{interacting}) ground state and $\left|\mathbf{k},\nu\right\rangle $ is the excited single-particle state with energy $E_{\mathbf{k},\nu}$.

\begin{figure}
	\includegraphics[scale=0.45]{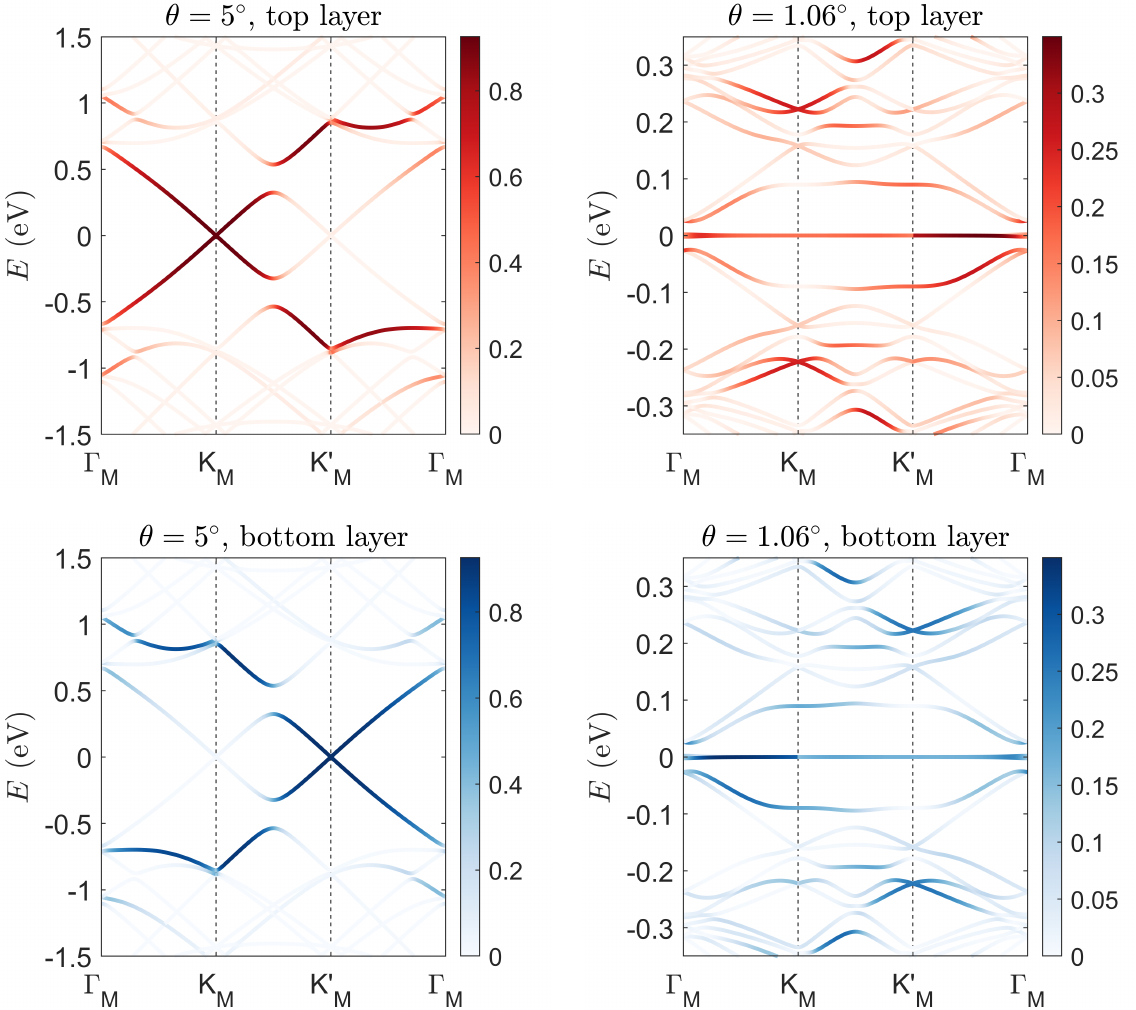}	
	\caption{Quasiparticle weight (QW) of the moiré bands for the $K$ valley, in the top and bottom layers, for twist angles $\theta=5^{\circ}$ and $\theta=1.06^{\circ}$ (magic angle). The QWs are calculated with Eq. \eqref{eq:QW} summing the contribution of electrons in both Dirac bands. At high twist angles the QW is mainly on the lowest moiré bands, while at lower twist angles the QW disperses to remote moiré bands.}\label{fig:QWlayers}
\end{figure}

Fig. \ref{fig:QWlayers} shows the quasiparticle weight (QW) in the top and bottom layers, calculated by summing Eq. \eqref{eq:QW} over both Dirac bands in each case. The strongest QW is always on the lowest (in energy) moiré bands, and around the $K_M$ ($K'_M$) for the top (bottom) layer, consistent with Fig. \ref{fig:TBG}. At relatively high twist angles only low-momentum couplings contribute to the self-energy, so the QW is almost entirely contained in those lowest moiré bands. This means that the quasiparticle renormalization around a Dirac point in one a layer is only influenced by the coupling to the nearest Dirac points in the other layer. At lower twist angles the QW disperses to higher moiré bands, resulting in a significant reduction in the weight of the lowest moiré bands. Around the magic angle ($\theta_{M}\sim1.06^{\circ}$) the QW of the lowest moiré bands are effectively reduced by more than a half, with higher bands having an increasingly higher weight. In addition, the QW of the lowest moiré bands in both layers becomes more uniformly distributed in momentum space, reflecting the high hybridization of the Dirac points. Breaking the layer degeneracy, for instance by an effective mass in one layer arising from a hBN substrate effect \cite{Long2023}, can thus result in a larger split of the narrow moiré bands as the twist angle decreases \cite{Long2022}.

\begin{figure}[t]
	\includegraphics[scale=0.465]{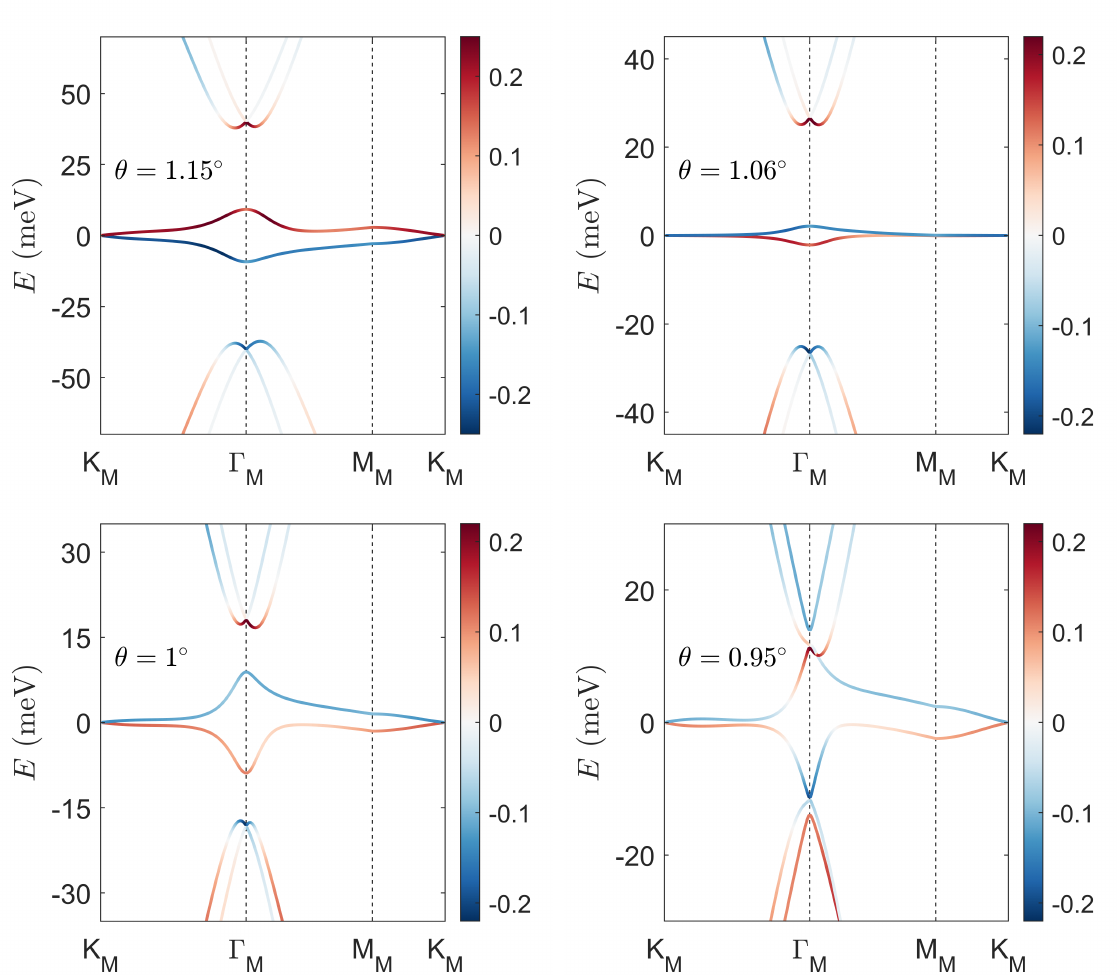}	
	\caption{Difference between the QW contribution of each Dirac band in the top layer, for the low-energy moiré bands in the $K$ valley around the magic angle. Color red (blue) indicate a QW coming mainly from electrons in the top (bottom) Dirac band. An inversion behavior in the lowest moiré bands takes place around the magic angle $\theta=1.06^{\circ}$.}\label{fig:QWdiff}
\end{figure}

By differentiating in each layer the QW contribution of electrons in each Dirac band, one observes that an interesting band-inversion behavior takes place around the magic angle \cite{Yu2023}. This is seen in Fig. \ref{fig:QWdiff}, where we show the QW difference $\Delta Z=Z_{\nu,\mathbf{k},+}-Z_{\nu,\mathbf{k},-}$ in the top layer, for the low-energy moiré bands. For $\theta>\theta_{M}$ the QW of the top (bottom) lowest moiré band comes mainly from electrons in the top (bottom) Dirac band. This behavior is expected for a low moiré coupling, or when interband transitions are nondominant. However, as the moiré bands flatten around the magic angle, a band-inversion behavior takes place, with the QW of the top lowest moiré band now coming mainly from electrons in the bottom Dirac band, and vice versa for the other lowest moiré band. As the twist angle is further decreased, one sees a non-uniform, momentum-dependent QW behavior in the lowest moiré bands, where the weight of electrons in the top Dirac band may dominate around the $\Gamma_M$ point, while that of electrons in the bottom Dirac band may dominate around the $K_M$ point (in the particular case of $\theta=0.95^{\circ}$ in Fig. \ref{fig:QWdiff}, such behavior arises because the remote bands touch the lowest bands at the $\Gamma_M$ point). In general, a band-inversion takes place when two moiré bands touch at a particular momentum $\mathbf{k}$, or when at such point the quasiparticle velocity vanish. This band-inversion behavior has been noted before by projecting the moiré eigenstates on the eigenstates of the decoupled layers \cite{Yu2023}.

The quasiparticle weight in each layer can be experimentally relevant in angle-resolved photoemission spectroscopy (ARPES) of moiré heterostructures \cite{Ohta2012,Nishi2017,Jones2020,Lisi2021}, particularly because then, by design, the topmost layer typically contributes the most to the signal. Within the simplest sudden approximation, the resolved photoemission intensity $I$ can be directly linked to the one-particle spectral density \cite{Lu2012,Berthod2018}
\begin{equation}
	I\sim\mathcal{M}A_{\mathbf{k}}\left(\omega\right)f\left(\omega\right),
\end{equation}
where $f\left(\omega\right)$ is the Fermi-Dirac distribution and $\mathcal{M}$ are the matrix elements accounting for the signal dependence on the incident light intensity, polarization and frequency (see, e.g., Refs. \cite{Pal2013,Amorim2018} for a detailed account). The above equation gives a direct intensity dependence on the spectral density and thus a signal pattern similar to that in Fig. \ref{fig:QWlayers}, which is in relatively good agreement with recent experiments \cite{Li2022,Nunn2023,Jiang2023}. We note that photoemission spectroscopy measurements have been previously explained by a band-unfolding approach \cite{Nishi2017,Li2022}, which in essence is similar to what Eq. \eqref{eq:QW2} measures.

\subsection{Renormalized velocity and flat bands}\label{subsec:Qvelocity}

The emergence of flat moiré bands is one of the most remarkable and intriguing aspects of the continuum model, prompting many works devoted to study their nature and origin \cite{Tarnopolsky2019,Bernevig2021,Wang2023,Ceferino2024}. Within the perturbation scheme developed here, the moiré-induced self-energy provides a direct account for the quasiparticle velocity renormalization and the origin of moiré flat bands. In line with the previous sections, in what follows we will consider the perturbation on electrons in the top Dirac band of the top layer, and for brevity omit those indices in the self-energy. Thus throughout this section it should be understood that $\Sigma{}_{+,\mathbf{k},s}\rightarrow\Sigma{}_{\mathbf{k}}$. The analysis done, however, is valid for both layers and Dirac bands. 

From Eq. \eqref{eq:Qenergies} it follows that the quasiparticle velocity vector $\mathbf{v}_{\mathbf{k},\nu}^{\star}=\nabla_{\mathbf{k}}E_{\mathbf{k},\nu}$ of the moiré band $\nu$ is given by
\begin{equation}
	\mathbf{v}_{\mathbf{k},\nu}^{\star}=Z_{\mathbf{k},\nu}\left[v\hat{\mathbf{k}}+\left(\nabla_{\mathbf{k}}\Sigma'_{\mathbf{k}}\right)_{\omega=E_{\mathbf{k},\nu}}\right],
\end{equation}
where $Z_{\mathbf{k},\nu}$ is the quasiparticle weight given by Eq. \eqref{eq:QW}, and $\nabla_{\mathbf{k}}\epsilon_{\mathbf{k}}=v\hat{\mathbf{k}}$ is the velocity vector in graphene. We are, in particular, interested in the renormalized velocity $\mathbf{v}^{\star}$ of the lowest (in energy) moiré bands at the Dirac point. To leading order in the perturbation, Eq. \eqref{eq:self1} gives $\Sigma'_{\mathbf{k}=0}\left(\omega=0\right)=0$ and $\nabla_{\mathbf{k}}\Sigma'_{\mathbf{k}}\left(k=0=\omega\right)\propto\hat{\mathbf{k}}$, and it can be numerically checked from Eq. \eqref{eq:self} that the same holds at higher orders. Consequently, $\mathbf{v}^{\star}=v^{\star}\hat{\mathbf{k}}$ with
\begin{equation}
	\frac{v^{\star}}{v}=\frac{1+\left(\partial\Sigma'_{\mathbf{k}}/\partial k\right)_{0}/v}{1-\left(\partial\Sigma'_{\mathbf{k}}/\partial\omega\right)_{0}},\label{eq:vstar}
\end{equation} 
where the subindex ``0" implies evaluating the derivatives at $k=0=\omega$. To leading order, Eq. \eqref{eq:self1} gives $\left(\partial\Sigma'_{\mathbf{k}}/\partial k\right)_{0}=-3 v\alpha_{1}^{2}$ and $\left(\partial\Sigma'_{\mathbf{k}}/\partial\omega\right)_{0}=-3\left(\alpha_{0}^{2}+\alpha_{1}^{2}\right)$, where $\alpha_{i}=w_{i}/vk_{\theta}$ are the moiré coupling strengths. Replacing in Eq. \eqref{eq:vstar} then leads to the well-known result $v^{\star}/v=\left(1-3\alpha_{1}^{2}\right)/\left(1+3\alpha_{0}^{2}+3\alpha_{1}^{2}\right)$ when the TBG Hamiltonian is truncated at the first shell \cite{Bistritzer2011,Tarnopolsky2019,Bernevig2021}. The computation of $v^{\star}$, using the self-energy, can be carried out at higher perturbation orders, cf. Appendix \ref{app:QVrenorm}.

\begin{figure}
	\includegraphics[scale=0.4]{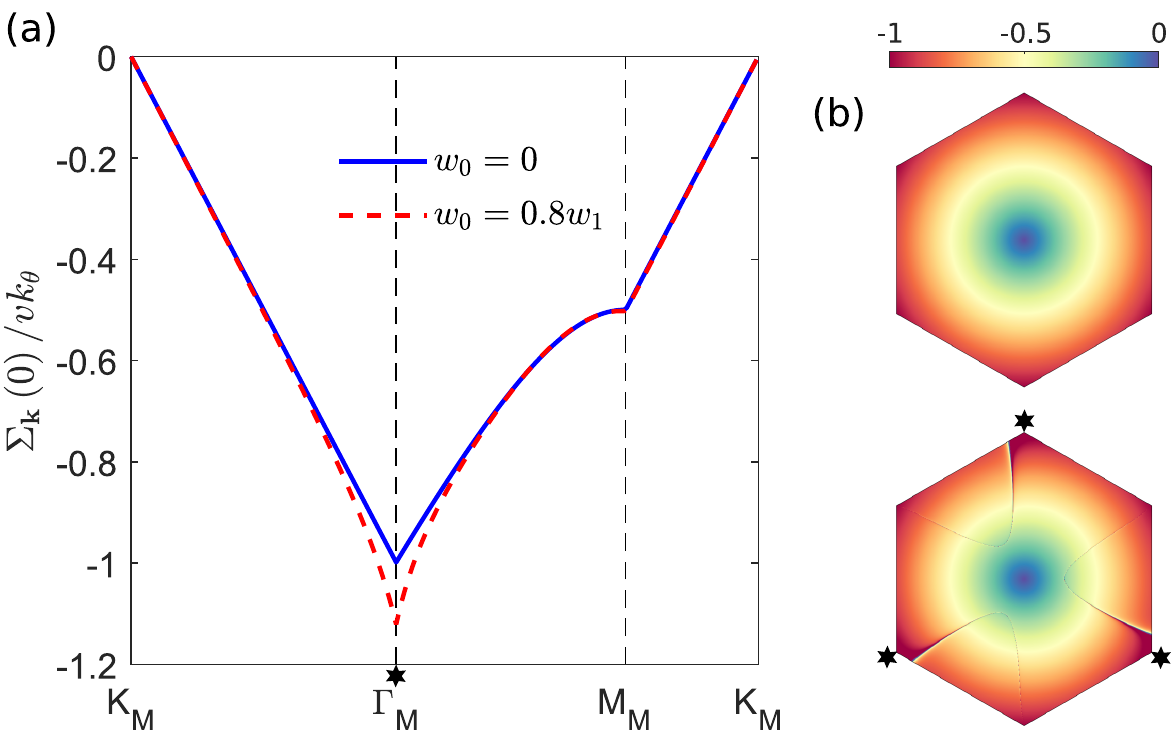}
	\caption{(a) Static self-energy at the magic angle $\theta=1.06^{\circ}$ for $w_{0}=0$ (chiral model) and $w_{0}=0.8w_{1}$. (b) Density plots for the uncoupled dispersion $-\epsilon_{\mathbf{k}}/vk_{\theta}$ (top), and the static self-energy $\Sigma_{\mathbf{k}}\left(0\right)/vk_{\theta}$ at the magic angle (bottom), centered at the Dirac point $K_{M}$. Except for small deviations around the $\Gamma_{M}$ point, the static self-energy converges to $\Sigma_{\mathbf{k}}\left(0\right)\sim-vk$, leading to zero-energy flat bands.}\label{fig:StaticSelf}
\end{figure}

When $v^{\star}$ vanishes the lowest moiré bands actually flatten across the whole moiré BZ \cite{Bistritzer2011,Bernevig2021}. To understand this overall behavior one needs to go beyond the analysis of the quasiparticle velocity at the Dirac point. The self-energy again provides a natural framework for this. Indeed, from Eq. \eqref{eq:Qenergies} it follows that low-energy ($\omega\sim0$) flat bands emerge when the condition
\begin{equation}
	\epsilon_{\mathbf{k}}+\Sigma_{\mathbf{k}}\left(0\right)\sim0\label{eq:FBc}
\end{equation}
is satisfied, which can only occur when the static self-energy behaves as $\Sigma\propto-k$ to leading order in $\mathbf{k}$. 

Considering the self-energy given by Eq. \eqref{eq:self1}, expanding in orders of $k/k_{\theta}$ (where $k_{\theta}=\left|\mathbf{q}_{j}\right|$), at $\omega=0$ one has
\begin{align}
	\Sigma_{\mathbf{k}}^{\left(1\right)}\left(0\right) & \simeq-3vk\left[\alpha_{1}^{2}-\alpha_{0}^{2}\sin\left(3\theta_{\mathbf{k}}\right)\left(k/k_{\theta}\right)\right.\nonumber \\
	& \quad\left.-2\alpha_{0}\alpha_{1}\cos\left(3\theta_{\mathbf{k}}\right)\left(k/k_{\theta}\right)^{2}+\mathcal{O}\left(k/k_{\theta}\right)^{3}\right].\label{eq:self1w0}
\end{align}
Up to order $\sim\left(k/k_{\theta}\right)$, the flat band condition $\epsilon_{\mathbf{k}}+\Sigma_{\mathbf{k}}^{\left(1\right)}\left(0\right)\sim0$ is then satisfied at the first magic angle where $\alpha_{1}^{2}\sim1/3$ and $v^{\star}\sim0$. Since the two leading order corrections in Eq. \eqref{eq:self1w0} scale with $\sim\alpha_0$, the moiré bands become more flat as the AA hopping energies decrease.

Equation \eqref{eq:self1w0} only gives a partial picture at the simplest first shell approximation. At higher perturbation orders (i.e., higher number of shells) one actually sees that at \textit{exactly} the magic angle the self-energy converges to $\Sigma_{\mathbf{k}}\left(0\right)\sim-vk$, only up to small deviations around the $\Gamma_{M}$ point, see Fig. \ref{fig:StaticSelf}. As a result, the uncoupled dispersion $\epsilon_{\mathbf{k}}=vk$ is renormalized to zero-energy moiré bands that satisfy the flat band condition $\epsilon_{\mathbf{k}}+\Sigma_{\mathbf{k}}\left(0\right)\sim0$ over almost the whole moiré Brillouin zone. In the chiral limit one has $\Sigma_{\mathbf{k}}\left(0\right)\sim-vk$ in \emph{all} the mBZ, leading to the formation of  \emph{absolute} flat bands \cite{Tarnopolsky2019}.

\subsection{Origin of flat bands}\label{subsec:OriginFB}

\begin{figure}
	\includegraphics[scale=0.36]{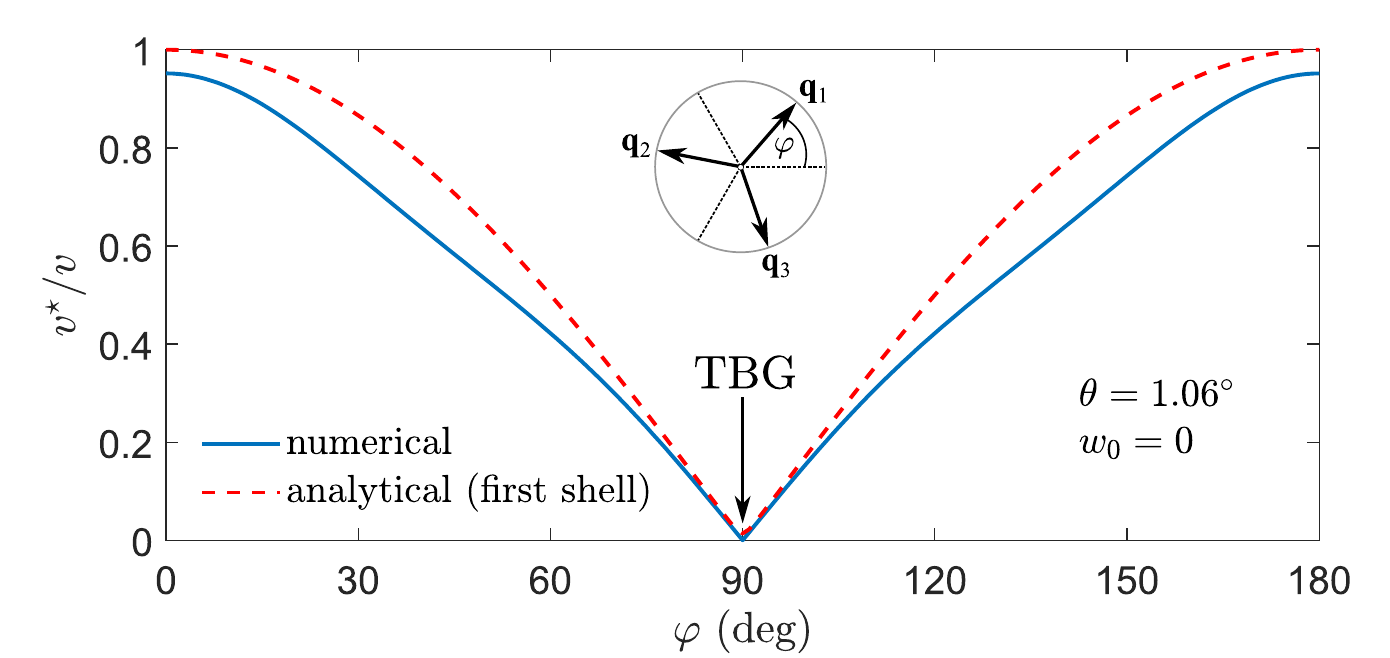}
	\caption{Renormalized Dirac velocity at the first magic angle in the chiral limit, as a function of the orientation $\varphi$ of the momentum transfer vectors $\mathbf{q}_{j}$ relative to the phases $\phi_{j}$ in the moiré potential (see inset). The solid line is the converged numerical result, while the dashed line is the analytical result at the first shell given by Eq. \eqref{eq:VelPhi}. The velocity only vanishes in the TBG case $\varphi=90^{\circ}$. }\label{fig:vphi}
\end{figure}

We now ask what properties of the moiré potential lead to those in the self energy that, in particular, allow the flat band condition \eqref{eq:FBc} to be satisfied. A first point to note is that the behavior of the moiré potential, as given by Eq. \eqref{eq:Tjband}, depends on both the phases $\phi_{j}$ of the moiré matrices and the momentum transfer vectors $\mathbf{q}_{j}$. In the continuum model the phases $\phi_{j}$ arise from the underlying hexagonal symmetry of the layers (technically, from phase factors $\sim e^{i\mathbf{b}\cdot\boldsymbol{\delta}}$ relating the reciprocal lattice vectors $\mathbf{b}$ and the basis vectors $\boldsymbol{\delta}$ of a honeycomb lattice \cite{Lopes dos Santos2007,Bistritzer2011,Moon2013}), and are therefore \emph{independent} of the geometrical properties of the moiré pattern. The momentum transfer vectors $\mathbf{q}_{j}$, on the other hand, depend on the orientation and shape of the moiré BZ \cite{Bi2019,Sinner2023,Koegl2023,Escudero2024}, which are determined by the difference between the reciprocal lattice vectors in each layer. 

To understand how both $\phi_{j}$ and $\mathbf{q}_{j}$ may influence the emergence of flat bands, we consider the leading order self-energy given by Eq. \eqref{eq:self1}, but keeping now explicitly its dependence on those parameters. At the origin point $k=0=\omega$ one has
\begin{equation}
	\Sigma_{\mathbf{0}}^{\left(1\right)}\left(0\right)=-6 vk_{\theta}\alpha_{0}\alpha_{1}\cos\varphi,\label{eq:selfPhi}
\end{equation}
where 
\begin{equation}
	\varphi=\theta_{\mathbf{q}_{j}}-\phi_{j}
\end{equation}
is the $j$-independent difference between the angles $\theta_{\mathbf{q}_{j}}$ of the momentum transfer vectors and the phases $\phi_{j}$. Therefore in general $\Sigma_{\mathbf{0}}\left(0\right)$ only vanishes in the TBG case $\varphi=90^{\circ}$ (see Fig. \ref{fig:TBG}), which implies that only then there is a zero energy solution to Eq. \eqref{eq:Qenergies} at $k=0$; otherwise the Dirac points are shifted in energy. 

The energy shift of the Dirac points modifies the computation of the corresponding quasiparticle velocity $v^{\star}$, since then the derivatives of the self-energy must be evaluated at $\omega$ such that $\omega-\Sigma_{\mathbf{0}}\left(\omega\right)=0$, which in general depends on $\varphi$. The situation simplifies in the particular chiral limit $\alpha_{0}=0$, where Eq. \eqref{eq:selfPhi} gives $\Sigma_{\mathbf{0}}\left(0\right)=0$ for all $\varphi$. In that case, the quasiparticle velocity in terms of $\varphi$ reads (cf. Appendix \ref{app:QVphi})
\begin{equation}
	\frac{v^{\star}}{v}=\frac{\sqrt{1+6\alpha_{1}^{2}\cos\left(2\varphi\right)+9\alpha_{1}^{4}}}{1+3\alpha_{1}^{2}}.\label{eq:VelPhi}
\end{equation}
Crucially, we see that $v^{\star}$ can only vanish in the TBG case where $\varphi=90^{\circ}$. The same behavior is numerically observed at higher perturbation orders, see Fig. \ref{fig:vphi}. 

\begin{figure}[t]
	\includegraphics[scale=0.5]{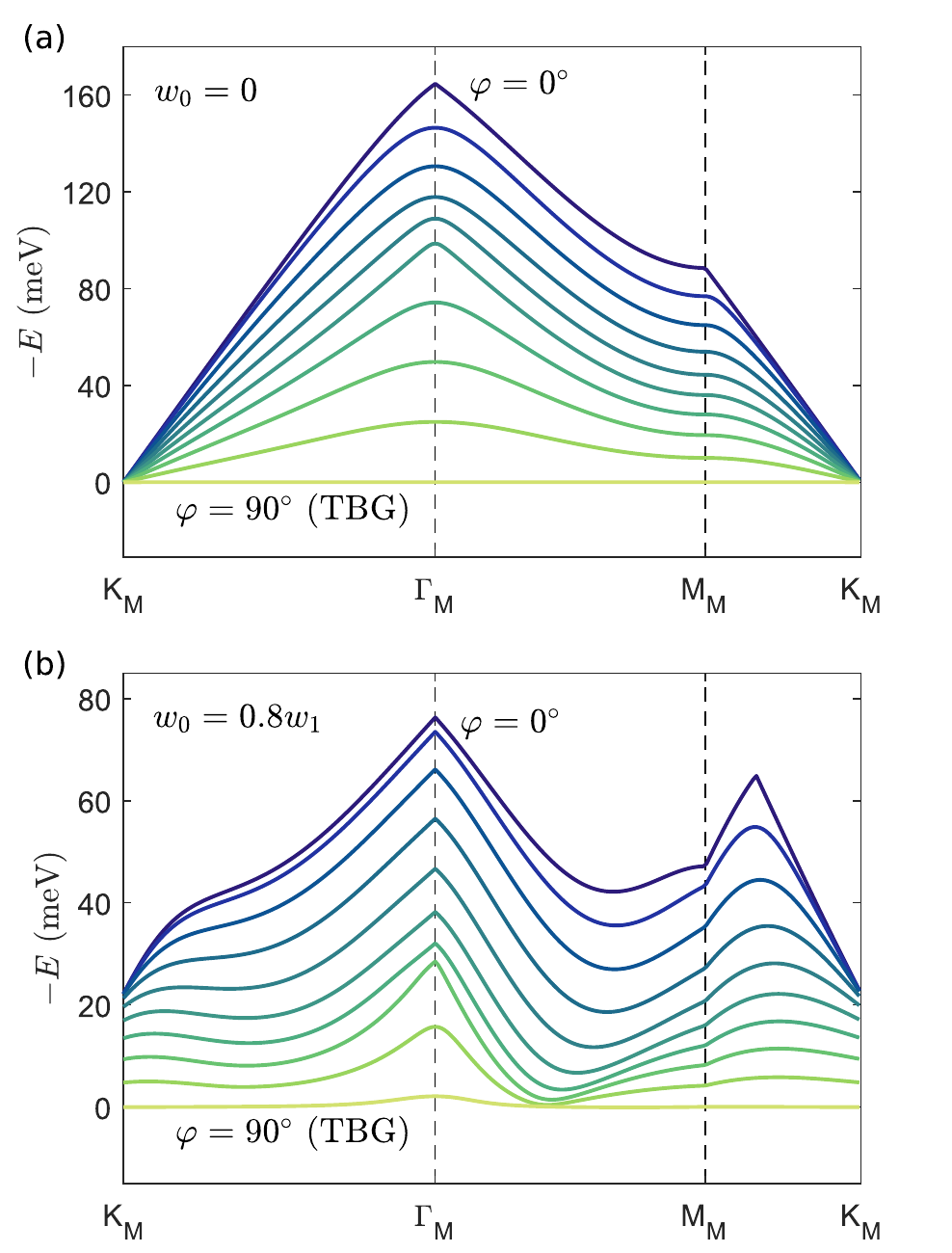}
	\caption{Bottom lowest moiré band for the magic angle $\theta=1.06^{\circ}$ and different orientations $\varphi$ of the momentum transfer vectors in the continuum model, in steps of $10^{\circ}$ from top to bottom (see inset in Fig. \ref{fig:vphi}), for the cases (a) $w_0=0$ (chiral model) and (b) $w_0=0.8w_1$.  Flat bands across the mBZ only occur in the TBG case $\varphi=90^{\circ}$.}\label{fig:CMphi}
\end{figure}

Equations \eqref{eq:selfPhi} and \eqref{eq:VelPhi} put the value of $\varphi$ as a key factor in the origin of magic angles. This observation can be tested by studying the behavior of the moiré bands for different orientations of the momentum transfer vectors $\mathbf{q}_{j}$ in the continuum model. The results for the bottom lowest moiré band is shown in Fig. \ref{fig:CMphi}, for both the chiral and nonchiral cases. In line with Eq. \eqref{eq:VelPhi}, at the magic angle only the TBG case $\varphi=90^{\circ}$ leads to the formation of flat bands across the whole mBZ. This behavior also occurs in the general nonchiral case, where in addition the Dirac points are effectively shifted in energy, leading to a stronger suppression of the flat bands as $\varphi$ changes. 

Figure \ref{fig:CMphi}, and in particular Eq. \eqref{eq:VelPhi}, incidentally give a simple explanation as to why flat bands in TBG seem to disappear in the presence of strain \cite{Huder2018,Bi2019,Escudero2024}, whose effect generally modifies the momentum transfer vectors. Even in those cases in which the hexagonal symmetry of the moiré pattern in preserved, as e.g. with a combination of twist and biaxial strain \cite{Escudero2024}, the rotation of the wave vectors $\mathbf{q}_j$ would generally suppress the formation of flat bands.    

\section{Conclusions}\label{sec:Conclusions}

We have developed a diagrammatic perturbation approach to study the properties of moiré bands in twisted bilayer graphene. This approach differs from previous studies on several aspects. First, it is build upon a perturbation of the two uncoupled Dirac bands in each monolayer. This provides a natural framework to deploy perturbation methods and gives a direct physical picture of the moiré-induced scattering processes. Second, our approach treats the perturbation by the moiré potential within the full framework of a many-body problem. In this scheme, the moiré potential acts as an effective one-body potential that transfers electrons from one layer to another, in all possible Dirac bands, by the exchange of the three wave vectors that determine the moiré Brillouin zone. The strength of this coupling is naturally accounted by the value of the twist angle; namely, the decreasing length of the momentum transfer vectors as the twist angle is lowered translates to an increase in the coupling between the Dirac points, and therefore in the perturbation strength. 

The developed perturbation approach leads to a direct diagrammatic representation of the scattering processes taking place. By a set of simple rules, one can obtain, describe or approximate, the moiré-induced self-energy of the electrons, and from it the moiré bands of interest. This not only brings the full power of Feynman diagrams into the problem, but also, and perhaps more importantly, provides a physical picture of the relevant scattering events. Furthermore, since the perturbation problem is quadratic in operators, one can naturally obtain, through a straightforward matrix diagonalization, a closed expression for the Green's function. As a result, there is a direct relation between the moiré bands properties at a given shell cutoff of the continuum model, and the diagrammatic expansion of the Dyson equation. 

The moiré-induced self-energy can be related to different one-particle properties of interest. In particular, we have obtained the quasiparticle weight of the moiré bands for electrons in each layer and Dirac band. The layer distinction can be particularly useful for interpreting photoemission spectroscopy experiments in moiré heterostructures, where usually the topmost layer gives the larger contribution to the measured signal. In line with previous studies, the perturbation approach naturally yields an increasingly weight of higher moiré bands as the twist angle decreases, reflecting the contribution of multiple high-momentum scatterings events at successive Dirac points. Around the magic angle, an interesting band-inversion behavior takes place: The contribution of electrons in the top and bottom Dirac bands to the quasiparticle weight of the flat bands is inverted.  

The self-energy also gives a natural account for the emergence of flat bands in twisted bilayer graphene. We have shown that exactly at the magic angle the static self-energy practically cancels out the uncoupled linear dispersion in each layer, except only for small deviations around the $\Gamma_M$ point in the nonchiral case, leading to zero-energy flat bands over almost the whole moiré Brillouin zone. By analyzing how the moiré potential influences such behavior in the self-energy, we have further shown that flat bands can emerge only under a particular orientation of the momentum transfer vectors, which is indeed satisfied in twisted bilayer graphene. For different orientations of the momentum transfer vectors the Dirac velocity  can never vanish. The geometrical properties of twisted bilayer graphene thus seem to play a crucial role in the formation of flat bands.   

Finally, we note that the developed diagrammatic approach has only taken into account the moiré potential as a perturbation. A more realistic treatment should also include electronic correlations, which can indeed become significant as the moiré bands flatten around the magic angle. Electron-electron interactions can be included in our model as an additional perturbation to the system, competing with that of the moiré potential. Both perturbations may then be tackled diagrammatically, in a similar approach as the one developed here. It would be interesting to study the competition of both perturbations as the twist angle decreases, and how this influences the properties of the moiré bands. We leave this for future work.

\begin{acknowledgments}
	I thank Pierre A. Pantaleón, Juan Sebastián Ardenghi and Francisco Guinea for discussions. This paper was partially supported by the CONICET (Argentina National Research Council) and Universidad Nacional del Sur (UNS). F. E. acknowledges a research fellowship from CONICET.
\end{acknowledgments}

\onecolumngrid
\appendix

\section{Continuum model Hamiltonian in the band basis}\label{app:CMband}

Here we give details on the construction of the TBG Hamiltonian in the band basis. We adopt the original Tripod model \cite{Bistritzer2011,Bernevig2021}, with the origin set a Dirac point (cf. Fig. \ref{fig:shell}). Following the convention in the main text, we take the origin Dirac point belonging to the top layer.

The continuum model Hamiltonian in the band basis, which we shall denote as $H_{B}$, can be directly obtained from the usual one in the sublattice basis $H$, Eq. \eqref{eq:H}, by the similarity transformation 
\begin{equation}
	H_{B}=\mathcal{C}^{-1}H_{\mathrm{}}\mathcal{C},\label{eq:Sim}
\end{equation}
where $\mathcal{C}$ is the matrix that diagonalizes the Dirac Hamiltonian $H_{0}$ (i.e., that without the moiré potential coupling). In momentum space $\mathcal{C}$ is a matrix whose columns are thus given by the eigenstates of $H_{0}$, 
\begin{equation}
	\mathcal{C}=\left(\begin{array}{ccccc}
		\Psi_{\mathbf{k}} & 0 & 0 & 0 & \cdots\\
		0 & \Psi_{\mathbf{k}+\mathbf{q}_{1}} & 0 & 0 & \cdots\\
		0 & 0 & \Psi_{\mathbf{k}+\mathbf{q}_{2}} & 0 & \cdots\\
		0 & 0 & 0 & \Psi_{\mathbf{k}+\mathbf{q}_{3}} & \cdots\\
		\vdots & \vdots & \vdots & \vdots & \ddots
	\end{array}\right),\label{eq:Csim}
\end{equation}
where [cf. Eq. \eqref{eq:psi}]
\begin{equation}
	\Psi_{\mathbf{k}}=\left(\begin{array}{cc}
		\psi_{\mathbf{k},-} & \psi_{\mathbf{k},+}\end{array}\right)=\frac{1}{\sqrt{2}}\left(\begin{array}{cc}
		1 & 1\\
		-e^{i\theta_{\mathbf{k}}} & e^{i\theta_{\mathbf{k}}}
	\end{array}\right).\label{eq:Mpsi}
\end{equation}
Note that $\mathcal{C}^{-1}=\mathcal{C}^{\dagger}$. The similarity transformation
\eqref{eq:Sim} leads to the moiré matrices in the band basis:
\begin{equation}
	\left\langle \mathbf{k}_{-} \right\vert \hat{T}_{j} \left\vert \mathbf{k}_{+}\right\rangle_B =\Psi_{\mathbf{k}_{-}}^{\dagger}T_{j}\Psi_{\mathbf{k}_{+}}\delta_{\mathbf{k}_{-},\mathbf{k}_{+}+\mathbf{q}_{j}},
\end{equation}
where $T_{j}$ is given by Eq. \eqref{eq:Tj}. The matrix components in the above equation lead to Eq. \eqref{eq:Tjband} in the main text. The above result can be also obtained by making a direct basis change of the moiré matrix elements, using the relation $\left|\mathbf{k},s\right\rangle =\left(\left|A\right\rangle +se^{i\theta_{\mathbf{k}}}\left|B\right\rangle \right)/\sqrt{2}$ between the band basis and the sublattice basis.  

The computation of the Green's function through Eq. \eqref{eq:gf2} is done by a matrix inversion of the continuum model Hamiltonian in the band basis. Given the relation \eqref{eq:Sim}, this inversion can be also directly obtained from the original TBG Hamiltonian $H$, Eq. \eqref{eq:H} (in momentum space), and a basis change:
\begin{equation}
	\mathcal{G}\left(i\omega_{n}\right)=\mathcal{C}^{\dagger}\left(i\omega_{n}\mathbb{I}-H\right)^{-1}\mathcal{C},\label{eq:gfC}
\end{equation}
In particular, for the two cases $\gamma=\gamma'=\left(+,\mathbf{k},s=\pm\right)$, the matrix components of $\left(i\omega_{n}\mathbb{I}-H\right)_{\gamma\gamma}^{-1}$, given our convention, are $\gamma\gamma=\left(1,1\right)$ and $\gamma\gamma=\left(2,2\right)$ for the bottom $\left(s=-1\right)$ and top $\left(s=1\right)$ Dirac band, respectively. Defining 
\begin{equation}
	\tilde{H}\left(i\omega_{n}\right)=i\omega_{n}\mathbb{I}-H,
\end{equation}
we thus have
\begin{align}
	\mathcal{G}_{+,\mathbf{k},-}\left(i\omega_{n}\right) & =\mathcal{C}_{1\mu}^{\dagger}\tilde{H}_{\mu\nu}^{-1}\left(i\omega_{n}\right)\mathcal{C}_{\nu1},\\
	\mathcal{G}_{+,\mathbf{k},+}\left(i\omega_{n}\right) & =\mathcal{C}_{2\mu}^{\dagger}\tilde{H}_{\mu\nu}^{-1}\left(i\omega_{n}\right)\mathcal{C}_{\nu2},
\end{align}
where repeated indices are to be summed over. Here it should be understand that $\tilde{H}_{\mu\nu}^{-1}\left(i\omega_{n}\right)$ are the $\left(\mu,\nu\right)$ components of $\tilde{H}^{-1}\left(i\omega_{n}\right)$. From Eqs. \eqref{eq:Csim} and \eqref{eq:Mpsi} it follows that
\begin{align}
	\mathcal{G}_{+,\mathbf{k},\pm}\left(i\omega_{n}\right) & =\frac{1}{2}\left(\tilde{H}_{11}^{-1}+\tilde{H}_{22}^{-1}\pm\tilde{H}_{12}^{-1}e^{i\theta_{\mathbf{k}}}\pm\tilde{H}_{21}^{-1}e^{-i\theta_{\mathbf{k}}}\right).\label{eq:gfCk}
\end{align}
This provides a direct way to obtain the Green's function by working entirely in the sublattice basis. From it we readily see that
\begin{align}
	\mathcal{G}_{+,\mathbf{k},+}\left(i\omega_{n}\right)+\mathcal{G}_{+,\mathbf{k},-}\left(i\omega_{n}\right) & =\tilde{H}_{11}^{-1}+\tilde{H}_{22}^{-1},\\
	\mathcal{G}_{+,\mathbf{k},+}\left(i\omega_{n}\right)-\mathcal{G}_{+,\mathbf{k},-}\left(i\omega_{n}\right) & =\tilde{H}_{12}^{-1}e^{i\theta_{\mathbf{k}}}+\tilde{H}_{21}^{-1}e^{-i\theta_{\mathbf{k}}}.
\end{align}
It should be noted that the dependence of $\mathcal{G}$ on the moiré potential $\mathcal{T}_{j,\mathbf{k},s,s'}$ in the band basis, when computed using Eq. \eqref{eq:gfCk}, would not be explicit; it would rather be implicitly contained in the couplings between the Hamiltonian $\sim\boldsymbol{\sigma}\cdot\mathbf{k}$ of different Dirac points. From a diagrammatic point of view, and physical interpretation, it is thus more convenient to compute $\mathcal{G}$ by directly working in the band representation.

\section{Perturbation expansion of the moiré potential}\label{app:PerExpansion}

By treating the moiré potential $\hat{T}$, Eq. \eqref{eq:Tpot}, as a perturbation, the Green's function given by Eq. \eqref{eq:gf}  can be expanded as \cite{Mahan1990,Jishi2013,Coleman2015} (we set $\hbar=1$)
\begin{equation}
	\mathcal{G}_{\gamma\gamma'}\left(\tau\right)=-\left\langle T_{\tau}\hat{c}_{\gamma}\left(\tau\right)\hat{c}_{\gamma'}^{\dagger}\left(0\right)\hat{U}\left(\beta,0\right)\right\rangle _{0,c},\label{eq:gfApp}
\end{equation}
where $\beta=1/k_{B}T$ and $\left\langle \cdots\right\rangle _{0,c}$ is the ensemble average over the noninteracting system, taking into account only the contribution of connected diagrams. The evolution operator reads
\begin{equation}
	\hat{U}\left(\tau,\tau'\right)=\sum_{l=0}^{\infty}\frac{1}{l!}\left(-1\right)^{l}\int_{\tau'}^{\tau}d\tau_{1}\ldots\int_{\tau'}^{\tau}d\tau_{l}T_{\tau}\left[\hat{T}\left(\tau_{1}\right)\ldots\hat{T}\left(\tau_{l}\right)\right].\label{eq:Gper}
\end{equation}
The perturbation expansion of the evolution operator naturally accounts for the shell truncation of the continuum model Hamiltonian. Following Sec. \ref{sec:Perturbation}, we are interested in the case $\gamma=\gamma'$ in Eq. \eqref{eq:gfApp}; we note $\mathcal{G}_{\gamma\gamma}\left(\tau\right)=\mathcal{G}_{\gamma}\left(\tau\right)$. In that situation, since the moiré potential $\hat{T}$ always exchanges one electron from one layer to another, only even orders in $l$ contribute to the expansion of the evolution operator. For conciseness we shall consider the particular case $\gamma=\left(+,\mathbf{k},s\right)$. The leading order correction to the Green's function (second order in $l$) reads 
\begin{equation}
	\mathcal{G}_{+,\mathbf{k},s}^{\left(2\right)}\left(\tau\right)=-\frac{1}{2!}\left(-1\right)^{2}\int_{0}^{\beta}d\tau_{1}\int_{0}^{\beta}d\tau_{2}\left\langle T_{\tau}\hat{c}_{+,\mathbf{k},s}\left(\tau\right)c_{+,\mathbf{k},s}^{\dagger}\left(0\right)\hat{T}\left(\tau_{1}\right)\hat{T}\left(\tau_{2}\right)\right\rangle _{0,c}.
\end{equation}
Replacing Eq. \eqref{eq:Tpot} leads to the following four kinds of ensemble averages:
\begin{align}
	C_{1} & =\left\langle T_{\tau}\hat{c}_{+,\mathbf{k},s}\left(\tau\right)\hat{c}_{+,\mathbf{k},s}^{\dagger}\left(0\right)\hat{c}_{-,\mathbf{k}_{1}+\mathbf{q}_{j_{1}},s_{1}'}^{\dagger}\left(\tau_{1}\right)\hat{c}_{+,\mathbf{k}_{1},s_{1}}\left(\tau_{1}\right)\hat{c}_{-,\mathbf{k}_{2}+\mathbf{q}_{j_{2}},s_{2}'}^{\dagger}\left(\tau_{2}\right)\hat{c}_{+,\mathbf{k}_{2},s_{2}}\left(\tau_{2}\right)\right\rangle _{0,c},\\
	C_{2} & =\left\langle T_{\tau}\hat{c}_{+,\mathbf{k},s}\left(\tau\right)\hat{c}_{+,\mathbf{k},s}^{\dagger}\left(0\right)\hat{c}_{-,\mathbf{k}_{1}+\mathbf{q}_{j_{1}},s_{1}'}^{\dagger}\left(\tau_{1}\right)\hat{c}_{+,\mathbf{k}_{1},s_{1}}\left(\tau_{1}\right)\hat{c}_{+,\mathbf{k}_{2},s_{2}}^{\dagger}\left(\tau_{2}\right)\hat{c}_{-,\mathbf{k}_{2}+\mathbf{q}_{j_{2}},s_{2}'}\left(\tau_{2}\right)\right\rangle _{0,c},\\
	C_{3} & =\left\langle T_{\tau}\hat{c}_{+,\mathbf{k},s}\left(\tau\right)\hat{c}_{+,\mathbf{k},s}^{\dagger}\left(0\right)\hat{c}_{+,\mathbf{k}_{1},s_{1}}^{\dagger}\left(\tau_{1}\right)\hat{c}_{-,\mathbf{k}_{1}+\mathbf{q}_{j_{1}},s_{1}'}\left(\tau_{1}\right)\hat{c}_{-,\mathbf{k}_{2}+\mathbf{q}_{j_{2}},s_{2}'}^{\dagger}\left(\tau_{2}\right)\hat{c}_{+,\mathbf{k}_{2},s_{2}}\left(\tau_{2}\right)\right\rangle _{0,c},\\
	C_{4} & =\left\langle T_{\tau}\hat{c}_{+,\mathbf{k},s}\left(\tau\right)\hat{c}_{+,\mathbf{k},s}^{\dagger}\left(0\right)\hat{c}_{+,\mathbf{k}_{1},s_{1}}^{\dagger}\left(\tau_{1}\right)\hat{c}_{-,\mathbf{k}_{1}+\mathbf{q}_{j_{1}},s_{1}'}\left(\tau_{1}\right)\hat{c}_{+,\mathbf{k}_{2},s_{2}}^{\dagger}\left(\tau_{2}\right)\hat{c}_{-,\mathbf{k}_{2}+\mathbf{q}_{j_{2}},s_{2}'}\left(\tau_{2}\right)\right\rangle _{0,c}.
\end{align}
Applying Wick's theorem, considering the anticommutation relation $\left\{ \hat{c}_{\ell,\mathbf{k},s},\hat{c}_{\ell',\mathbf{k}',s'}^{\dagger}\right\} =\delta_{\ell,\ell'}\delta_{\mathbf{k},\mathbf{k}'}\delta_{s,s'}$ (all others anticommutators vanish), and taking into account only those contractions that represent connected diagrams, yields $C_{1}=0=C_{4}$ and
\begin{align}
	C_{2} & =-\mathcal{G}_{\mathbf{k},s}^{0}\left(\tau-\tau_{2}\right)\delta_{\mathbf{k},\mathbf{k}_{2}}\delta_{s,s_{2}}\mathcal{G}_{\mathbf{k},s}^{0}\left(\tau_{1}\right)\delta_{\mathbf{k},\mathbf{k}_{1}}\delta_{s,s_{1}}\mathcal{G}_{\mathbf{k}+\mathbf{q}_{j_{1}},s_{1}'}^{0}\left(\tau_{2}-\tau_{1}\right)\delta_{\mathbf{k}_{2}+\mathbf{q}_{j_{2}},\mathbf{k}_{1}+\mathbf{q}_{j_{1}}}\delta_{s'_{2},s'_{1}},\\
	C_{3} & =-\mathcal{G}_{\mathbf{k},s}^{0}\left(\tau-\tau_{1}\right)\delta_{\mathbf{k},\mathbf{k}_{1}}\delta_{s,s_{1}}\mathcal{G}_{\mathbf{k},s}^{0}\left(\tau_{2}\right)\delta_{\mathbf{k},\mathbf{k}_{2}}\delta_{s,s_{2}}\mathcal{G}_{\mathbf{k}+\mathbf{q}_{j_{2}},s_{2}'}^{0}\left(\tau_{1}-\tau_{2}\right)\delta_{\mathbf{k}_{2}+\mathbf{q}_{j_{2}},\mathbf{k}_{1}+\mathbf{q}_{j_{1}}}\delta_{s'_{2},s'_{1}},
\end{align}
where
\begin{equation}
	\left\langle T_{\tau}\hat{c}_{\ell,\mathbf{k},s}\left(\tau\right)\hat{c}_{\ell',\mathbf{k}',s'}^{\dagger}\left(\tau'\right)\right\rangle _{0}=-\delta_{\ell,\ell'}\delta_{\mathbf{k},\mathbf{k}'}\delta_{s,s'}\mathcal{G}_{\mathbf{k},s}^{0}\left(\tau-\tau'\right)
\end{equation}
is the noninteracting Green's function. Both $C_{2}$ and $C_{3}$ give topologically equivalent diagrams. Thus 
\begin{equation}
	\mathcal{G}_{+,\mathbf{k},s}^{\left(2\right)}\left(\tau\right)=\sum_{j,s'}\left|\mathcal{T}_{j,\mathbf{k},s,s'}\right|^{2}\int_{0}^{\beta}d\tau_{1}\int_{0}^{\beta}d\tau_{2}\mathcal{G}_{\mathbf{k},s}^{0}\left(\tau-\tau_{2}\right)\mathcal{G}_{\mathbf{k},s}^{0}\left(\tau_{1}\right)\mathcal{G}_{\mathbf{k}+\mathbf{q}_{j},s'}^{0}\left(\tau_{2}-\tau_{1}\right).
\end{equation}
Using the Fourier transform
\begin{equation}
	\mathcal{G}_{\mathbf{k},s}^{0}\left(\tau\right)=\frac{1}{\beta}\sum_{n}\mathcal{G}_{\mathbf{k},s}^{0}\left(i\omega_{n}\right)e^{-i\omega_{n}\tau},\qquad\omega_{n}=\left(2n+1\right)\pi/\beta,
\end{equation}
and the relation 
\begin{equation}
	\int_{0}^{\beta}d\tau\exp\left[i\left(\omega_{n_{j}}-\omega_{n_{j'}}\right)\tau\right]=\beta\delta_{\omega_{n_{j}},\omega_{n_{j'}}},
\end{equation}
we get
\begin{equation}
	\mathcal{G}_{+,\mathbf{k},s}^{\left(2\right)}\left(i\omega_{n}\right)=\mathcal{G}_{\mathbf{k},s}^{0}\left(i\omega_{n}\right)\mathcal{G}_{\mathbf{k},s}^{0}\left(i\omega_{n}\right)\sum_{j,s'}\left|\mathcal{T}_{j,\mathbf{k},s,s'}\right|^{2}\mathcal{G}_{\mathbf{k}+\mathbf{q}_{j},s'}^{0}\left(i\omega_{n}\right),\label{eq:gf2App}
\end{equation}
where
\begin{equation}
	\mathcal{G}_{\mathbf{k},s}^{0}\left(i\omega_{n}\right)=\frac{1}{i\omega_{n}-\epsilon_{\mathbf{k},s}}.
\end{equation}
Since the perturbation is quadratic in the operators, there is no dependence on the statistic of the particles. One readily sees that Eq. \eqref{eq:gf2App} represents the leading correction to the Green's function coming from the \emph{intraband} contribution to the leading order self-energy, cf. Eq. \eqref{eq:self1intra} (the other \textit{interband} contribution comes from higher orders in the perturbation expansion). 

From the above derivation one can easily infer the Feynman rules for the system, see Fig. \ref{fig:Frules}. As an example of applying the Feynman rules, the contribution to the self-energy of the five diagrams in Fig. \ref{fig:SelfDiag} (in the same order as shown) read
\begin{align}
	\Sigma_{\mathbf{k},s}^{\left(l=2\right)}\left(i\omega_{n}\right) & =\sum_{j}\sum_{r}\mathcal{T}_{j,\mathbf{k},s,r}\mathcal{T}_{j,\mathbf{k},s,r}^{*}\mathcal{G}_{\mathbf{k}+\mathbf{q}_{j},r}^{0}\left(i\omega_{n}\right),\\
	\Sigma_{\mathbf{k},s}^{\left(l=4,a\right)}\left(i\omega_{n}\right) & =\sum_{j}\sum_{j'}\sum_{r_{1},r_{2}}\mathcal{T}_{j,\mathbf{k},s,r_{1}}\mathcal{T}_{j,\mathbf{k},-s,r_{1}}^{*}\mathcal{T}_{j',\mathbf{k},-s,r_{2}}\mathcal{T}_{j',\mathbf{k},s,r_{2}}^{*}\mathcal{G}_{\mathbf{k}+\mathbf{q}_{j},r_{1}}^{0}\left(i\omega_{n}\right)\mathcal{G}_{\mathbf{k},-s}^{0}\left(i\omega_{n}\right)\mathcal{G}_{\mathbf{k}+\mathbf{q}_{j'},r_{2}}^{0}\left(i\omega_{n}\right),\\
	\Sigma_{\mathbf{k},s}^{\left(l=4,b\right)}\left(i\omega_{n}\right) & =\sum_{j}\sum_{j'\neq j}\sum_{r_{1},r_{2}}\sum_{s_{1}}\mathcal{T}_{j,\mathbf{k},s,r_{1}}\mathcal{T}_{j',\mathbf{k}+\mathbf{q}_{j}-\mathbf{q}_{j'},s_{1},r_{1}}^{*}\mathcal{T}_{j',\mathbf{k}+\mathbf{q}_{j}-\mathbf{q}_{j'},s_{1},r_{2}}\mathcal{T}_{j,\mathbf{k},s,r_{2}}^{*}\nonumber \\
	& \qquad\times\mathcal{G}_{\mathbf{k}+\mathbf{q}_{j},r_{1}}^{0}\left(i\omega_{n}\right)\mathcal{G}_{\mathbf{k}+\mathbf{q}_{j}-\mathbf{q}_{j'},s_{1}}^{0}\left(i\omega_{n}\right)\mathcal{G}_{\mathbf{k}+\mathbf{q}_{j},r_{2}}^{0}\left(i\omega_{n}\right),\\
	\Sigma_{\mathbf{k},s}^{\left(l=6,a\right)}\left(i\omega_{n}\right) & =\sum_{j}\sum_{j'\neq j}\sum_{j''\neq j'}\sum_{r_{1},r_{2},r_{3}}\sum_{s_{1},s_{2}}\mathcal{T}_{j,\mathbf{k},s,r_{1}}\mathcal{T}_{j',\mathbf{k}+\mathbf{q}_{j}-\mathbf{q}_{j'},s_{1},r_{1}}^{*}\mathcal{T}_{j'',\mathbf{k}+\mathbf{q}_{j}-\mathbf{q}_{j'},s_{1},r_{2}}\nonumber \\
	& \times\mathcal{T}_{j'',\mathbf{k}+\mathbf{q}_{j}-\mathbf{q}_{j'},s_{2},r_{2}}^{*}\mathcal{T}_{j',\mathbf{k}+\mathbf{q}_{j}-\mathbf{q}_{j'},s_{2},r_{3}}\mathcal{T}_{j,\mathbf{k},s,r_{3}}^{*}\nonumber \\
	& \times\mathcal{G}_{\mathbf{k}+\mathbf{q}_{j},r_{1}}^{0}\left(i\omega_{n}\right)\mathcal{G}_{\mathbf{k}+\mathbf{q}_{j}-\mathbf{q}_{j'},s_{1}}^{0}\left(i\omega_{n}\right)\mathcal{G}_{\mathbf{k}+\mathbf{q}_{j}-\mathbf{q}_{j'}+\mathbf{q}_{j''},r_{2}}^{0}\left(i\omega_{n}\right)\mathcal{G}_{\mathbf{k}+\mathbf{q}_{j}-\mathbf{q}_{j'},s_{2}}^{0}\left(i\omega_{n}\right)\mathcal{G}_{\mathbf{k}+\mathbf{q}_{j},r_{3}}^{0}\left(i\omega_{n}\right),\\
	\Sigma_{\mathbf{k},s}^{\left(l=6,b\right)}\left(i\omega_{n}\right) & =\sum_{j}\sum_{j'\neq j}\sum_{j''\neq j,j'}\sum_{r_{1},r_{2},r_{3}}\sum_{s_{1},s_{2}}\mathcal{T}_{j,\mathbf{k},s,r_{1}}\mathcal{T}_{j',\mathbf{k}+\mathbf{q}_{j}-\mathbf{q}_{j'},s_{1},r_{1}}^{*}\mathcal{T}_{j'',\mathbf{k}+\mathbf{q}_{j}-\mathbf{q}_{j'},s_{1},r_{2}}\nonumber \\
	& \times\mathcal{T}_{j,\mathbf{k}-\mathbf{q}_{j'}+\mathbf{q}_{j''},s_{2},r_{2}}^{*}\mathcal{T}_{j',\mathbf{k}-\mathbf{q}_{j'}+\mathbf{q}_{j''},s_{2},r_{3}}\mathcal{T}_{j'',\mathbf{k},s,r_{3}}^{*}\nonumber \\
	& \times\mathcal{G}_{\mathbf{k}+\mathbf{q}_{j},r_{1}}^{0}\left(i\omega_{n}\right)\mathcal{G}_{\mathbf{k}+\mathbf{q}_{j}-\mathbf{q}_{j'},s_{1}}^{0}\left(i\omega_{n}\right)\mathcal{G}_{\mathbf{k}+\mathbf{q}_{j}-\mathbf{q}_{j'}+\mathbf{q}_{j''},r_{2}}^{0}\left(i\omega_{n}\right)\mathcal{G}_{\mathbf{k}-\mathbf{q}_{j'}+\mathbf{q}_{j''},s_{2}}^{0}\left(i\omega_{n}\right)\mathcal{G}_{\mathbf{k}+\mathbf{q}_{j''},r_{3}}^{0}\left(i\omega_{n}\right).
\end{align}

\section{Diagrammatic computation of the self-energy at second shell}\label{app:SEsecond}

As noted in the main text, the self-energy for a momentum truncation at the second shell (see Fig. \ref{fig:shell}) can be directly computed by inverting the corresponding $20\times20$ TBG Hamiltonian, but this leads to a highly cumbersome expression, which is hard to work with and gives little physical insight. For this reason, here we rather compute the self-energy by means of only diagrammatic methods. This approach highlights the involved scattering processes and provides a workable analytical expression from which one can further study different properties of the moiré bands, such as the quasiparticle velocity renormalization (Appendix \ref{app:QVrenorm}).

\begin{figure*}
	\includegraphics[scale=0.8]{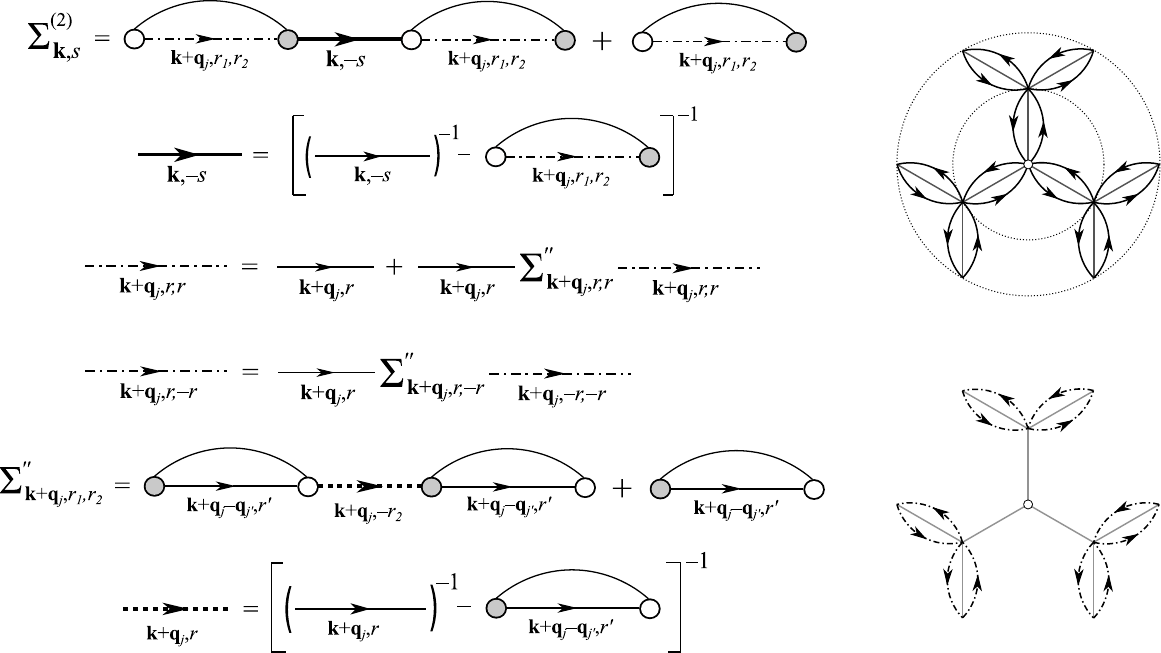}	
	\caption{Left: Diagrammatic representation of the moiré-induced self-energy in the top layer, up to a momentum truncation at the second shell (cf. Fig. \ref{fig:shell}). Dot-dashed Green's functions represent the dressed propagator for an electron in the bottom layer, with momentum $\mathbf{k}+\mathbf{q}_{j}$, by the coupling to an electron in the top layer with momentum $\mathbf{k}+\mathbf{q}_{j}-\mathbf{q}_{j'}$, where $\mathbf{q}_{j'}\neq\mathbf{q}_{j}$. Right: schematic hexagonal picture of all the scattering processes up to a second shell truncation (top panel), and the dressed dot-dashed Green's function (bottom panel).}\label{fig:A2}
\end{figure*}

We consider, as in the main text, the self-energy for the top layer. The diagrammatic representation that accounts for all scattering processes with momentum exchange up to $\Delta\mathbf{k}\sim\left|\mathbf{q}_{1}-\mathbf{q}_{2}\right|$ (second shell in Fig. \ref{fig:shell}) is shown in Fig. \ref{fig:A2}. Applying the Feynman rules we get
\begin{align}
	\Sigma_{+,\mathbf{k},s}^{\left(2\right)}\left(i\omega_{n}\right) & =\sum_{j,r_{1},r_{2}}\mathcal{T}_{j,\mathbf{k},s,r_{1}}\mathcal{T}_{j,\mathbf{k},-s,r_{2}}^{*}\tilde{\mathcal{G}}_{\mathbf{k}+\mathbf{q}_{j},r_{1},r_{2}}\left(i\omega_{n}\right)\sum_{j,r_{1},r_{2}}\mathcal{T}_{j,\mathbf{k},-s,r_{2}}\mathcal{T}_{j,\mathbf{k},s,r_{1}}^{*}\tilde{\mathcal{G}}_{\mathbf{k}+\mathbf{q}_{j},r_{1},r_{2}}\left(i\omega_{n}\right)\nonumber \\
	& \qquad\times\left(i\omega_{n}-\epsilon_{\mathbf{k},-s}-\sum_{j,r_{1},r_{2}}\mathcal{T}_{j,\mathbf{k},-s,r_{1}}\mathcal{T}_{j,\mathbf{k},-s,r_{2}}^{*}\tilde{\mathcal{G}}_{\mathbf{k}+\mathbf{q}_{j},r_{1},r_{2}}\left(i\omega_{n}\right)\right)^{-1}\nonumber \\
	& \qquad\qquad+\sum_{j,r_{1},r_{2}}\mathcal{T}_{j,\mathbf{k},s,r_{1}}\mathcal{T}_{j,\mathbf{k},s,r_{2}}^{*}\tilde{\mathcal{G}}_{\mathbf{k}+\mathbf{q}_{j},r_{1},r_{2}}\left(i\omega_{n}\right). \label{eq:self2Ap}
\end{align}
Here we defined $\tilde{\mathcal{G}}_{\mathbf{k}+\mathbf{q}_{j},r_{1},r_{2}}\left(i\omega_{n}\right)$ as the Green's function for the propagation of $\left|-,\mathbf{k}+\mathbf{q}_{j},r_{1}\right\rangle $ to $\left|-,\mathbf{k}+\mathbf{q}_{j},r_{2}\right\rangle $, taking into account only the interactions with the other layer with $\mathbf{q}_{j'}\neq\mathbf{q}_{j}$. It can be obtained diagrammatically by differentiating the cases $r_{1}=r_{2}$ and $r_{1}=-r_{2}$ (cf. Fig. \ref{fig:A2}), with the result
\begin{align}
	\tilde{\mathcal{G}}_{\mathbf{k}+\mathbf{q}_{j},r,r}\left(i\omega_{n}\right) & =\frac{1}{i\omega_{n}-\epsilon_{\mathbf{k}+\mathbf{q}_{j},r}-\Sigma''_{\mathbf{k}+\mathbf{q}_{j},r,r}\left(i\omega_{n}\right)},\label{eq:Gtilde1}\\
	\tilde{\mathcal{G}}_{\mathbf{k}+\mathbf{q}_{j},r,-r}\left(i\omega_{n}\right) & =\frac{1}{i\omega_{n}-\epsilon_{\mathbf{k}+\mathbf{q}_{j},-r}-\Sigma''_{\mathbf{k}+\mathbf{q}_{j},-r,-r}\left(i\omega_{n}\right)}\frac{\Sigma''_{\mathbf{k}+\mathbf{q}_{j},r,-r}\left(i\omega_{n}\right)}{i\omega_n-\epsilon_{\mathbf{k}+\mathbf{q}_{j},r}}.\label{eq:Gtilde2}
\end{align}
where
\begin{align}
	\Sigma''_{\mathbf{k}+\mathbf{q}_{j},r_{1},r_{2}}\left(i\omega_{n}\right) & =\sum_{j'\neq j,r'}\frac{\mathcal{T}_{j',\mathbf{k}+\mathbf{q}_{j}-\mathbf{q}_{j'},r',r_{1}}^{*}\mathcal{T}_{j',\mathbf{k}+\mathbf{q}_{j}-\mathbf{q}_{j'},r',-r_{2}}}{i\omega_{n}-\epsilon_{\mathbf{k}+\mathbf{q}_{j}-\mathbf{q}_{j'},r'}}\sum_{j'\neq j,r'}\frac{\mathcal{T}_{j',\mathbf{k}+\mathbf{q}_{j}-\mathbf{q}_{j'},r',-r_{2}}^{*}\mathcal{T}_{j',\mathbf{k}+\mathbf{q}_{j}-\mathbf{q}_{j'},r',r_{2}}}{i\omega_{n}-\epsilon_{\mathbf{k}+\mathbf{q}_{j}-\mathbf{q}_{j'},r'}}\nonumber \\
	& \times\left(i\omega_n-\epsilon_{\mathbf{k}+\mathbf{q}_{j},-r_{2}}-\sum_{j'\neq j,r'}\frac{\left|\mathcal{T}_{j',\mathbf{k}+\mathbf{q}_{j}-\mathbf{q}_{j'},r',-r_{2}}\right|^{2}}{i\omega_{n}-\epsilon_{\mathbf{k}+\mathbf{q}_{j}-\mathbf{q}_{j'},r'}}\right)^{-1}+\sum_{j'\neq j,r'}\frac{\mathcal{T}_{j',\mathbf{k}+\mathbf{q}_{j}-\mathbf{q}_{j'},r',r_{1}}^{*}\mathcal{T}_{j',\mathbf{k}+\mathbf{q}_{j}-\mathbf{q}_{j'},r',r_{2}}}{i\omega_{n}-\epsilon_{\mathbf{k}+\mathbf{q}_{j}-\mathbf{q}_{j'},r'}}.\label{eq:self2p}
\end{align}
It is important to note that $j'\neq j$ in the summation over $j'$, in order to not overcount scattering processes. We have checked numerically that the self-energy given by Eq. \eqref{eq:self2Ap} agrees with the one obtained from the general expression given by Eq. \eqref{eq:self}, for a momentum cutoff at the second shell. 

Note that Eq. \eqref{eq:self2Ap} reduces to Eq. \eqref{eq:self1} when the momentum is truncated at the first shell and therefore one does not consider the dressing of the propagator with momentum $\mathbf{k}+\mathbf{q}_{j}$ by the interaction with higher order Dirac points, i.e., when $\tilde{\mathcal{G}}_{\mathbf{k}+\mathbf{q}_{j},r_{1},r_{2}}\rightarrow\mathcal{G}_{\mathbf{k}+\mathbf{q}_{j},r_{1}}^{0}\delta_{r_{1},r_{2}}$.

\section{Dirac velocity renormalization}\label{app:QVrenorm}

The renormalized velocity $v^{\star}$ of the lowest moiré bands at the Dirac point can be directly computed from the self-energy (Sec. \ref{subsec:Qvelocity}). Here we give, for completeness and reference, details of the analytical calculation of $v^{\star}$ up to the second shell momentum cutoff, using Eqs. \eqref{eq:self1} and \eqref{eq:self2Ap}. Of course, this approach complements and yields the same results as usually obtained by traditional perturbation methods applied to a truncated TBG Hamiltonian \cite{Bistritzer2011,Tarnopolsky2019,Bernevig2021}. As we are interested in the $k$ and $\omega$ derivatives, around the zero point $k=0=\omega$, of the real part of the retarded self-energy, we shall directly replace $i\omega_{n}\rightarrow\omega$ when doing the analytical continuation. 

Following Sec. \ref{subsec:Qvelocity} in the main text, we consider the self-energy for the top layer, but keeping here in general both bands $s$. The generalization of Eq. \eqref{eq:vstar} for both bands is
\begin{equation}
	\frac{v^{\star}}{v}=\frac{1+s\left(\partial\Sigma'_{\mathbf{k},s}/\partial k\right)_{0}/v}{1-\left(\partial\Sigma'_{\mathbf{k},s}/\partial\omega\right)_{0}}.\label{eq:vfs}
\end{equation}
To condense the notation we will note $\Sigma'_{\mathbf{k},s}=\Sigma$.

\subsection{First shell}

The self-energy when the momentum is truncated at the first shell is given by Eq. \eqref{eq:self1}. From Eq. \eqref{eq:Tjband} one has
\begin{align}
	\left|\mathcal{T}_{j,\mathbf{k},s,r}\right|^{2} & =w_{0}^{2}\frac{1+sr\cos\left(\theta_{\mathbf{k}}-\theta_{\mathbf{k}+\mathbf{q}_{j}}\right)}{2}+w_{1}^{2}\frac{1+sr\cos\left(\theta_{\mathbf{k}}+\theta_{\mathbf{k}+\mathbf{q}_{j}}-2\phi_{j}\right)}{2}\nonumber \\
	& \qquad+sw_{0}w_{1}\left[\cos\left(\theta_{\mathbf{k}}-\phi_{j}\right)+sr\cos\left(\theta_{\mathbf{k}+\mathbf{q}_{j}}-\phi_{j}\right)\right],\label{eq:Texp1}\\
	\mathcal{T}_{j,\mathbf{k},s,r}\mathcal{T}_{j,\mathbf{k},-s,r}^{*} & =iw_{0}^{2}\frac{sr\sin\left(\theta_{\mathbf{k}}-\theta_{\mathbf{k}+\mathbf{q}_{j}}\right)}{2}+iw_{1}^{2}\frac{sr\sin\left(\theta_{\mathbf{k}}+\theta_{\mathbf{k}+\mathbf{q}_{j}}-2\phi_{j}\right)}{2}\nonumber \\
	& \qquad+isw_{0}w_{1}\sin\left(\theta_{\mathbf{k}}-\phi_{j}\right).\label{eq:Texp2}
\end{align}
It follows that
\begin{equation}
	\left(\sum_{j,r}\frac{\mathcal{T}_{j,\mathbf{k},s,r}\mathcal{T}_{j,\mathbf{k},-s,r}^{*}}{\omega-\epsilon_{\mathbf{k}+\mathbf{q}_{j},r}}\right)_{0}=\left(\frac{\partial}{\partial k}\sum_{j,r}\frac{\mathcal{T}_{j,\mathbf{k},s,r}\mathcal{T}_{j,\mathbf{k},-s,r}^{*}}{\omega-\epsilon_{\mathbf{k}+\mathbf{q}_{j},r}}\right)_{0}=\left(\frac{\partial}{\partial\omega}\sum_{j,r}\frac{\mathcal{T}_{j,\mathbf{k},s,r}\mathcal{T}_{j,\mathbf{k},-s,r}^{*}}{\omega-\epsilon_{\mathbf{k}+\mathbf{q}_{j},r}}\right)_{0}=0,
\end{equation}
where the subindex ``0" implies evaluation at $k=0=\omega$. Consequently the interband term given by Eq. \eqref{eq:self1inter} makes no contribution to the first order derivatives $\left(\partial\Sigma/\partial k\right)_{0}$ and $\left(\partial\Sigma/\partial\omega\right)_{0}$. We thus continue considering only the intraband contribution given by Eq. \eqref{eq:self1intra}, which yields
\begin{align}
	\left(\frac{\partial\Sigma}{\partial k}\right)_{0} & =-\sum_{j=1}^{3}\sum_{r=\pm1}\left(\frac{\partial\left|\mathcal{T}_{j,\mathbf{k},s,r}\right|^{2}}{\partial k}\right)_{0}\frac{1}{rvk_{\theta}}+\sum_{j=1}^{3}\sum_{r=\pm1}\frac{\left|\mathcal{T}_{j,\mathbf{0},s,r}\right|^{2}}{\left(vk_{\theta}\right)^{2}}\left(\frac{\partial\epsilon_{\mathbf{k}+\mathbf{q}_{j},r}}{\partial k}\right)_{0}.
\end{align}
where $\mathcal{T}_{j,\mathbf{0},s,r}=\mathcal{T}_{j,\mathbf{k}=\mathbf{0},s,r}$ and we replaced $\epsilon_{\mathbf{q}_{j}}=vk_{\theta}$, which is independent of $j$ ($\left|\mathbf{q}_{j}\right|=k_{\theta}$ being the moiré BZ length). We get
\begin{align}
	\sum_{j=1}^{3}\sum_{r=\pm1}\left(\frac{\partial\left|\mathcal{T}_{j,\mathbf{k},s,r}\right|^{2}}{\partial k}\right)_{0}\frac{1}{rvk_{\theta}} & =\frac{3}{2}s v\left(\alpha_{0}^{2}+\alpha_{1}^{2}\right),\label{eq:derv1}\\
	\sum_{j=1}^{3}\sum_{r=\pm1}\frac{\left|\mathcal{T}_{j,\mathbf{0},s,r}\right|^{2}}{\left(vk_{\theta}\right)^{2}}\left(\frac{\partial\epsilon_{\mathbf{k}+\mathbf{q}_{j},r}}{\partial k}\right)_{0} & =\frac{3}{2}s v\left(\alpha_{0}^{2}-\alpha_{1}^{2}\right),\label{eq:derv2}
\end{align}
where $\alpha_{i}=w_{i}/vk_{\theta}$. Thus, 
\begin{align}
	\left(\frac{\partial\Sigma}{\partial k}\right)_{0} & =-3sv\alpha_{1}^{2}.
\end{align}
For the $\omega$-derivative we get
\begin{equation}
	\left(\frac{\partial\Sigma}{\partial\omega}\right)_{0}=-\sum_{j=1}^{3}\sum_{r=\pm1}\frac{\left|\mathcal{T}_{j,\mathbf{0},s,r}\right|^{2}}{\left(vk_{\theta}\right)^{2}}=-3\left(\alpha_{0}^{2}+\alpha_{1}^{2}\right).
\end{equation}
Replacing in Eq. \eqref{eq:vfs} leads to the well-known leading order renormalization
\begin{equation}
	\frac{v^{\star}}{v}=\frac{1-3\alpha_{1}^{2}}{1+3\alpha_{0}^{2}+3\alpha_{1}^{2}}.\label{eq:vfirst}
\end{equation}
The case $w_{0}=w_{1}$ gives the original result first obtained in Ref. \cite{Bistritzer2011}.

\subsection{Second shell}

Truncating the momentum at the second shell yields the self-energy obtained in Appendix \ref{app:SEsecond}. The moiré potentials in Eq. \eqref{eq:self2p} are given by Eqs. \eqref{eq:Texp1} and \eqref{eq:Texp2}, with the replacements $j\rightarrow j'$, $s\rightarrow r'$ and $\mathbf{k}\rightarrow\mathbf{k}+\mathbf{q}_{j}-\mathbf{q}_{j'}$. From Eq. \eqref{eq:self2p} it follows that,
at $k=0=\omega$,
\begin{align}
	\Sigma''_{\mathbf{q}_{j},r,r}\left(0\right) & =-rvk_{\theta}\alpha_{0}^{2},\label{eq:self2p1}\\
	\Sigma''_{\mathbf{q}_{j},r,-r}\left(0\right) & =0,\label{eq:self2p2}
\end{align}
and therefore, at the same point, Eqs. \eqref{eq:Gtilde1} and \eqref{eq:Gtilde2} reduce to
\begin{align}
	\tilde{\mathcal{G}}_{\mathbf{q}_{j},r,r}\left(0\right) & =-\frac{r}{vk_{\theta}}\frac{1}{1-\alpha_{0}^{2}},\label{eq:gtilde}\\
	\tilde{\mathcal{G}}_{\mathbf{q}_{j},r,-r}\left(0\right) & =0.
\end{align}
Moreover, one has
\begin{align}
	\sum_{j,r}\mathcal{T}_{j,\mathbf{0},s,r}\mathcal{T}_{j,\mathbf{0},-s,r}^{*}\tilde{\mathcal{G}}_{\mathbf{q}_{j},r,r}\left(0\right) & =0.
\end{align}
Given these relations, from Eq. \eqref{eq:self2Ap} we have
\begin{align}
	\left(\frac{\partial\Sigma}{\partial k}\right)_{0} & =\sum_{j,r}\left(\frac{\partial\left|\mathcal{T}_{j,\mathbf{k},s,r}\right|^{2}}{\partial k}\right)_{0}\tilde{\mathcal{G}}_{\mathbf{q}_{j},r,r}\left(0\right)+\sum_{j,r_{1},r_{2}}\mathcal{T}_{j,\mathbf{0},s,r_{1}}\mathcal{T}_{j,\mathbf{0},s,r_{2}}^{*}\left(\frac{\partial\tilde{\mathcal{G}}_{\mathbf{k}+\mathbf{q}_{j},r_1,r_2}}{\partial k}\right)_{0}.\label{eq:deriv2}
\end{align}
The first term is the same as the one appearing in the first shell case, Eq. \eqref{eq:derv1}, but with the inclusion of the factor $\sim\left(1-\alpha_{0}^{2}\right)^{-1}$ in Eq. \eqref{eq:gtilde}, so
\begin{align}
	\sum_{j,r}\left(\frac{\partial\left|\mathcal{T}_{j,\mathbf{k},s,r}\right|^{2}}{\partial k}\right)_{0}\tilde{\mathcal{G}}_{\mathbf{q}_{j},r,r}\left(0\right) & =-\frac{3}{2}sv\frac{\alpha_{0}^{2}+\alpha_{1}^{2}}{1-\alpha_{0}^{2}}.
\end{align}
For the second term in Eq. \eqref{eq:deriv2}, using Eqs. \eqref{eq:Gtilde1} and \eqref{eq:Gtilde2} we get
\begin{align}
	\sum_{j,r_{1},r_{2}}\mathcal{T}_{j,\mathbf{0},s,r_{1}}\mathcal{T}_{j,\mathbf{0},s,r_{2}}^{*}\left(\frac{\partial\tilde{\mathcal{G}}_{\mathbf{k}+\mathbf{q}_{j},r_{1},r_{2}}}{\partial k}\right)_{0} & =\frac{1}{\left(1-\alpha_{0}^{2}\right)^{2}}\sum_{j,r}\frac{\left|\mathcal{T}_{j,\mathbf{0},s,r}\right|^{2}}{\left(vk_{\theta}\right)^{2}}\left[\left(\frac{\partial\epsilon_{\mathbf{k}+\mathbf{q}_{j},r}}{\partial k}\right)_{0}+\left(\frac{\partial\Sigma''_{\mathbf{k}+\mathbf{q}_{j},r,r}}{\partial k}\right)_{0}\right]\nonumber \\
	& \qquad-\frac{1}{1-\alpha_{0}^{2}}\sum_{j,r}\frac{\mathcal{T}_{j,\mathbf{0},s,r}\mathcal{T}_{j,\mathbf{0},s,-r}^{*}}{\left(vk_{\theta}\right)^{2}}\left(\frac{\partial\Sigma''_{\mathbf{k}+\mathbf{q}_{j},r,-r}}{\partial k}\right)_{0}.
\end{align}
The first summation with the term proportional to $\left(\partial\epsilon_{\mathbf{k}+\mathbf{q}_{j},r}/\partial k\right)_{0}$ is given by Eq. \eqref{eq:derv2}. For the other terms, a long but straightforward calculation using Eq. \eqref{eq:self2p} yields
\begin{align}
	\sum_{j,r}\frac{\left|\mathcal{T}_{j,\mathbf{0},s,r}\right|^{2}}{\left(vk_{\theta}\right)^{2}}\left(\frac{\partial\Sigma''_{\mathbf{k}+\mathbf{q}_{j},r,r}}{\partial k}\right)_{0} & =\frac{1}{2}sv\left[\left(\alpha_{0}^{2}-\alpha_{1}^{2}\right)^{2}+8\alpha_{0}^{2}\alpha_{1}^{2}\right],\\
	\sum_{j,r}\frac{\mathcal{T}_{j,\mathbf{0},s,r}\mathcal{T}_{j,\mathbf{0},s,-r}^{*}}{\left(vk_{\theta}\right)^{2}}\left(\frac{\partial\Sigma''_{\mathbf{k}+\mathbf{q}_{j},r,-r}}{\partial k}\right)_{0} & =-\frac{1}{2}sv\frac{\left(\alpha_{1}^{2}-2\alpha_{0}^{2}\right)\left(\alpha_{0}^{2}+\alpha_{1}^{2}\right)}{1-\alpha_{0}^{2}}.
\end{align}
Combining the previous results and simplifying, for the $k$-derivative we have
\begin{equation}
	\left(\frac{\partial\Sigma}{\partial k}\right)_{0}=-s v\frac{3\alpha_{1}^{2}-\alpha_{0}^{4}-\alpha_{1}^{4}-4\alpha_{0}^{2}\alpha_{1}^{2}}{\left(1-\alpha_{0}^{2}\right)^{2}}.
\end{equation}
For the other derivative over $\omega$, proceeding as with the momentum derivatives we get
\begin{equation}
	\left(\frac{\partial\Sigma}{\partial \omega}\right)_{0}=-\frac{3\left(\alpha_{0}^{2}+\alpha_{1}^{2}\right)+2\left(\alpha_{0}^{4}+\alpha_{1}^{4}+4\alpha_{1}^{2}\alpha_{0}^{2}\right)}{\left(1-\alpha_{0}^{2}\right)^{2}}.
\end{equation}
Finally, replacing in Eq. \eqref{eq:vfs} yields 
\begin{equation}
	\frac{v^{\star}}{v}=\frac{\left(1-\alpha_{0}^{2}\right)^{2}-3\alpha_{1}^{2}+\alpha_{0}^{4}+\alpha_{1}^{4}+4\alpha_{0}^{2}\alpha_{1}^{2}}{\left(1-\alpha_{0}^{2}\right)^{2}+3\left(\alpha_{0}^{2}+\alpha_{1}^{2}\right)+2\left(\alpha_{0}^{4}+\alpha_{1}^{4}+4\alpha_{0}^{2}\alpha_{1}^{2}\right)}.
\end{equation}
In the chiral case $\alpha_{0}=0$, the above reduces to $v^{\star}/v=\left(1-3\alpha_{1}^{2}+\alpha_{1}^{4}\right)/\left(1+3\alpha_{1}^{2}+2\alpha_{1}^{4}\right)$, in agreement with Ref. \cite{Tarnopolsky2019} up to $\sim\alpha_{1}^{4}$.

\section{First-shell velocity $v^{\star}$ as function of $\phi_j$ and $\mathbf{q}_j$}\label{app:QVphi}

Here we give details of the calculation of $v^{\star}$ at the first shell using Eq. \eqref{eq:self1}, but keeping explicitly the dependence on the phases $\phi_j$ and momentum transfer vectors $\mathbf{q}_j$ as they enter in the moiré potential given by Eq. \eqref{eq:Tjband}; cf. Sec. \ref{subsec:OriginFB}. As before, we focus on the perturbation on the top layer, and for simplicity note throughout $\Sigma_{+,\mathbf{k},s}^{\left(1\right)}\rightarrow\Sigma_{\mathbf{k},s}$. We will, in particular, consider the chiral limit $\alpha_{0}=0$, which simplifies the calculation because then the Dirac points are never shifted in energy [cf. Eq. \eqref{eq:selfPhi}], and thus $v^{\star}$ can be obtained, as before, by simply evaluating the derivatives of the self-energy at $k=\omega=0$.

The calculation of $v^{\star}$ can be done by following the same procedure as in the particular TBG case in which $\theta_{\mathbf{q}_{j}}-\phi_{j}=90^{\circ}$, where $\theta_{\mathbf{q}_{j}}$ is the angle of the momentum transfer vectors $\mathbf{q}_{j}$ (see Fig. \ref{fig:TBG}). However, since for $\theta_{\mathbf{q}_{j}}-\phi_{j}\neq90^{\circ}$ the derivatives of the interband contribution to the self-energy do not vanish, one must be careful with the divergences of the last term in Eq. \eqref{eq:self1inter} when evaluated at $k=0=\omega$. One way to circumvent that is to rather start from Eq. \eqref{eq:Qenergies}, 
\begin{equation}
	\omega-\epsilon_{\mathbf{k},s}-\left(\Sigma_{\mathbf{k},s}^{+}+\Sigma_{\mathbf{k},s}^{\mathrm{-}}\right)=0,\label{eq:QWphi}
\end{equation}
where we have explicitly separated the intraband and interband contributions to the self-energy, Eqs. \eqref{eq:self1intra} and \eqref{eq:self1inter}, and to simplify the notation we have noted
\begin{equation}
	\Sigma_{\mathbf{k},s}^{+}=\Sigma_{\mathbf{k},s}^{\mathrm{intra}},\quad\Sigma_{\mathbf{k},s}^{-}=\Sigma_{\mathbf{k},s}^{\mathrm{inter}}.
\end{equation}
Now we multiply Eq. \eqref{eq:QWphi} by the last factor in Eq. \eqref{eq:self1inter}, so that
\begin{equation}
	\left[\omega-\epsilon_{\mathbf{k},s}-\left(\Sigma_{\mathbf{k},s}^{+}+\Sigma_{\mathbf{k},s}^{\mathrm{-}}\right)\right]\left(\omega-\epsilon_{\mathbf{k},-s}-\Sigma_{\mathbf{k},-s}^{+}\right)=0.
\end{equation}
Using $\epsilon_{\mathbf{k},-s}=-\epsilon_{\mathbf{k},s}$ we get
\begin{equation}
	\omega^{2}-\epsilon_{\mathbf{k},s}^{2}-\left(\omega-\epsilon_{\mathbf{k},s}\right)\Sigma_{\mathbf{k},-s}^{+}-\left(\omega+\epsilon_{\mathbf{k},s}\right)\Sigma_{\mathbf{k},s}^{+}+\Sigma_{\mathbf{k},s}^{+}\Sigma_{\mathbf{k},-s}^{+}-\Sigma_{\mathbf{k},s}^{-}\left(\omega+\epsilon_{\mathbf{k},s}-\Sigma_{\mathbf{k},-s}^{+}\right)=0.\label{eq:QWphi2}
\end{equation}
The last factor $\Sigma_{\mathbf{k},s}^{-}\left(\omega+\epsilon_{\mathbf{k},s}-\Sigma_{\mathbf{k},-s}^{+}\right)$ gets effectively rid of the divergences of $\Sigma_{\mathbf{k},s}^{-}$ at $k=0=\omega$. It is convenient to define
\begin{equation}
	\tilde{\Sigma}_{\mathbf{k},s}=\sum_{j,r}\frac{\mathcal{T}_{j,\mathbf{k},s,r}\mathcal{T}_{j,\mathbf{k},-s,r}^{*}}{\omega-\epsilon_{\mathbf{k}+\mathbf{q}_{j},r}},
\end{equation}
so that $\Sigma_{\mathbf{k},s}^{-}\left(\omega+\epsilon_{\mathbf{k},s}-\Sigma_{\mathbf{k},-s}^{+}\right)=\tilde{\Sigma}_{\mathbf{k},s}\tilde{\Sigma}_{\mathbf{k},s}^{*}$, cf. Eq. \eqref{eq:self1inter}. To obtain the quasiparticle velocity $v^{\star}=\left(\partial\omega/\partial k\right)_{0}$ at the Dirac point, we now derive Eq. \eqref{eq:QWphi2} twice in momentum, evaluate all at $k=0=\omega$, and use the chiral limit results 
\begin{align}
	\left(\sum_{j,r}\frac{\left|\mathcal{T}_{j,\mathbf{k},s,r}\right|^{2}}{\omega-\epsilon_{\mathbf{k}+\mathbf{q}_{j},r}}\right)_{0} & =\left(\sum_{j,r}\frac{\mathcal{T}_{j,\mathbf{k},s,r}\mathcal{T}_{j,\mathbf{k},-s,r}^{*}}{\omega-\epsilon_{\mathbf{k}+\mathbf{q}_{j},r}}\right)_{0}=0,\\
	\left(\frac{\partial}{\partial k}\sum_{j,r}\frac{\left|\mathcal{T}_{j,\mathbf{k},s,r}\right|^{2}}{\omega-\epsilon_{\mathbf{k}+\mathbf{q}_{j},r}}\right)_{0} & =3sv\alpha_{1}^{2}\cos\left(2\varphi\right),\qquad\left(\frac{\partial}{\partial k}\sum_{j,r}\frac{\mathcal{T}_{j,\mathbf{k},s,r}\mathcal{T}_{j,\mathbf{k},-s,r}^{*}}{\omega-\epsilon_{\mathbf{k}+\mathbf{q}_{j},r}}\right)_{0}=i3sv\alpha_{1}^{2}\sin\left(2\varphi\right),\\
	\left(\frac{\partial}{\partial\omega}\sum_{j,r}\frac{\left|\mathcal{T}_{j,\mathbf{k},s,r}\right|^{2}}{\omega-\epsilon_{\mathbf{k}+\mathbf{q}_{j},r}}\right)_{0} & =-3\alpha_{1}^{2},\qquad\left(\frac{\partial}{\partial\omega}\sum_{j,r}\frac{\mathcal{T}_{j,\mathbf{k},s,r}\mathcal{T}_{j,\mathbf{k},-s,r}^{*}}{\omega-\epsilon_{\mathbf{k}+\mathbf{q}_{j},r}}\right)_{0}=0,\\
	\left(\frac{\partial^{2}}{\partial k^{2}}\sum_{j,r}\frac{\left|\mathcal{T}_{j,\mathbf{k},s,r}\right|^{2}}{\omega-\epsilon_{\mathbf{k}+\mathbf{q}_{j},r}}\right)_{0} & =\left(\frac{\partial^{2}}{\partial\omega^{2}}\sum_{j,r}\frac{\left|\mathcal{T}_{j,\mathbf{k},s,r}\right|^{2}}{\omega-\epsilon_{\mathbf{k}+\mathbf{q}_{j},r}}\right)_{0}=\left(\frac{\partial^{2}}{\partial k\partial\omega}\sum_{j,r}\frac{\left|\mathcal{T}_{j,\mathbf{k},s,r}\right|^{2}}{\omega-\epsilon_{\mathbf{k}+\mathbf{q}_{j},r}}\right)_{0}=0,\\
	\left(\frac{\partial^{2}}{\partial k^{2}}\sum_{j,r}\frac{\mathcal{T}_{j,\mathbf{k},s,r}\mathcal{T}_{j,\mathbf{k},-s,r}^{*}}{\omega-\epsilon_{\mathbf{k}+\mathbf{q}_{j},r}}\right)_{0} & =\left(\frac{\partial^{2}}{\partial\omega^{2}}\sum_{j,r}\frac{\mathcal{T}_{j,\mathbf{k},s,r}\mathcal{T}_{j,\mathbf{k},-s,r}^{*}}{\omega-\epsilon_{\mathbf{k}+\mathbf{q}_{j},r}}\right)_{0}=\left(\frac{\partial^{2}}{\partial k\partial\omega}\sum_{j,r}\frac{\mathcal{T}_{j,\mathbf{k},s,r}\mathcal{T}_{j,\mathbf{k},-s,r}^{*}}{\omega-\epsilon_{\mathbf{k}+\mathbf{q}_{j},r}}\right)_{0}=0,
\end{align}
where $\varphi=\theta_{\mathbf{q}_{j}}-\phi_{j}$ (which is $j$-independent). The last two rows imply, by Eq. \eqref{eq:QWphi}, that also $\left(\partial^{2}\omega/\partial k^{2}\right)_{0}=0$. Then we get
\begin{align}
	0 & =\left(v^{\star}\right)^{2}\left(1-\partial_{\omega}^{0}\Sigma_{\mathbf{k},s}^{+}-\partial_{\omega}^{0}\Sigma_{\mathbf{k},-s}^{+}+\partial_{\omega}^{0}\Sigma_{\mathbf{k},s}^{+}\partial_{\omega}^{0}\Sigma_{\mathbf{k},-s}^{\mathrm{+}}\right)\nonumber \\
	& \quad-v^{\star}\left(\partial_{k}^{0}\Sigma_{\mathbf{k},s}^{\mathrm{+}}+\partial_{k}^{0}\Sigma_{\mathbf{k},-s}^{\mathrm{+}}+v\partial_{\omega}^{0}\Sigma_{\mathbf{k},s}^{+}-v\partial_{\omega}^{0}\Sigma_{\mathbf{k},-s}^{+}-\partial_{k}^{0}\Sigma_{\mathbf{k},s}^{\mathrm{+}}\partial_{\omega}^{0}\Sigma_{\mathbf{k},-s}^{\mathrm{+}}-\partial_{k}^{0}\Sigma_{\mathbf{k},-s}^{\mathrm{+}}\partial_{\omega}^{0}\Sigma_{\mathbf{k},s}^{\mathrm{+}}\right)\nonumber \\
	& \quad-v^{2}-v\left(\partial_{k}^{0}\Sigma_{\mathbf{k},s}^{\mathrm{+}}-\partial_{k}^{0}\Sigma_{\mathbf{k},-s}^{+}\right)+\partial_{k}^{0}\Sigma_{\mathbf{k},s}^{+}\partial_{k}^{0}\Sigma_{\mathbf{k},-s}^{+}-\partial_{k}^{0}\tilde{\Sigma}_{\mathbf{k},s}\partial_{k}^{0}\tilde{\Sigma}_{\mathbf{k},s}^{*},\label{eq:vphiA}
\end{align}
where, to condense the notation, we have noted $\partial_{k}^{0}\rightarrow\left(\partial/\partial k\right)_{0}$ and $\partial_{\omega}^{0}\rightarrow\left(\partial/\partial\omega\right)_{0}$. Replacing the derivatives yields
\begin{equation}
	0=\left(v^{\star}\right)^{2}\left(1+6\alpha_{1}^{2}+9\alpha_{1}^{4}\right)-v^{2}-sv\left[6sv\alpha_{1}^{2}\cos\left(2\varphi\right)\right]-\left[3v\alpha_{1}^{2}\cos\left(2\varphi\right)\right]^{2}-\left[3v\alpha_{1}^{2}\sin\left(2\varphi\right)\right]^{2},
\end{equation}
which leads to Eq. \eqref{eq:VelPhi} in the main text. Note that the interband contribution comes only from the last term in Eq. \eqref{eq:vphiA}, which vanishes when $\varphi=90^{\circ}$. This means that moiré-induced interband scatterings, within each layer, do not contribute to $v^{\star}$ only for the specific orientation of the momentum transfer vectors in TBG; otherwise they influence $v^{\star}$ and can actually prevent it from vanishing.


\begin{thebibliography}{73}%
	\makeatletter
	\providecommand \@ifxundefined [1]{%
		\@ifx{#1\undefined}
	}%
	\providecommand \@ifnum [1]{%
		\ifnum #1\expandafter \@firstoftwo
		\else \expandafter \@secondoftwo
		\fi
	}%
	\providecommand \@ifx [1]{%
		\ifx #1\expandafter \@firstoftwo
		\else \expandafter \@secondoftwo
		\fi
	}%
	\providecommand \natexlab [1]{#1}%
	\providecommand \enquote  [1]{``#1''}%
	\providecommand \bibnamefont  [1]{#1}%
	\providecommand \bibfnamefont [1]{#1}%
	\providecommand \citenamefont [1]{#1}%
	\providecommand \href@noop [0]{\@secondoftwo}%
	\providecommand \href [0]{\begingroup \@sanitize@url \@href}%
	\providecommand \@href[1]{\@@startlink{#1}\@@href}%
	\providecommand \@@href[1]{\endgroup#1\@@endlink}%
	\providecommand \@sanitize@url [0]{\catcode `\\12\catcode `\$12\catcode
		`\&12\catcode `\#12\catcode `\^12\catcode `\_12\catcode `\%12\relax}%
	\providecommand \@@startlink[1]{}%
	\providecommand \@@endlink[0]{}%
	\providecommand \url  [0]{\begingroup\@sanitize@url \@url }%
	\providecommand \@url [1]{\endgroup\@href {#1}{\urlprefix }}%
	\providecommand \urlprefix  [0]{URL }%
	\providecommand \Eprint [0]{\href }%
	\providecommand \doibase [0]{https://doi.org/}%
	\providecommand \selectlanguage [0]{\@gobble}%
	\providecommand \bibinfo  [0]{\@secondoftwo}%
	\providecommand \bibfield  [0]{\@secondoftwo}%
	\providecommand \translation [1]{[#1]}%
	\providecommand \BibitemOpen [0]{}%
	\providecommand \bibitemStop [0]{}%
	\providecommand \bibitemNoStop [0]{.\EOS\space}%
	\providecommand \EOS [0]{\spacefactor3000\relax}%
	\providecommand \BibitemShut  [1]{\csname bibitem#1\endcsname}%
	\let\auto@bib@innerbib\@empty
	%</preamble>
	\bibitem [{\citenamefont {Cao}\ \emph {et~al.}(2018{\natexlab{a}})\citenamefont
		{Cao}, \citenamefont {Fatemi}, \citenamefont {Fang}, \citenamefont
		{Watanabe}, \citenamefont {Taniguchi}, \citenamefont {Kaxiras},\ and\
		\citenamefont {Jarillo-Herrero}}]{cao_unconventional_2018}%
	\BibitemOpen
	\bibfield  {author} {\bibinfo {author} {\bibfnamefont {Y.}~\bibnamefont
			{Cao}}, \bibinfo {author} {\bibfnamefont {V.}~\bibnamefont {Fatemi}},
		\bibinfo {author} {\bibfnamefont {S.}~\bibnamefont {Fang}}, \bibinfo {author}
		{\bibfnamefont {K.}~\bibnamefont {Watanabe}}, \bibinfo {author}
		{\bibfnamefont {T.}~\bibnamefont {Taniguchi}}, \bibinfo {author}
		{\bibfnamefont {E.}~\bibnamefont {Kaxiras}},\ and\ \bibinfo {author}
		{\bibfnamefont {P.}~\bibnamefont {Jarillo-Herrero}},\ }\bibfield  {title}
	{\bibinfo {title} {Unconventional superconductivity in magic-angle graphene
			superlattices},\ }\href {https://doi.org/10.1038/nature26160} {\bibfield
		{journal} {\bibinfo  {journal} {Nature}\ }\textbf {\bibinfo {volume} {556}},\
		\bibinfo {pages} {43} (\bibinfo {year} {2018}{\natexlab{a}})}\BibitemShut
	{NoStop}%
	\bibitem [{\citenamefont {Cao}\ \emph {et~al.}(2018{\natexlab{b}})\citenamefont
		{Cao}, \citenamefont {Fatemi}, \citenamefont {Demir}, \citenamefont {Fang},
		\citenamefont {Tomarken}, \citenamefont {Luo}, \citenamefont
		{Sanchez-Yamagishi}, \citenamefont {Watanabe}, \citenamefont {Taniguchi},
		\citenamefont {Kaxiras}, \citenamefont {Ashoori},\ and\ \citenamefont
		{Jarillo-Herrero}}]{Cao2018}%
	\BibitemOpen
	\bibfield  {author} {\bibinfo {author} {\bibfnamefont {Y.}~\bibnamefont
			{Cao}}, \bibinfo {author} {\bibfnamefont {V.}~\bibnamefont {Fatemi}},
		\bibinfo {author} {\bibfnamefont {A.}~\bibnamefont {Demir}}, \bibinfo
		{author} {\bibfnamefont {S.}~\bibnamefont {Fang}}, \bibinfo {author}
		{\bibfnamefont {S.~L.}\ \bibnamefont {Tomarken}}, \bibinfo {author}
		{\bibfnamefont {J.~Y.}\ \bibnamefont {Luo}}, \bibinfo {author} {\bibfnamefont
			{J.~D.}\ \bibnamefont {Sanchez-Yamagishi}}, \bibinfo {author} {\bibfnamefont
			{K.}~\bibnamefont {Watanabe}}, \bibinfo {author} {\bibfnamefont
			{T.}~\bibnamefont {Taniguchi}}, \bibinfo {author} {\bibfnamefont
			{E.}~\bibnamefont {Kaxiras}}, \bibinfo {author} {\bibfnamefont {R.~C.}\
			\bibnamefont {Ashoori}},\ and\ \bibinfo {author} {\bibfnamefont
			{P.}~\bibnamefont {Jarillo-Herrero}},\ }\bibfield  {title} {\bibinfo {title}
		{Correlated insulator behaviour at half-filling in magic-angle graphene
			superlattices},\ }\href {https://doi.org/10.1038/nature26154} {\bibfield
		{journal} {\bibinfo  {journal} {Nature}\ }\textbf {\bibinfo {volume} {556}},\
		\bibinfo {pages} {80} (\bibinfo {year} {2018}{\natexlab{b}})}\BibitemShut
	{NoStop}%
	\bibitem [{\citenamefont {Kerelsky}\ \emph {et~al.}(2019)\citenamefont
		{Kerelsky}, \citenamefont {McGilly}, \citenamefont {Kennes}, \citenamefont
		{Xian}, \citenamefont {Yankowitz}, \citenamefont {Chen}, \citenamefont
		{Watanabe}, \citenamefont {Taniguchi}, \citenamefont {Hone}, \citenamefont
		{Dean}, \citenamefont {Rubio},\ and\ \citenamefont
		{Pasupathy}}]{Kerelsky2019}%
	\BibitemOpen
	\bibfield  {author} {\bibinfo {author} {\bibfnamefont {A.}~\bibnamefont
			{Kerelsky}}, \bibinfo {author} {\bibfnamefont {L.~J.}\ \bibnamefont
			{McGilly}}, \bibinfo {author} {\bibfnamefont {D.~M.}\ \bibnamefont {Kennes}},
		\bibinfo {author} {\bibfnamefont {L.}~\bibnamefont {Xian}}, \bibinfo {author}
		{\bibfnamefont {M.}~\bibnamefont {Yankowitz}}, \bibinfo {author}
		{\bibfnamefont {S.}~\bibnamefont {Chen}}, \bibinfo {author} {\bibfnamefont
			{K.}~\bibnamefont {Watanabe}}, \bibinfo {author} {\bibfnamefont
			{T.}~\bibnamefont {Taniguchi}}, \bibinfo {author} {\bibfnamefont
			{J.}~\bibnamefont {Hone}}, \bibinfo {author} {\bibfnamefont {C.}~\bibnamefont
			{Dean}}, \bibinfo {author} {\bibfnamefont {A.}~\bibnamefont {Rubio}},\ and\
		\bibinfo {author} {\bibfnamefont {A.~N.}\ \bibnamefont {Pasupathy}},\
	}\bibfield  {title} {\bibinfo {title} {Maximized electron interactions at the
			magic angle in twisted bilayer graphene},\ }\href
	{https://doi.org/10.1038/s41586-019-1431-9} {\bibfield  {journal} {\bibinfo
			{journal} {Nature}\ }\textbf {\bibinfo {volume} {572}},\ \bibinfo {pages}
		{95} (\bibinfo {year} {2019})}\BibitemShut {NoStop}%
	\bibitem [{\citenamefont {Oh}\ \emph {et~al.}(2021)\citenamefont {Oh},
		\citenamefont {Nuckolls}, \citenamefont {Wong}, \citenamefont {Lee},
		\citenamefont {Liu}, \citenamefont {Watanabe}, \citenamefont {Taniguchi},\
		and\ \citenamefont {Yazdani}}]{Oh2021}%
	\BibitemOpen
	\bibfield  {author} {\bibinfo {author} {\bibfnamefont {M.}~\bibnamefont
			{Oh}}, \bibinfo {author} {\bibfnamefont {K.~P.}\ \bibnamefont {Nuckolls}},
		\bibinfo {author} {\bibfnamefont {D.}~\bibnamefont {Wong}}, \bibinfo {author}
		{\bibfnamefont {R.~L.}\ \bibnamefont {Lee}}, \bibinfo {author} {\bibfnamefont
			{X.}~\bibnamefont {Liu}}, \bibinfo {author} {\bibfnamefont {K.}~\bibnamefont
			{Watanabe}}, \bibinfo {author} {\bibfnamefont {T.}~\bibnamefont
			{Taniguchi}},\ and\ \bibinfo {author} {\bibfnamefont {A.}~\bibnamefont
			{Yazdani}},\ }\bibfield  {title} {\bibinfo {title} {Evidence for
			unconventional superconductivity in twisted bilayer graphene},\ }\href
	{https://doi.org/10.1038/s41586-021-04121-x} {\bibfield  {journal} {\bibinfo
			{journal} {Nature}\ }\textbf {\bibinfo {volume} {600}},\ \bibinfo {pages}
		{240} (\bibinfo {year} {2021})}\BibitemShut {NoStop}%
	\bibitem [{\citenamefont {Jiang}\ \emph {et~al.}(2019)\citenamefont {Jiang},
		\citenamefont {Lai}, \citenamefont {Watanabe}, \citenamefont {Taniguchi},
		\citenamefont {Haule}, \citenamefont {Mao},\ and\ \citenamefont
		{Andrei}}]{Jiang2019}%
	\BibitemOpen
	\bibfield  {author} {\bibinfo {author} {\bibfnamefont {Y.}~\bibnamefont
			{Jiang}}, \bibinfo {author} {\bibfnamefont {X.}~\bibnamefont {Lai}}, \bibinfo
		{author} {\bibfnamefont {K.}~\bibnamefont {Watanabe}}, \bibinfo {author}
		{\bibfnamefont {T.}~\bibnamefont {Taniguchi}}, \bibinfo {author}
		{\bibfnamefont {K.}~\bibnamefont {Haule}}, \bibinfo {author} {\bibfnamefont
			{J.}~\bibnamefont {Mao}},\ and\ \bibinfo {author} {\bibfnamefont {E.~Y.}\
			\bibnamefont {Andrei}},\ }\bibfield  {title} {\bibinfo {title} {Charge order
			and broken rotational symmetry in magic-angle twisted bilayer graphene},\
	}\href {https://doi.org/10.1038/s41586-019-1460-4} {\bibfield  {journal}
		{\bibinfo  {journal} {Nature}\ }\textbf {\bibinfo {volume} {573}},\ \bibinfo
		{pages} {91} (\bibinfo {year} {2019})}\BibitemShut {NoStop}%
	\bibitem [{\citenamefont {Andrei}\ and\ \citenamefont
		{MacDonald}(2020)}]{Andrei2020}%
	\BibitemOpen
	\bibfield  {author} {\bibinfo {author} {\bibfnamefont {E.~Y.}\ \bibnamefont
			{Andrei}}\ and\ \bibinfo {author} {\bibfnamefont {A.~H.}\ \bibnamefont
			{MacDonald}},\ }\bibfield  {title} {\bibinfo {title} {Graphene bilayers with
			a twist},\ }\href {https://doi.org/10.1038/s41563-020-00840-0} {\bibfield
		{journal} {\bibinfo  {journal} {Nature Materials}\ }\textbf {\bibinfo
			{volume} {19}},\ \bibinfo {pages} {1265} (\bibinfo {year}
		{2020})}\BibitemShut {NoStop}%
	\bibitem [{\citenamefont {Nimbalkar}\ and\ \citenamefont
		{Kim}(2020)}]{Nimbalkar2020}%
	\BibitemOpen
	\bibfield  {author} {\bibinfo {author} {\bibfnamefont {A.}~\bibnamefont
			{Nimbalkar}}\ and\ \bibinfo {author} {\bibfnamefont {H.}~\bibnamefont
			{Kim}},\ }\bibfield  {title} {\bibinfo {title} {Opportunities and
			{Challenges} in {Twisted} {Bilayer} {Graphene}: {A} {Review}},\ }\href
	{https://doi.org/10.1007/s40820-020-00464-8} {\bibfield  {journal} {\bibinfo
			{journal} {Nano-Micro Letters}\ }\textbf {\bibinfo {volume} {12}},\ \bibinfo
		{pages} {126} (\bibinfo {year} {2020})}\BibitemShut {NoStop}%
	\bibitem [{\citenamefont {Aggarwal}\ \emph {et~al.}(2023)\citenamefont
		{Aggarwal}, \citenamefont {Narula},\ and\ \citenamefont
		{Ghosh}}]{Aggarwal2023}%
	\BibitemOpen
	\bibfield  {author} {\bibinfo {author} {\bibfnamefont {D.}~\bibnamefont
			{Aggarwal}}, \bibinfo {author} {\bibfnamefont {R.}~\bibnamefont {Narula}},\
		and\ \bibinfo {author} {\bibfnamefont {S.}~\bibnamefont {Ghosh}},\ }\bibfield
	{title} {\bibinfo {title} {A primer on twistronics: a massless {Dirac}
			fermion’s journey to moiré patterns and flat bands in twisted bilayer
			graphene},\ }\href {https://doi.org/10.1088/1361-648X/acb984} {\bibfield
		{journal} {\bibinfo  {journal} {Journal of Physics: Condensed Matter}\
		}\textbf {\bibinfo {volume} {35}},\ \bibinfo {pages} {143001} (\bibinfo
		{year} {2023})}\BibitemShut {NoStop}%
	\bibitem [{\citenamefont {Lopes dos Santos}\ \emph
		{et~al.}(2007)\citenamefont {Lopes dos Santos}, \citenamefont {Peres},\
		and\ \citenamefont {Castro Neto}}]{Lopes dos Santos2007}%
	\BibitemOpen
	\bibfield  {author} {\bibinfo {author} {\bibfnamefont {J.~M.~B.}\
			\bibnamefont {Lopes dos Santos}}, \bibinfo {author} {\bibfnamefont
			{N.~M.~R.}\ \bibnamefont {Peres}},\ and\ \bibinfo {author} {\bibfnamefont
			{A.~H.}\ \bibnamefont {Castro Neto}},\ }\bibfield  {title} {\bibinfo {title}
		{Graphene {Bilayer} with a {Twist}: {Electronic} {Structure}},\ }\href
	{https://doi.org/10.1103/PhysRevLett.99.256802} {\bibfield  {journal}
		{\bibinfo  {journal} {Physical Review Letters}\ }\textbf {\bibinfo {volume}
			{99}},\ \bibinfo {pages} {256802} (\bibinfo {year} {2007})}\BibitemShut
	{NoStop}%
	\bibitem [{\citenamefont {Andrei}\ \emph {et~al.}(2021)\citenamefont {Andrei},
		\citenamefont {Efetov}, \citenamefont {Jarillo-Herrero}, \citenamefont
		{MacDonald}, \citenamefont {Mak}, \citenamefont {Senthil}, \citenamefont
		{Tutuc}, \citenamefont {Yazdani},\ and\ \citenamefont {Young}}]{Andrei2021}%
	\BibitemOpen
	\bibfield  {author} {\bibinfo {author} {\bibfnamefont {E.~Y.}\ \bibnamefont
			{Andrei}}, \bibinfo {author} {\bibfnamefont {D.~K.}\ \bibnamefont {Efetov}},
		\bibinfo {author} {\bibfnamefont {P.}~\bibnamefont {Jarillo-Herrero}},
		\bibinfo {author} {\bibfnamefont {A.~H.}\ \bibnamefont {MacDonald}}, \bibinfo
		{author} {\bibfnamefont {K.~F.}\ \bibnamefont {Mak}}, \bibinfo {author}
		{\bibfnamefont {T.}~\bibnamefont {Senthil}}, \bibinfo {author} {\bibfnamefont
			{E.}~\bibnamefont {Tutuc}}, \bibinfo {author} {\bibfnamefont
			{A.}~\bibnamefont {Yazdani}},\ and\ \bibinfo {author} {\bibfnamefont {A.~F.}\
			\bibnamefont {Young}},\ }\bibfield  {title} {\bibinfo {title} {The marvels of
			moiré materials},\ }\href {https://doi.org/10.1038/s41578-021-00284-1}
	{\bibfield  {journal} {\bibinfo  {journal} {Nature Reviews Materials}\
		}\textbf {\bibinfo {volume} {6}},\ \bibinfo {pages} {201} (\bibinfo {year}
		{2021})}\BibitemShut {NoStop}%
	\bibitem [{\citenamefont {Bistritzer}\ and\ \citenamefont
		{MacDonald}(2011)}]{Bistritzer2011}%
	\BibitemOpen
	\bibfield  {author} {\bibinfo {author} {\bibfnamefont {R.}~\bibnamefont
			{Bistritzer}}\ and\ \bibinfo {author} {\bibfnamefont {A.~H.}\ \bibnamefont
			{MacDonald}},\ }\bibfield  {title} {\bibinfo {title} {Moiré bands in twisted
			double-layer graphene},\ }\href {https://doi.org/10.1073/pnas.1108174108}
	{\bibfield  {journal} {\bibinfo  {journal} {Proceedings of the National
				Academy of Sciences}\ }\textbf {\bibinfo {volume} {108}},\ \bibinfo {pages}
		{12233} (\bibinfo {year} {2011})}\BibitemShut {NoStop}%
	\bibitem [{\citenamefont {Yankowitz}\ \emph {et~al.}(2019)\citenamefont
		{Yankowitz}, \citenamefont {Chen}, \citenamefont {Polshyn}, \citenamefont
		{Zhang}, \citenamefont {Watanabe}, \citenamefont {Taniguchi}, \citenamefont
		{Graf}, \citenamefont {Young},\ and\ \citenamefont {Dean}}]{Yankowitz2019}%
	\BibitemOpen
	\bibfield  {author} {\bibinfo {author} {\bibfnamefont {M.}~\bibnamefont
			{Yankowitz}}, \bibinfo {author} {\bibfnamefont {S.}~\bibnamefont {Chen}},
		\bibinfo {author} {\bibfnamefont {H.}~\bibnamefont {Polshyn}}, \bibinfo
		{author} {\bibfnamefont {Y.}~\bibnamefont {Zhang}}, \bibinfo {author}
		{\bibfnamefont {K.}~\bibnamefont {Watanabe}}, \bibinfo {author}
		{\bibfnamefont {T.}~\bibnamefont {Taniguchi}}, \bibinfo {author}
		{\bibfnamefont {D.}~\bibnamefont {Graf}}, \bibinfo {author} {\bibfnamefont
			{A.~F.}\ \bibnamefont {Young}},\ and\ \bibinfo {author} {\bibfnamefont
			{C.~R.}\ \bibnamefont {Dean}},\ }\bibfield  {title} {\bibinfo {title} {Tuning
			superconductivity in twisted bilayer graphene},\ }\href
	{https://doi.org/10.1126/science.aav1910} {\bibfield  {journal} {\bibinfo
			{journal} {Science}\ }\textbf {\bibinfo {volume} {363}},\ \bibinfo {pages}
		{1059} (\bibinfo {year} {2019})}\BibitemShut {NoStop}%
	\bibitem [{\citenamefont {Taillefer}(2010)}]{Taillefer2010}%
	\BibitemOpen
	\bibfield  {author} {\bibinfo {author} {\bibfnamefont {L.}~\bibnamefont
			{Taillefer}},\ }\bibfield  {title} {\bibinfo {title} {Scattering and pairing
			in cuprate superconductors},\ }\href
	{https://doi.org/10.1146/annurev-conmatphys-070909-104117} {\bibfield
		{journal} {\bibinfo  {journal} {Annual Review of Condensed Matter Physics}\
		}\textbf {\bibinfo {volume} {1}},\ \bibinfo {pages} {51} (\bibinfo {year}
		{2010})}\BibitemShut {NoStop}%
	\bibitem [{\citenamefont {Singh}(2021)}]{Singh2021}%
	\BibitemOpen
	\bibfield  {author} {\bibinfo {author} {\bibfnamefont {N.}~\bibnamefont
			{Singh}},\ }\bibfield  {title} {\bibinfo {title} {Leading theories of the
			cuprate superconductivity: {A} critique},\ }\href
	{https://doi.org/10.1016/j.physc.2020.1353782} {\bibfield  {journal}
		{\bibinfo  {journal} {Physica C: Superconductivity and its Applications}\
		}\textbf {\bibinfo {volume} {580}},\ \bibinfo {pages} {1353782} (\bibinfo
		{year} {2021})}\BibitemShut {NoStop}%
	\bibitem [{\citenamefont {Reich}\ \emph {et~al.}(2002)\citenamefont {Reich},
		\citenamefont {Maultzsch}, \citenamefont {Thomsen},\ and\ \citenamefont
		{Ordejón}}]{Reich2002}%
	\BibitemOpen
	\bibfield  {author} {\bibinfo {author} {\bibfnamefont {S.}~\bibnamefont
			{Reich}}, \bibinfo {author} {\bibfnamefont {J.}~\bibnamefont {Maultzsch}},
		\bibinfo {author} {\bibfnamefont {C.}~\bibnamefont {Thomsen}},\ and\ \bibinfo
		{author} {\bibfnamefont {P.}~\bibnamefont {Ordejón}},\ }\bibfield  {title}
	{\bibinfo {title} {Tight-binding description of graphene},\ }\href
	{https://doi.org/10.1103/PhysRevB.66.035412} {\bibfield  {journal} {\bibinfo
			{journal} {Physical Review B}\ }\textbf {\bibinfo {volume} {66}},\ \bibinfo
		{pages} {035412} (\bibinfo {year} {2002})}\BibitemShut {NoStop}%
	\bibitem [{\citenamefont {Suárez~Morell}\ \emph {et~al.}(2010)\citenamefont
		{Suárez~Morell}, \citenamefont {Correa}, \citenamefont {Vargas},
		\citenamefont {Pacheco},\ and\ \citenamefont
		{Barticevic}}]{SuarezMorell2010}%
	\BibitemOpen
	\bibfield  {author} {\bibinfo {author} {\bibfnamefont {E.}~\bibnamefont
			{Suárez~Morell}}, \bibinfo {author} {\bibfnamefont {J.~D.}\ \bibnamefont
			{Correa}}, \bibinfo {author} {\bibfnamefont {P.}~\bibnamefont {Vargas}},
		\bibinfo {author} {\bibfnamefont {M.}~\bibnamefont {Pacheco}},\ and\ \bibinfo
		{author} {\bibfnamefont {Z.}~\bibnamefont {Barticevic}},\ }\bibfield  {title}
	{\bibinfo {title} {Flat bands in slightly twisted bilayer graphene:
			{Tight}-binding calculations},\ }\href
	{https://doi.org/10.1103/PhysRevB.82.121407} {\bibfield  {journal} {\bibinfo
			{journal} {Physical Review B}\ }\textbf {\bibinfo {volume} {82}},\ \bibinfo
		{pages} {121407} (\bibinfo {year} {2010})}\BibitemShut {NoStop}%
	\bibitem [{\citenamefont {Sboychakov}\ \emph {et~al.}(2015)\citenamefont
		{Sboychakov}, \citenamefont {Rakhmanov}, \citenamefont {Rozhkov},\ and\
		\citenamefont {Nori}}]{Sboychakov2015}%
	\BibitemOpen
	\bibfield  {author} {\bibinfo {author} {\bibfnamefont {A.~O.}\ \bibnamefont
			{Sboychakov}}, \bibinfo {author} {\bibfnamefont {A.~L.}\ \bibnamefont
			{Rakhmanov}}, \bibinfo {author} {\bibfnamefont {A.~V.}\ \bibnamefont
			{Rozhkov}},\ and\ \bibinfo {author} {\bibfnamefont {F.}~\bibnamefont
			{Nori}},\ }\bibfield  {title} {\bibinfo {title} {Electronic spectrum of
			twisted bilayer graphene},\ }\href
	{https://doi.org/10.1103/physrevb.92.075402} {\bibfield  {journal} {\bibinfo
			{journal} {Physical Review B}\ }\textbf {\bibinfo {volume} {92}},\ \bibinfo
		{pages} {075402} (\bibinfo {year} {2015})}\BibitemShut {NoStop}%
	\bibitem [{\citenamefont {Lin}\ and\ \citenamefont {Tománek}(2018)}]{Lin2018}%
	\BibitemOpen
	\bibfield  {author} {\bibinfo {author} {\bibfnamefont {X.}~\bibnamefont
			{Lin}}\ and\ \bibinfo {author} {\bibfnamefont {D.}~\bibnamefont {Tománek}},\
	}\bibfield  {title} {\bibinfo {title} {Minimum model for the electronic
			structure of twisted bilayer graphene and related structures},\ }\href
	{https://doi.org/10.1103/PhysRevB.98.081410} {\bibfield  {journal} {\bibinfo
			{journal} {Physical Review B}\ }\textbf {\bibinfo {volume} {98}},\ \bibinfo
		{pages} {081410} (\bibinfo {year} {2018})}\BibitemShut {NoStop}%
	\bibitem [{\citenamefont {Carr}\ \emph
		{et~al.}(2019{\natexlab{a}})\citenamefont {Carr}, \citenamefont {Fang},
		\citenamefont {Po}, \citenamefont {Vishwanath},\ and\ \citenamefont
		{Kaxiras}}]{Carr2019}%
	\BibitemOpen
	\bibfield  {author} {\bibinfo {author} {\bibfnamefont {S.}~\bibnamefont
			{Carr}}, \bibinfo {author} {\bibfnamefont {S.}~\bibnamefont {Fang}}, \bibinfo
		{author} {\bibfnamefont {H.~C.}\ \bibnamefont {Po}}, \bibinfo {author}
		{\bibfnamefont {A.}~\bibnamefont {Vishwanath}},\ and\ \bibinfo {author}
		{\bibfnamefont {E.}~\bibnamefont {Kaxiras}},\ }\bibfield  {title} {\bibinfo
		{title} {Derivation of {Wannier} orbitals and minimal-basis tight-binding
			{Hamiltonians} for twisted bilayer graphene: {First}-principles approach},\
	}\href {https://doi.org/10.1103/PhysRevResearch.1.033072} {\bibfield
		{journal} {\bibinfo  {journal} {Physical Review Research}\ }\textbf {\bibinfo
			{volume} {1}},\ \bibinfo {pages} {033072} (\bibinfo {year}
		{2019}{\natexlab{a}})}\BibitemShut {NoStop}%
	\bibitem [{\citenamefont {Lopes~dos Santos}\ \emph {et~al.}(2012)\citenamefont
		{Lopes~dos Santos}, \citenamefont {Peres},\ and\ \citenamefont
		{Castro~Neto}}]{LopesdosSantos2012}%
	\BibitemOpen
	\bibfield  {author} {\bibinfo {author} {\bibfnamefont {J.~M.~B.}\
			\bibnamefont {Lopes~dos Santos}}, \bibinfo {author} {\bibfnamefont
			{N.~M.~R.}\ \bibnamefont {Peres}},\ and\ \bibinfo {author} {\bibfnamefont
			{A.~H.}\ \bibnamefont {Castro~Neto}},\ }\bibfield  {title} {\bibinfo {title}
		{Continuum model of the twisted graphene bilayer},\ }\href
	{https://doi.org/10.1103/PhysRevB.86.155449} {\bibfield  {journal} {\bibinfo
			{journal} {Physical Review B}\ }\textbf {\bibinfo {volume} {86}},\ \bibinfo
		{pages} {155449} (\bibinfo {year} {2012})}\BibitemShut {NoStop}%
	\bibitem [{\citenamefont {Moon}\ and\ \citenamefont
		{Koshino}(2013)}]{Moon2013}%
	\BibitemOpen
	\bibfield  {author} {\bibinfo {author} {\bibfnamefont {P.}~\bibnamefont
			{Moon}}\ and\ \bibinfo {author} {\bibfnamefont {M.}~\bibnamefont {Koshino}},\
	}\bibfield  {title} {\bibinfo {title} {Optical absorption in twisted bilayer
			graphene},\ }\href {https://doi.org/10.1103/PhysRevB.87.205404} {\bibfield
		{journal} {\bibinfo  {journal} {Physical Review B}\ }\textbf {\bibinfo
			{volume} {87}},\ \bibinfo {pages} {205404} (\bibinfo {year}
		{2013})}\BibitemShut {NoStop}%
	\bibitem [{\citenamefont {Mele}(2010)}]{Mele2010}%
	\BibitemOpen
	\bibfield  {author} {\bibinfo {author} {\bibfnamefont {E.~J.}\ \bibnamefont
			{Mele}},\ }\bibfield  {title} {\bibinfo {title} {Commensuration and
			interlayer coherence in twisted bilayer graphene},\ }\href
	{https://doi.org/10.1103/PhysRevB.81.161405} {\bibfield  {journal} {\bibinfo
			{journal} {Physical Review B}\ }\textbf {\bibinfo {volume} {81}},\ \bibinfo
		{pages} {161405} (\bibinfo {year} {2010})}\BibitemShut {NoStop}%
	\bibitem [{\citenamefont {Shallcross}\ \emph {et~al.}(2010)\citenamefont
		{Shallcross}, \citenamefont {Sharma}, \citenamefont {Kandelaki},\ and\
		\citenamefont {Pankratov}}]{Shallcross2010}%
	\BibitemOpen
	\bibfield  {author} {\bibinfo {author} {\bibfnamefont {S.}~\bibnamefont
			{Shallcross}}, \bibinfo {author} {\bibfnamefont {S.}~\bibnamefont {Sharma}},
		\bibinfo {author} {\bibfnamefont {E.}~\bibnamefont {Kandelaki}},\ and\
		\bibinfo {author} {\bibfnamefont {O.~A.}\ \bibnamefont {Pankratov}},\
	}\bibfield  {title} {\bibinfo {title} {Electronic structure of turbostratic
			graphene},\ }\href {https://doi.org/10.1103/PhysRevB.81.165105} {\bibfield
		{journal} {\bibinfo  {journal} {Physical Review B}\ }\textbf {\bibinfo
			{volume} {81}},\ \bibinfo {pages} {165105} (\bibinfo {year}
		{2010})}\BibitemShut {NoStop}%
	\bibitem [{\citenamefont {Carr}\ \emph {et~al.}(2020)\citenamefont {Carr},
		\citenamefont {Fang},\ and\ \citenamefont {Kaxiras}}]{Carr2020}%
	\BibitemOpen
	\bibfield  {author} {\bibinfo {author} {\bibfnamefont {S.}~\bibnamefont
			{Carr}}, \bibinfo {author} {\bibfnamefont {S.}~\bibnamefont {Fang}},\ and\
		\bibinfo {author} {\bibfnamefont {E.}~\bibnamefont {Kaxiras}},\ }\bibfield
	{title} {\bibinfo {title} {Electronic-structure methods for twisted moiré
			layers},\ }\href {https://doi.org/10.1038/s41578-020-0214-0} {\bibfield
		{journal} {\bibinfo  {journal} {Nature Reviews Materials}\ }\textbf {\bibinfo
			{volume} {5}},\ \bibinfo {pages} {748} (\bibinfo {year} {2020})}\BibitemShut
	{NoStop}%
	\bibitem [{\citenamefont {Leconte}\ \emph {et~al.}(2022)\citenamefont
		{Leconte}, \citenamefont {Javvaji}, \citenamefont {An}, \citenamefont
		{Samudrala},\ and\ \citenamefont {Jung}}]{Leconte2022}%
	\BibitemOpen
	\bibfield  {author} {\bibinfo {author} {\bibfnamefont {N.}~\bibnamefont
			{Leconte}}, \bibinfo {author} {\bibfnamefont {S.}~\bibnamefont {Javvaji}},
		\bibinfo {author} {\bibfnamefont {J.}~\bibnamefont {An}}, \bibinfo {author}
		{\bibfnamefont {A.}~\bibnamefont {Samudrala}},\ and\ \bibinfo {author}
		{\bibfnamefont {J.}~\bibnamefont {Jung}},\ }\bibfield  {title} {\bibinfo
		{title} {Relaxation effects in twisted bilayer graphene: A multiscale
			approach},\ }\href {https://doi.org/10.1103/physrevb.106.115410} {\bibfield
		{journal} {\bibinfo  {journal} {Physical Review B}\ }\textbf {\bibinfo
			{volume} {106}},\ \bibinfo {pages} {115410} (\bibinfo {year}
		{2022})}\BibitemShut {NoStop}%
	\bibitem [{\citenamefont {Haddadi}\ \emph {et~al.}(2020)\citenamefont
		{Haddadi}, \citenamefont {Wu}, \citenamefont {Kruchkov},\ and\ \citenamefont
		{Yazyev}}]{Haddadi2020}%
	\BibitemOpen
	\bibfield  {author} {\bibinfo {author} {\bibfnamefont {F.}~\bibnamefont
			{Haddadi}}, \bibinfo {author} {\bibfnamefont {Q.}~\bibnamefont {Wu}},
		\bibinfo {author} {\bibfnamefont {A.~J.}\ \bibnamefont {Kruchkov}},\ and\
		\bibinfo {author} {\bibfnamefont {O.~V.}\ \bibnamefont {Yazyev}},\ }\bibfield
	{title} {\bibinfo {title} {Moiré {Flat} {Bands} in {Twisted} {Double}
			{Bilayer} {Graphene}},\ }\href {https://doi.org/10.1021/acs.nanolett.9b05117}
	{\bibfield  {journal} {\bibinfo  {journal} {Nano Letters}\ }\textbf {\bibinfo
			{volume} {20}},\ \bibinfo {pages} {2410} (\bibinfo {year}
		{2020})}\BibitemShut {NoStop}%
	\bibitem [{\citenamefont {Koshino}(2015)}]{Koshino2015}%
	\BibitemOpen
	\bibfield  {author} {\bibinfo {author} {\bibfnamefont {M.}~\bibnamefont
			{Koshino}},\ }\bibfield  {title} {\bibinfo {title} {Interlayer interaction in
			general incommensurate atomic layers},\ }\href
	{https://doi.org/10.1088/1367-2630/17/1/015014} {\bibfield  {journal}
		{\bibinfo  {journal} {New Journal of Physics}\ }\textbf {\bibinfo {volume}
			{17}},\ \bibinfo {pages} {015014} (\bibinfo {year} {2015})}\BibitemShut
	{NoStop}%
	\bibitem [{\citenamefont {Carr}\ \emph
		{et~al.}(2019{\natexlab{b}})\citenamefont {Carr}, \citenamefont {Fang},
		\citenamefont {Zhu},\ and\ \citenamefont {Kaxiras}}]{Carr2019a}%
	\BibitemOpen
	\bibfield  {author} {\bibinfo {author} {\bibfnamefont {S.}~\bibnamefont
			{Carr}}, \bibinfo {author} {\bibfnamefont {S.}~\bibnamefont {Fang}}, \bibinfo
		{author} {\bibfnamefont {Z.}~\bibnamefont {Zhu}},\ and\ \bibinfo {author}
		{\bibfnamefont {E.}~\bibnamefont {Kaxiras}},\ }\bibfield  {title} {\bibinfo
		{title} {Exact continuum model for low-energy electronic states of twisted
			bilayer graphene},\ }\href {https://doi.org/10.1103/PhysRevResearch.1.013001}
	{\bibfield  {journal} {\bibinfo  {journal} {Physical Review Research}\
		}\textbf {\bibinfo {volume} {1}},\ \bibinfo {pages} {013001} (\bibinfo {year}
		{2019}{\natexlab{b}})}\BibitemShut {NoStop}%
	\bibitem [{\citenamefont {Guinea}\ and\ \citenamefont
		{Walet}(2019)}]{Guinea2019}%
	\BibitemOpen
	\bibfield  {author} {\bibinfo {author} {\bibfnamefont {F.}~\bibnamefont
			{Guinea}}\ and\ \bibinfo {author} {\bibfnamefont {N.~R.}\ \bibnamefont
			{Walet}},\ }\bibfield  {title} {\bibinfo {title} {Continuum models for
			twisted bilayer graphene: {Effect} of lattice deformation and hopping
			parameters},\ }\href {https://doi.org/10.1103/PhysRevB.99.205134} {\bibfield
		{journal} {\bibinfo  {journal} {Physical Review B}\ }\textbf {\bibinfo
			{volume} {99}},\ \bibinfo {pages} {205134} (\bibinfo {year}
		{2019})}\BibitemShut {NoStop}%
	\bibitem [{\citenamefont {Bernevig}\ \emph {et~al.}(2021)\citenamefont
		{Bernevig}, \citenamefont {Song}, \citenamefont {Regnault},\ and\
		\citenamefont {Lian}}]{Bernevig2021}%
	\BibitemOpen
	\bibfield  {author} {\bibinfo {author} {\bibfnamefont {B.~A.}\ \bibnamefont
			{Bernevig}}, \bibinfo {author} {\bibfnamefont {Z.-D.}\ \bibnamefont {Song}},
		\bibinfo {author} {\bibfnamefont {N.}~\bibnamefont {Regnault}},\ and\
		\bibinfo {author} {\bibfnamefont {B.}~\bibnamefont {Lian}},\ }\bibfield
	{title} {\bibinfo {title} {Twisted bilayer graphene. I. {Matrix} elements,
			approximations, perturbation theory, and a
			$k\cdot p$ two-band model},\ }\href {https://doi.org/10.1103/PhysRevB.103.205411}
	{\bibfield  {journal} {\bibinfo  {journal} {Physical Review B}\ }\textbf
		{\bibinfo {volume} {103}},\ \bibinfo {pages} {205411} (\bibinfo {year}
		{2021})}\BibitemShut {NoStop}%
	\bibitem [{\citenamefont {Kang}\ and\ \citenamefont {Vafek}(2023)}]{Kang2023}%
	\BibitemOpen
	\bibfield  {author} {\bibinfo {author} {\bibfnamefont {J.}~\bibnamefont
			{Kang}}\ and\ \bibinfo {author} {\bibfnamefont {O.}~\bibnamefont {Vafek}},\
	}\bibfield  {title} {\bibinfo {title} {Pseudomagnetic fields, particle-hole
			asymmetry, and microscopic effective continuum {Hamiltonians} of twisted
			bilayer graphene},\ }\href {https://doi.org/10.1103/PhysRevB.107.075408}
	{\bibfield  {journal} {\bibinfo  {journal} {Physical Review B}\ }\textbf
		{\bibinfo {volume} {107}},\ \bibinfo {pages} {075408} (\bibinfo {year}
		{2023})}\BibitemShut {NoStop}%
	\bibitem [{\citenamefont {Miao}\ \emph {et~al.}()\citenamefont {Miao},
		\citenamefont {Li}, \citenamefont {Han}, \citenamefont {Pan},\ and\
		\citenamefont {Dai}}]{Miao2023}%
	\BibitemOpen
	\bibfield  {author} {\bibinfo {author} {\bibfnamefont {W.}~\bibnamefont
			{Miao}}, \bibinfo {author} {\bibfnamefont {C.}~\bibnamefont {Li}}, \bibinfo
		{author} {\bibfnamefont {X.}~\bibnamefont {Han}}, \bibinfo {author}
		{\bibfnamefont {D.}~\bibnamefont {Pan}},\ and\ \bibinfo {author}
		{\bibfnamefont {X.}~\bibnamefont {Dai}},\ }\bibfield  {title} {\bibinfo
		{title} {Truncated atomic plane wave method for subband structure
			calculations of moiré systems},\ }\href
	{https://doi.org/10.1103/PhysRevB.107.125112} {\bibfield  {journal} {\bibinfo
			{journal} {Physical Review B}\ }\textbf {\bibinfo {volume} {107}},\ \bibinfo
		{pages} {125112} (\bibinfo {year} {2023})}\BibitemShut {NoStop}%
	\bibitem [{\citenamefont {Ceferino}\ and\ \citenamefont
		{Guinea}(2024)}]{Ceferino2024}%
	\BibitemOpen
	\bibfield  {author} {\bibinfo {author} {\bibfnamefont {A.}~\bibnamefont
			{Ceferino}}\ and\ \bibinfo {author} {\bibfnamefont {F.}~\bibnamefont
			{Guinea}},\ }\bibfield  {title} {\bibinfo {title} {Pseudomagnetic fields in
			fully relaxed twisted bilayer and trilayer graphene},\ }
	\href {https://doi.org/10.1088/2053-1583/ad3b0e}
	{\bibfield  {journal} {\bibinfo  {journal} {2D Materials}\ }\textbf {\bibinfo
			{volume} {11}},\ \bibinfo {pages} {035015} (\bibinfo {year}
		{2024})}\BibitemShut {NoStop}%
	\bibitem [{\citenamefont {Becker}\ \emph {et~al.}(2022)\citenamefont {Becker},
		\citenamefont {Embree}, \citenamefont {Wittsten},\ and\ \citenamefont
		{Zworski}}]{Becker2022}%
	\BibitemOpen
	\bibfield  {author} {\bibinfo {author} {\bibfnamefont {S.}~\bibnamefont
			{Becker}}, \bibinfo {author} {\bibfnamefont {M.}~\bibnamefont {Embree}},
		\bibinfo {author} {\bibfnamefont {J.}~\bibnamefont {Wittsten}},\ and\
		\bibinfo {author} {\bibfnamefont {M.}~\bibnamefont {Zworski}},\ }\bibfield
	{title} {\bibinfo {title} {Mathematics of magic angles in a model of twisted
			bilayer graphene},\ }\href {https://doi.org/10.2140/pmp.2022.3.69} {\bibfield
		{journal} {\bibinfo  {journal} {Probability and Mathematical Physics}\
		}\textbf {\bibinfo {volume} {3}},\ \bibinfo {pages} {69} (\bibinfo {year}
		{2022})}\BibitemShut {NoStop}%
	\bibitem [{\citenamefont {Becker}\ \emph {et~al.}(2023)\citenamefont {Becker},
		\citenamefont {Humbert},\ and\ \citenamefont {Zworski}}]{Becker2023}%
	\BibitemOpen
	\bibfield  {author} {\bibinfo {author} {\bibfnamefont {S.}~\bibnamefont
			{Becker}}, \bibinfo {author} {\bibfnamefont {T.}~\bibnamefont {Humbert}},\
		and\ \bibinfo {author} {\bibfnamefont {M.}~\bibnamefont {Zworski}},\
	}\bibfield  {title} {\bibinfo {title} {Integrability in the {Chiral} {Model}
			of {Magic} {Angles}},\ }\href {https://doi.org/10.1007/s00220-023-04814-6}
	{\bibfield  {journal} {\bibinfo  {journal} {Communications in Mathematical
				Physics}\ }\textbf {\bibinfo {volume} {403}},\ \bibinfo {pages} {1153}
		(\bibinfo {year} {2023})}\BibitemShut {NoStop}%
	\bibitem [{\citenamefont {Watson}\ \emph {et~al.}(2023)\citenamefont {Watson},
		\citenamefont {Kong}, \citenamefont {MacDonald},\ and\ \citenamefont
		{Luskin}}]{Watson2023}%
	\BibitemOpen
	\bibfield  {author} {\bibinfo {author} {\bibfnamefont {A.~B.}\ \bibnamefont
			{Watson}}, \bibinfo {author} {\bibfnamefont {T.}~\bibnamefont {Kong}},
		\bibinfo {author} {\bibfnamefont {A.~H.}\ \bibnamefont {MacDonald}},\ and\
		\bibinfo {author} {\bibfnamefont {M.}~\bibnamefont {Luskin}},\ }\bibfield
	{title} {\bibinfo {title} {Bistritzer–{MacDonald} dynamics in twisted
			bilayer graphene},\ }\href {https://doi.org/10.1063/5.0115771} {\bibfield
		{journal} {\bibinfo  {journal} {Journal of Mathematical Physics}\ }\textbf
		{\bibinfo {volume} {64}},\ \bibinfo {pages} {031502} (\bibinfo {year}
		{2023})}\BibitemShut {NoStop}%
	\bibitem [{\citenamefont {San-Jose}\ and\ \citenamefont
		{Prada}(2013)}]{SanJose2013}%
	\BibitemOpen
	\bibfield  {author} {\bibinfo {author} {\bibfnamefont {P.}~\bibnamefont
			{San-Jose}}\ and\ \bibinfo {author} {\bibfnamefont {E.}~\bibnamefont
			{Prada}},\ }\bibfield  {title} {\bibinfo {title} {Helical networks in twisted
			bilayer graphene under interlayer bias},\ }\href
	{https://doi.org/10.1103/PhysRevB.88.121408} {\bibfield  {journal} {\bibinfo
			{journal} {Physical Review B}\ }\textbf {\bibinfo {volume} {88}},\ \bibinfo
		{pages} {121408} (\bibinfo {year} {2013})}\BibitemShut {NoStop}%
	\bibitem [{\citenamefont {Efimkin}\ and\ \citenamefont
		{MacDonald}(2018)}]{Efimkin2018}%
	\BibitemOpen
	\bibfield  {author} {\bibinfo {author} {\bibfnamefont {D.~K.}\ \bibnamefont
			{Efimkin}}\ and\ \bibinfo {author} {\bibfnamefont {A.~H.}\ \bibnamefont
			{MacDonald}},\ }\bibfield  {title} {\bibinfo {title} {Helical network model
			for twisted bilayer graphene},\ }\href
	{https://doi.org/10.1103/PhysRevB.98.035404} {\bibfield  {journal} {\bibinfo
			{journal} {Physical Review B}\ }\textbf {\bibinfo {volume} {98}},\ \bibinfo
		{pages} {035404} (\bibinfo {year} {2018})}\BibitemShut {NoStop}%
	\bibitem [{\citenamefont {De~Beule}\ \emph {et~al.}(2021)\citenamefont
		{De~Beule}, \citenamefont {Dominguez},\ and\ \citenamefont
		{Recher}}]{DeBeule2021}%
	\BibitemOpen
	\bibfield  {author} {\bibinfo {author} {\bibfnamefont {C.}~\bibnamefont
			{De~Beule}}, \bibinfo {author} {\bibfnamefont {F.}~\bibnamefont
			{Dominguez}},\ and\ \bibinfo {author} {\bibfnamefont {P.}~\bibnamefont
			{Recher}},\ }\bibfield  {title} {\bibinfo {title} {Network model and
			four-terminal transport in minimally twisted bilayer graphene},\ }\href
	{https://doi.org/10.1103/PhysRevB.104.195410} {\bibfield  {journal} {\bibinfo
			{journal} {Physical Review B}\ }\textbf {\bibinfo {volume} {104}},\ \bibinfo
		{pages} {195410} (\bibinfo {year} {2021})}\BibitemShut {NoStop}%
	\bibitem [{\citenamefont {Chou}\ and\ \citenamefont
		{Das~Sarma}(2023)}]{Chou2023}%
	\BibitemOpen
	\bibfield  {author} {\bibinfo {author} {\bibfnamefont {Y.-Z.}\ \bibnamefont
			{Chou}}\ and\ \bibinfo {author} {\bibfnamefont {S.}~\bibnamefont
			{Das~Sarma}},\ }\bibfield  {title} {\bibinfo {title} {Kondo {Lattice} {Model}
			in {Magic}-{Angle} {Twisted} {Bilayer} {Graphene}},\ }\href
	{https://doi.org/10.1103/PhysRevLett.131.026501} {\bibfield  {journal}
		{\bibinfo  {journal} {Physical Review Letters}\ }\textbf {\bibinfo {volume}
			{131}},\ \bibinfo {pages} {026501} (\bibinfo {year} {2023})}\BibitemShut
	{NoStop}%
	\bibitem [{\citenamefont {Hu}\ \emph {et~al.}(2023)\citenamefont {Hu},
		\citenamefont {Bernevig},\ and\ \citenamefont {Tsvelik}}]{Hu2023}%
	\BibitemOpen
	\bibfield  {author} {\bibinfo {author} {\bibfnamefont {H.}~\bibnamefont
			{Hu}}, \bibinfo {author} {\bibfnamefont {B.~A.}\ \bibnamefont {Bernevig}},\
		and\ \bibinfo {author} {\bibfnamefont {A.~M.}\ \bibnamefont {Tsvelik}},\
	}\bibfield  {title} {\bibinfo {title} {Kondo {Lattice} {Model} of
			{Magic}-{Angle} {Twisted}-{Bilayer} {Graphene}: {Hund}'s {Rule},
			{Local}-{Moment} {Fluctuations}, and {Low}-{Energy} {Effective} {Theory}},\
	}\href {https://doi.org/10.1103/PhysRevLett.131.026502} {\bibfield  {journal}
		{\bibinfo  {journal} {Physical Review Letters}\ }\textbf {\bibinfo {volume}
			{131}},\ \bibinfo {pages} {026502} (\bibinfo {year} {2023})}\BibitemShut
	{NoStop}%
	\bibitem [{\citenamefont {Bennett}\ \emph {et~al.}(2024)\citenamefont
		{Bennett}, \citenamefont {Larson}, \citenamefont {Sharma}, \citenamefont
		{Carr},\ and\ \citenamefont {Kaxiras}}]{Bennett2024}%
	\BibitemOpen
	\bibfield  {author} {\bibinfo {author} {\bibfnamefont {D.}~\bibnamefont
			{Bennett}}, \bibinfo {author} {\bibfnamefont {D.~T.}\ \bibnamefont {Larson}},
		\bibinfo {author} {\bibfnamefont {L.}~\bibnamefont {Sharma}}, \bibinfo
		{author} {\bibfnamefont {S.}~\bibnamefont {Carr}},\ and\ \bibinfo {author}
		{\bibfnamefont {E.}~\bibnamefont {Kaxiras}},\ }\bibfield  {title} {\bibinfo
		{title} {Twisted bilayer graphene revisited: Minimal two-band model for low-energy bands},\ }\href
	{https://doi.org/10.1103/PhysRevB.109.155422} {\bibfield  {journal} {\bibinfo
			{journal} {Physical Review B}\ }\textbf {\bibinfo {volume} {109}},\ \bibinfo
		{pages} {155422} (\bibinfo {year} {2024})}\BibitemShut {NoStop}%
	\bibitem [{\citenamefont {Song}\ and\ \citenamefont
		{Bernevig}(2022)}]{Song2022}%
	\BibitemOpen
	\bibfield  {author} {\bibinfo {author} {\bibfnamefont {Z.-D.}\ \bibnamefont
			{Song}}\ and\ \bibinfo {author} {\bibfnamefont {B.~A.}\ \bibnamefont
			{Bernevig}},\ }\bibfield  {title} {\bibinfo {title} {Magic-{Angle} {Twisted}
			{Bilayer} {Graphene} as a {Topological} {Heavy} {Fermion} {Problem}},\ }\href
	{https://doi.org/10.1103/PhysRevLett.129.047601} {\bibfield  {journal}
		{\bibinfo  {journal} {Physical Review Letters}\ }\textbf {\bibinfo {volume}
			{129}},\ \bibinfo {pages} {047601} (\bibinfo {year} {2022})}\BibitemShut
	{NoStop}%
	\bibitem [{\citenamefont {Shi}\ and\ \citenamefont {Dai}(2022)}]{Shi2022}%
	\BibitemOpen
	\bibfield  {author} {\bibinfo {author} {\bibfnamefont {H.}~\bibnamefont
			{Shi}}\ and\ \bibinfo {author} {\bibfnamefont {X.}~\bibnamefont {Dai}},\
	}\bibfield  {title} {\bibinfo {title} {Heavy-fermion representation for
			twisted bilayer graphene systems},\ }\href
	{https://doi.org/10.1103/PhysRevB.106.245129} {\bibfield  {journal} {\bibinfo
			{journal} {Physical Review B}\ }\textbf {\bibinfo {volume} {106}},\ \bibinfo
		{pages} {245129} (\bibinfo {year} {2022})}\BibitemShut {NoStop}%
	\bibitem [{\citenamefont {Castro~Neto}\ \emph {et~al.}(2009)\citenamefont
		{Castro~Neto}, \citenamefont {Guinea}, \citenamefont {Peres}, \citenamefont
		{Novoselov},\ and\ \citenamefont {Geim}}]{CastroNeto2009}%
	\BibitemOpen
	\bibfield  {author} {\bibinfo {author} {\bibfnamefont {A.~H.}\ \bibnamefont
			{Castro~Neto}}, \bibinfo {author} {\bibfnamefont {F.}~\bibnamefont {Guinea}},
		\bibinfo {author} {\bibfnamefont {N.~M.~R.}\ \bibnamefont {Peres}}, \bibinfo
		{author} {\bibfnamefont {K.~S.}\ \bibnamefont {Novoselov}},\ and\ \bibinfo
		{author} {\bibfnamefont {A.~K.}\ \bibnamefont {Geim}},\ }\bibfield  {title}
	{\bibinfo {title} {The electronic properties of graphene},\ }\href
	{https://doi.org/10.1103/RevModPhys.81.109} {\bibfield  {journal} {\bibinfo
			{journal} {Reviews of Modern Physics}\ }\textbf {\bibinfo {volume} {81}},\
		\bibinfo {pages} {109} (\bibinfo {year} {2009})}\BibitemShut {NoStop}%
	\bibitem [{\citenamefont {Das~Sarma}\ \emph {et~al.}(2011)\citenamefont
		{Das~Sarma}, \citenamefont {Adam}, \citenamefont {Hwang},\ and\ \citenamefont
		{Rossi}}]{DasSarma2011}%
	\BibitemOpen
	\bibfield  {author} {\bibinfo {author} {\bibfnamefont {S.}~\bibnamefont
			{Das~Sarma}}, \bibinfo {author} {\bibfnamefont {S.}~\bibnamefont {Adam}},
		\bibinfo {author} {\bibfnamefont {E.~H.}\ \bibnamefont {Hwang}},\ and\
		\bibinfo {author} {\bibfnamefont {E.}~\bibnamefont {Rossi}},\ }\bibfield
	{title} {\bibinfo {title} {Electronic transport in two-dimensional
			graphene},\ }\href {https://doi.org/10.1103/revmodphys.83.407} {\bibfield
		{journal} {\bibinfo  {journal} {Reviews of Modern Physics}\ }\textbf
		{\bibinfo {volume} {83}},\ \bibinfo {pages} {407} (\bibinfo {year}
		{2011})}\BibitemShut {NoStop}%
	\bibitem [{\citenamefont {Song}\ \emph {et~al.}(2021)\citenamefont {Song},
		\citenamefont {Lian}, \citenamefont {Regnault},\ and\ \citenamefont
		{Bernevig}}]{Song2021}%
	\BibitemOpen
	\bibfield  {author} {\bibinfo {author} {\bibfnamefont {Z.-D.}\ \bibnamefont
			{Song}}, \bibinfo {author} {\bibfnamefont {B.}~\bibnamefont {Lian}}, \bibinfo
		{author} {\bibfnamefont {N.}~\bibnamefont {Regnault}},\ and\ \bibinfo
		{author} {\bibfnamefont {B.~A.}\ \bibnamefont {Bernevig}},\ }\bibfield
	{title} {\bibinfo {title} {Twisted bilayer graphene. {II}. {Stable} symmetry
			anomaly},\ }\href {https://doi.org/10.1103/PhysRevB.103.205412} {\bibfield
		{journal} {\bibinfo  {journal} {Physical Review B}\ }\textbf {\bibinfo
			{volume} {103}},\ \bibinfo {pages} {205412} (\bibinfo {year}
		{2021})}\BibitemShut {NoStop}%
	\bibitem [{\citenamefont {Jain}\ \emph {et~al.}(2016)\citenamefont {Jain},
		\citenamefont {Juričić},\ and\ \citenamefont {Barkema}}]{Jain2016}%
	\BibitemOpen
	\bibfield  {author} {\bibinfo {author} {\bibfnamefont {S.~K.}\ \bibnamefont
			{Jain}}, \bibinfo {author} {\bibfnamefont {V.}~\bibnamefont {Juričić}},\
		and\ \bibinfo {author} {\bibfnamefont {G.~T.}\ \bibnamefont {Barkema}},\
	}\bibfield  {title} {\bibinfo {title} {Structure of twisted and buckled
			bilayer graphene},\ }\href {https://doi.org/10.1088/2053-1583/4/1/015018}
	{\bibfield  {journal} {\bibinfo  {journal} {2D Materials}\ }\textbf {\bibinfo
			{volume} {4}},\ \bibinfo {pages} {015018} (\bibinfo {year}
		{2016})}\BibitemShut {NoStop}%
	\bibitem [{\citenamefont {Koshino}\ and\ \citenamefont
		{Nam}(2020)}]{Koshino2020}%
	\BibitemOpen
	\bibfield  {author} {\bibinfo {author} {\bibfnamefont {M.}~\bibnamefont
			{Koshino}}\ and\ \bibinfo {author} {\bibfnamefont {N.~N.~T.}\ \bibnamefont
			{Nam}},\ }\bibfield  {title} {\bibinfo {title} {Effective continuum model for
			relaxed twisted bilayer graphene and moiré electron-phonon
			interaction},\ }\href {https://doi.org/10.1103/PhysRevB.101.195425}
	{\bibfield  {journal} {\bibinfo  {journal} {Physical Review B}\ }\textbf
		{\bibinfo {volume} {101}},\ \bibinfo {pages} {195425} (\bibinfo {year}
		{2020})}\BibitemShut {NoStop}%
	\bibitem [{\citenamefont {Tarnopolsky}\ \emph {et~al.}(2019)\citenamefont
		{Tarnopolsky}, \citenamefont {Kruchkov},\ and\ \citenamefont
		{Vishwanath}}]{Tarnopolsky2019}%
	\BibitemOpen
	\bibfield  {author} {\bibinfo {author} {\bibfnamefont {G.}~\bibnamefont
			{Tarnopolsky}}, \bibinfo {author} {\bibfnamefont {A.~J.}\ \bibnamefont
			{Kruchkov}},\ and\ \bibinfo {author} {\bibfnamefont {A.}~\bibnamefont
			{Vishwanath}},\ }\bibfield  {title} {\bibinfo {title} {Origin of {Magic}
			{Angles} in {Twisted} {Bilayer} {Graphene}},\ }\href
	{https://doi.org/10.1103/PhysRevLett.122.106405} {\bibfield  {journal}
		{\bibinfo  {journal} {Physical Review Letters}\ }\textbf {\bibinfo {volume}
			{122}},\ \bibinfo {pages} {106405} (\bibinfo {year} {2019})}\BibitemShut
	{NoStop}%
	\bibitem [{\citenamefont {Mahan}(1990)}]{Mahan1990}%
	\BibitemOpen
	\bibfield  {author} {\bibinfo {author} {\bibfnamefont {G.~D.}\ \bibnamefont
			{Mahan}},\ }\href {https://doi.org/10.1007/978-1-4613-1469-1} {\emph
		{\bibinfo {title} {Many-Particle Physics}}}\ (\bibinfo  {publisher} {Springer
		US},\ \bibinfo {year} {1990})\BibitemShut {NoStop}%
	\bibitem [{\citenamefont {Jishi}(2013)}]{Jishi2013}%
	\BibitemOpen
	\bibfield  {author} {\bibinfo {author} {\bibfnamefont {R.~A.}\ \bibnamefont
			{Jishi}},\ }\href {https://doi.org/10.1017/cbo9781139177771} {\emph {\bibinfo
			{title} {Feynman Diagram Techniques in Condensed Matter Physics}}}\ (\bibinfo
	{publisher} {Cambridge University Press},\ \bibinfo {year}
	{2013})\BibitemShut {NoStop}%
	\bibitem [{\citenamefont {Coleman}(2015)}]{Coleman2015}%
	\BibitemOpen
	\bibfield  {author} {\bibinfo {author} {\bibfnamefont {P.}~\bibnamefont
			{Coleman}},\ }\href {https://doi.org/10.1017/cbo9781139020916} {\emph
		{\bibinfo {title} {Introduction to Many-Body Physics}}}\ (\bibinfo
	{publisher} {Cambridge University Press},\ \bibinfo {year}
	{2015})\BibitemShut {NoStop}%
	\bibitem [{\citenamefont {Berthod}(2018)}]{Berthod2018}%
	\BibitemOpen
	\bibfield  {author} {\bibinfo {author} {\bibfnamefont {C.}~\bibnamefont
			{Berthod}},\ }\href
	{https://iopscience.iop.org/book/mono/978-0-7503-1741-2.pdf} {\emph {\bibinfo
			{title} {Spectroscopic {Probes} of {Quantum} {Matter}}}}\ (\bibinfo
	{publisher} {IOP Publishing},\ \bibinfo {year} {2018})\BibitemShut {NoStop}%
	\bibitem [{\citenamefont {Long}\ \emph {et~al.}(2023)\citenamefont {Long},
		\citenamefont {Zhan}, \citenamefont {Pantaleón}, \citenamefont
		{Silva-Guillén}, \citenamefont {Guinea},\ and\ \citenamefont
		{Yuan}}]{Long2023}%
	\BibitemOpen
	\bibfield  {author} {\bibinfo {author} {\bibfnamefont {M.}~\bibnamefont
			{Long}}, \bibinfo {author} {\bibfnamefont {Z.}~\bibnamefont {Zhan}}, \bibinfo
		{author} {\bibfnamefont {P.~A.}\ \bibnamefont {Pantaleón}}, \bibinfo
		{author} {\bibfnamefont {J.}\ \bibnamefont {Silva-Guillén}}, \bibinfo
		{author} {\bibfnamefont {F.}~\bibnamefont {Guinea}},\ and\ \bibinfo {author}
		{\bibfnamefont {S.}~\bibnamefont {Yuan}},\ }\bibfield  {title} {\bibinfo
		{title} {Electronic properties of twisted bilayer graphene suspended and
			encapsulated with hexagonal boron nitride},\ }\href
	{https://doi.org/10.1103/PhysRevB.107.115140} {\bibfield  {journal} {\bibinfo
			{journal} {Physical Review B}\ }\textbf {\bibinfo {volume} {107}},\ \bibinfo
		{pages} {115140} (\bibinfo {year} {2023})}\BibitemShut {NoStop}%
	\bibitem [{\citenamefont {Long}\ \emph {et~al.}(2022)\citenamefont {Long},
		\citenamefont {Pantaleón}, \citenamefont {Zhan}, \citenamefont {Guinea},
		\citenamefont {Silva-Guillén},\ and\ \citenamefont {Yuan}}]{Long2022}%
	\BibitemOpen
	\bibfield  {author} {\bibinfo {author} {\bibfnamefont {M.}~\bibnamefont
			{Long}}, \bibinfo {author} {\bibfnamefont {P.~A.}\ \bibnamefont
			{Pantaleón}}, \bibinfo {author} {\bibfnamefont {Z.}~\bibnamefont {Zhan}},
		\bibinfo {author} {\bibfnamefont {F.}~\bibnamefont {Guinea}}, \bibinfo
		{author} {\bibfnamefont {J.}\ \bibnamefont {Silva-Guillén}},\ and\
		\bibinfo {author} {\bibfnamefont {S.}~\bibnamefont {Yuan}},\ }\bibfield
	{title} {\bibinfo {title} {An atomistic approach for the structural and
			electronic properties of twisted bilayer graphene-boron nitride
			heterostructures},\ }\href {https://doi.org/10.1038/s41524-022-00763-1}
	{\bibfield  {journal} {\bibinfo  {journal} {npj Computational Materials}\
		}\textbf {\bibinfo {volume} {8}},\ \bibinfo {pages} {1} (\bibinfo {year}
		{2022})}\BibitemShut {NoStop}%
	\bibitem [{\citenamefont {Yu}\ \emph {et~al.}(2023)\citenamefont {Yu},
		\citenamefont {Wang}, \citenamefont {Katsnelson},\ and\ \citenamefont
		{Yuan}}]{Yu2023}%
	\BibitemOpen
	\bibfield  {author} {\bibinfo {author} {\bibfnamefont {G.}~\bibnamefont
			{Yu}}, \bibinfo {author} {\bibfnamefont {Y.}~\bibnamefont {Wang}}, \bibinfo
		{author} {\bibfnamefont {M.~I.}\ \bibnamefont {Katsnelson}},\ and\ \bibinfo
		{author} {\bibfnamefont {S.}~\bibnamefont {Yuan}},\ }\bibfield  {title}
	{\bibinfo {title} {Origin of the magic angle in twisted bilayer graphene from
			hybridization of valence and conduction bands},\ }\href
	{https://doi.org/10.1103/PhysRevB.108.045138} {\bibfield  {journal} {\bibinfo
			{journal} {Physical Review B}\ }\textbf {\bibinfo {volume} {108}},\ \bibinfo
		{pages} {045138} (\bibinfo {year} {2023})}\BibitemShut {NoStop}%
	\bibitem [{\citenamefont {Ohta}\ \emph {et~al.}(2012)\citenamefont {Ohta},
		\citenamefont {Robinson}, \citenamefont {Feibelman}, \citenamefont
		{Bostwick}, \citenamefont {Rotenberg},\ and\ \citenamefont
		{Beechem}}]{Ohta2012}%
	\BibitemOpen
	\bibfield  {author} {\bibinfo {author} {\bibfnamefont {T.}~\bibnamefont
			{Ohta}}, \bibinfo {author} {\bibfnamefont {J.~T.}\ \bibnamefont {Robinson}},
		\bibinfo {author} {\bibfnamefont {P.~J.}\ \bibnamefont {Feibelman}}, \bibinfo
		{author} {\bibfnamefont {A.}~\bibnamefont {Bostwick}}, \bibinfo {author}
		{\bibfnamefont {E.}~\bibnamefont {Rotenberg}},\ and\ \bibinfo {author}
		{\bibfnamefont {T.~E.}\ \bibnamefont {Beechem}},\ }\bibfield  {title}
	{\bibinfo {title} {Evidence for {Interlayer} {Coupling} and
			{Moiré} {Periodic} {Potentials} in {Twisted} {Bilayer}
			{Graphene}},\ }\href {https://doi.org/10.1103/PhysRevLett.109.186807}
	{\bibfield  {journal} {\bibinfo  {journal} {Physical Review Letters}\
		}\textbf {\bibinfo {volume} {109}},\ \bibinfo {pages} {186807} (\bibinfo
		{year} {2012})}\BibitemShut {NoStop}%
	\bibitem [{\citenamefont {Nishi}\ \emph {et~al.}(2017)\citenamefont {Nishi},
		\citenamefont {Matsushita},\ and\ \citenamefont {Oshiyama}}]{Nishi2017}%
	\BibitemOpen
	\bibfield  {author} {\bibinfo {author} {\bibfnamefont {H.}~\bibnamefont
			{Nishi}}, \bibinfo {author} {\bibfnamefont {Y.-i.}\ \bibnamefont
			{Matsushita}},\ and\ \bibinfo {author} {\bibfnamefont {A.}~\bibnamefont
			{Oshiyama}},\ }\bibfield  {title} {\bibinfo {title} {Band-unfolding approach
			to moiré-induced band-gap opening and {Fermi} level velocity
			reduction in twisted bilayer graphene},\ }\href
	{https://doi.org/10.1103/PhysRevB.95.085420} {\bibfield  {journal} {\bibinfo
			{journal} {Physical Review B}\ }\textbf {\bibinfo {volume} {95}},\ \bibinfo
		{pages} {085420} (\bibinfo {year} {2017})}\BibitemShut {NoStop}%
	\bibitem [{\citenamefont {Jones}\ \emph {et~al.}(2020)\citenamefont {Jones},
		\citenamefont {Muzzio}, \citenamefont {Majchrzak}, \citenamefont {Pakdel},
		\citenamefont {Curcio}, \citenamefont {Volckaert}, \citenamefont {Biswas},
		\citenamefont {Gobbo}, \citenamefont {Singh}, \citenamefont {Robinson},
		\citenamefont {Watanabe}, \citenamefont {Taniguchi}, \citenamefont {Kim},
		\citenamefont {Cacho}, \citenamefont {Lanata}, \citenamefont {Miwa},
		\citenamefont {Hofmann}, \citenamefont {Katoch},\ and\ \citenamefont
		{Ulstrup}}]{Jones2020}%
	\BibitemOpen
	\bibfield  {author} {\bibinfo {author} {\bibfnamefont {A.~J.~H.}\
			\bibnamefont {Jones}}, \bibinfo {author} {\bibfnamefont {R.}~\bibnamefont
			{Muzzio}}, \bibinfo {author} {\bibfnamefont {P.}~\bibnamefont {Majchrzak}},
		\bibinfo {author} {\bibfnamefont {S.}~\bibnamefont {Pakdel}}, \bibinfo
		{author} {\bibfnamefont {D.}~\bibnamefont {Curcio}}, \bibinfo {author}
		{\bibfnamefont {K.}~\bibnamefont {Volckaert}}, \bibinfo {author}
		{\bibfnamefont {D.}~\bibnamefont {Biswas}}, \bibinfo {author} {\bibfnamefont
			{J.}~\bibnamefont {Gobbo}}, \bibinfo {author} {\bibfnamefont
			{S.}~\bibnamefont {Singh}}, \bibinfo {author} {\bibfnamefont {J.~T.}\
			\bibnamefont {Robinson}}, \bibinfo {author} {\bibfnamefont {K.}~\bibnamefont
			{Watanabe}}, \bibinfo {author} {\bibfnamefont {T.}~\bibnamefont {Taniguchi}},
		\bibinfo {author} {\bibfnamefont {T.~K.}\ \bibnamefont {Kim}}, \bibinfo
		{author} {\bibfnamefont {C.}~\bibnamefont {Cacho}}, \bibinfo {author}
		{\bibfnamefont {N.}~\bibnamefont {Lanata}}, \bibinfo {author} {\bibfnamefont
			{J.~A.}\ \bibnamefont {Miwa}}, \bibinfo {author} {\bibfnamefont
			{P.}~\bibnamefont {Hofmann}}, \bibinfo {author} {\bibfnamefont
			{J.}~\bibnamefont {Katoch}},\ and\ \bibinfo {author} {\bibfnamefont
			{S.}~\bibnamefont {Ulstrup}},\ }\bibfield  {title} {\bibinfo {title}
		{Observation of {Electrically} {Tunable} van {Hove} {Singularities} in
			{Twisted} {Bilayer} {Graphene} from {NanoARPES}},\ }\href
	{https://doi.org/10.1002/adma.202001656} {\bibfield  {journal} {\bibinfo
			{journal} {Advanced Materials}\ }\textbf {\bibinfo {volume} {32}},\ \bibinfo
		{pages} {2001656} (\bibinfo {year} {2020})}\BibitemShut {NoStop}%
	\bibitem [{\citenamefont {Lisi}\ \emph {et~al.}(2021)\citenamefont {Lisi},
		\citenamefont {Lu}, \citenamefont {Benschop}, \citenamefont {de~Jong},
		\citenamefont {Stepanov}, \citenamefont {Duran}, \citenamefont {Margot},
		\citenamefont {Cucchi}, \citenamefont {Cappelli}, \citenamefont {Hunter},
		\citenamefont {Tamai}, \citenamefont {Kandyba}, \citenamefont {Giampietri},
		\citenamefont {Barinov}, \citenamefont {Jobst}, \citenamefont {Stalman},
		\citenamefont {Leeuwenhoek}, \citenamefont {Watanabe}, \citenamefont
		{Taniguchi}, \citenamefont {Rademaker}, \citenamefont {van~der Molen},
		\citenamefont {Allan}, \citenamefont {Efetov},\ and\ \citenamefont
		{Baumberger}}]{Lisi2021}%
	\BibitemOpen
	\bibfield  {author} {\bibinfo {author} {\bibfnamefont {S.}~\bibnamefont
			{Lisi}}, \bibinfo {author} {\bibfnamefont {X.}~\bibnamefont {Lu}}, \bibinfo
		{author} {\bibfnamefont {T.}~\bibnamefont {Benschop}}, \bibinfo {author}
		{\bibfnamefont {T.~A.}\ \bibnamefont {de~Jong}}, \bibinfo {author}
		{\bibfnamefont {P.}~\bibnamefont {Stepanov}}, \bibinfo {author}
		{\bibfnamefont {J.~R.}\ \bibnamefont {Duran}}, \bibinfo {author}
		{\bibfnamefont {F.}~\bibnamefont {Margot}}, \bibinfo {author} {\bibfnamefont
			{I.}~\bibnamefont {Cucchi}}, \bibinfo {author} {\bibfnamefont
			{E.}~\bibnamefont {Cappelli}}, \bibinfo {author} {\bibfnamefont
			{A.}~\bibnamefont {Hunter}}, \bibinfo {author} {\bibfnamefont
			{A.}~\bibnamefont {Tamai}}, \bibinfo {author} {\bibfnamefont
			{V.}~\bibnamefont {Kandyba}}, \bibinfo {author} {\bibfnamefont
			{A.}~\bibnamefont {Giampietri}}, \bibinfo {author} {\bibfnamefont
			{A.}~\bibnamefont {Barinov}}, \bibinfo {author} {\bibfnamefont
			{J.}~\bibnamefont {Jobst}}, \bibinfo {author} {\bibfnamefont
			{V.}~\bibnamefont {Stalman}}, \bibinfo {author} {\bibfnamefont
			{M.}~\bibnamefont {Leeuwenhoek}}, \bibinfo {author} {\bibfnamefont
			{K.}~\bibnamefont {Watanabe}}, \bibinfo {author} {\bibfnamefont
			{T.}~\bibnamefont {Taniguchi}}, \bibinfo {author} {\bibfnamefont
			{L.}~\bibnamefont {Rademaker}}, \bibinfo {author} {\bibfnamefont {S.~J.}\
			\bibnamefont {van~der Molen}}, \bibinfo {author} {\bibfnamefont {M.~P.}\
			\bibnamefont {Allan}}, \bibinfo {author} {\bibfnamefont {D.~K.}\ \bibnamefont
			{Efetov}},\ and\ \bibinfo {author} {\bibfnamefont {F.}~\bibnamefont
			{Baumberger}},\ }\bibfield  {title} {\bibinfo {title} {Observation of flat
			bands in twisted bilayer graphene},\ }\href
	{https://doi.org/10.1038/s41567-020-01041-x} {\bibfield  {journal} {\bibinfo
			{journal} {Nature Physics}\ }\textbf {\bibinfo {volume} {17}},\ \bibinfo
		{pages} {189} (\bibinfo {year} {2021})}\BibitemShut {NoStop}%
	\bibitem [{\citenamefont {Lu}\ \emph {et~al.}(2012)\citenamefont {Lu},
		\citenamefont {Vishik}, \citenamefont {Yi}, \citenamefont {Chen},
		\citenamefont {Moore},\ and\ \citenamefont {Shen}}]{Lu2012}%
	\BibitemOpen
	\bibfield  {author} {\bibinfo {author} {\bibfnamefont {D.}~\bibnamefont
			{Lu}}, \bibinfo {author} {\bibfnamefont {I.~M.}\ \bibnamefont {Vishik}},
		\bibinfo {author} {\bibfnamefont {M.}~\bibnamefont {Yi}}, \bibinfo {author}
		{\bibfnamefont {Y.}~\bibnamefont {Chen}}, \bibinfo {author} {\bibfnamefont
			{R.~G.}\ \bibnamefont {Moore}},\ and\ \bibinfo {author} {\bibfnamefont
			{Z.-X.}\ \bibnamefont {Shen}},\ }\bibfield  {title} {\bibinfo {title}
		{Angle-{Resolved} {Photoemission} {Studies} of {Quantum} {Materials}},\
	}\href {https://doi.org/10.1146/annurev-conmatphys-020911-125027} {\bibfield
		{journal} {\bibinfo  {journal} {Annual Review of Condensed Matter Physics}\
		}\textbf {\bibinfo {volume} {3}},\ \bibinfo {pages} {129} (\bibinfo {year}
		{2012})}\BibitemShut {NoStop}%
	\bibitem [{\citenamefont {Pal}\ and\ \citenamefont {Mele}(2013)}]{Pal2013}%
	\BibitemOpen
	\bibfield  {author} {\bibinfo {author} {\bibfnamefont {A.}~\bibnamefont
			{Pal}}\ and\ \bibinfo {author} {\bibfnamefont {E.~J.}\ \bibnamefont {Mele}},\
	}\bibfield  {title} {\bibinfo {title} {Nodal surfaces in photoemission from
			twisted bilayer graphene},\ }\href
	{https://doi.org/10.1103/physrevb.87.205444} {\bibfield  {journal} {\bibinfo
			{journal} {Physical Review B}\ }\textbf {\bibinfo {volume} {87}},\ \bibinfo
		{pages} {205444} (\bibinfo {year} {2013})}\BibitemShut {NoStop}%
	\bibitem [{\citenamefont {Amorim}(2018)}]{Amorim2018}%
	\BibitemOpen
	\bibfield  {author} {\bibinfo {author} {\bibfnamefont {B.}~\bibnamefont
			{Amorim}},\ }\bibfield  {title} {\bibinfo {title} {General theoretical
			description of angle-resolved photoemission spectroscopy of van der {Waals}
			structures},\ }\href {https://doi.org/10.1103/PhysRevB.97.165414} {\bibfield
		{journal} {\bibinfo  {journal} {Physical Review B}\ }\textbf {\bibinfo
			{volume} {97}},\ \bibinfo {pages} {165414} (\bibinfo {year}
		{2018})}\BibitemShut {NoStop}%
	\bibitem [{\citenamefont {Li}\ \emph {et~al.}(2022)\citenamefont {Li},
		\citenamefont {Zhang}, \citenamefont {Chen}, \citenamefont {Wei},
		\citenamefont {Zhang}, \citenamefont {Xiao}, \citenamefont {Gao},
		\citenamefont {Chen}, \citenamefont {Liang}, \citenamefont {Pei},
		\citenamefont {Xu}, \citenamefont {Watanabe}, \citenamefont {Taniguchi},
		\citenamefont {Yang}, \citenamefont {Miao}, \citenamefont {Liu},
		\citenamefont {Cheng}, \citenamefont {Wang}, \citenamefont {Chen},\ and\
		\citenamefont {Liu}}]{Li2022}%
	\BibitemOpen
	\bibfield  {author} {\bibinfo {author} {\bibfnamefont {Y.}~\bibnamefont
			{Li}}, \bibinfo {author} {\bibfnamefont {S.}~\bibnamefont {Zhang}}, \bibinfo
		{author} {\bibfnamefont {F.}~\bibnamefont {Chen}}, \bibinfo {author}
		{\bibfnamefont {L.}~\bibnamefont {Wei}}, \bibinfo {author} {\bibfnamefont
			{Z.}~\bibnamefont {Zhang}}, \bibinfo {author} {\bibfnamefont
			{H.}~\bibnamefont {Xiao}}, \bibinfo {author} {\bibfnamefont {H.}~\bibnamefont
			{Gao}}, \bibinfo {author} {\bibfnamefont {M.}~\bibnamefont {Chen}}, \bibinfo
		{author} {\bibfnamefont {S.}~\bibnamefont {Liang}}, \bibinfo {author}
		{\bibfnamefont {D.}~\bibnamefont {Pei}}, \bibinfo {author} {\bibfnamefont
			{L.}~\bibnamefont {Xu}}, \bibinfo {author} {\bibfnamefont {K.}~\bibnamefont
			{Watanabe}}, \bibinfo {author} {\bibfnamefont {T.}~\bibnamefont {Taniguchi}},
		\bibinfo {author} {\bibfnamefont {L.}~\bibnamefont {Yang}}, \bibinfo {author}
		{\bibfnamefont {F.}~\bibnamefont {Miao}}, \bibinfo {author} {\bibfnamefont
			{J.}~\bibnamefont {Liu}}, \bibinfo {author} {\bibfnamefont {B.}~\bibnamefont
			{Cheng}}, \bibinfo {author} {\bibfnamefont {M.}~\bibnamefont {Wang}},
		\bibinfo {author} {\bibfnamefont {Y.}~\bibnamefont {Chen}},\ and\ \bibinfo
		{author} {\bibfnamefont {Z.}~\bibnamefont {Liu}},\ }\bibfield  {title}
	{\bibinfo {title} {Observation of coexisting dirac bands and moiré flat
			bands in magic‐angle twisted trilayer graphene},\ }\bibfield  {journal}
	{\bibinfo  {journal} {Advanced Materials}\ }\textbf {\bibinfo {volume}
		{34}},\ \href {https://doi.org/10.1002/adma.202205996}
	{10.1002/adma.202205996} (\bibinfo {year} {2022})\BibitemShut {NoStop}%
	\bibitem [{\citenamefont {Nunn}\ \emph {et~al.}(2023)\citenamefont {Nunn},
		\citenamefont {McEllistrim}, \citenamefont {Weston}, \citenamefont
		{Garcia-Ruiz}, \citenamefont {Watson}, \citenamefont {Mucha-Kruczynski},
		\citenamefont {Cacho}, \citenamefont {Gorbachev}, \citenamefont {Fal’ko},\
		and\ \citenamefont {Wilson}}]{Nunn2023}%
	\BibitemOpen
	\bibfield  {author} {\bibinfo {author} {\bibfnamefont {J.~E.}\ \bibnamefont
			{Nunn}}, \bibinfo {author} {\bibfnamefont {A.}~\bibnamefont {McEllistrim}},
		\bibinfo {author} {\bibfnamefont {A.}~\bibnamefont {Weston}}, \bibinfo
		{author} {\bibfnamefont {A.}~\bibnamefont {Garcia-Ruiz}}, \bibinfo {author}
		{\bibfnamefont {M.~D.}\ \bibnamefont {Watson}}, \bibinfo {author}
		{\bibfnamefont {M.}~\bibnamefont {Mucha-Kruczynski}}, \bibinfo {author}
		{\bibfnamefont {C.}~\bibnamefont {Cacho}}, \bibinfo {author} {\bibfnamefont
			{R.~V.}\ \bibnamefont {Gorbachev}}, \bibinfo {author} {\bibfnamefont {V.~I.}\
			\bibnamefont {Fal’ko}},\ and\ \bibinfo {author} {\bibfnamefont {N.~R.}\
			\bibnamefont {Wilson}},\ }\bibfield  {title} {\bibinfo {title} {{ARPES}
			{Signatures} of {Few}-{Layer} {Twistronic} {Graphenes}},\ }\href
	{https://doi.org/10.1021/acs.nanolett.3c01173} {\bibfield  {journal}
		{\bibinfo  {journal} {Nano Letters}\ }\textbf {\bibinfo {volume} {23}},\
		\bibinfo {pages} {5201} (\bibinfo {year} {2023})}\BibitemShut {NoStop}%
	\bibitem [{\citenamefont {Jiang}\ \emph {et~al.}(2023)\citenamefont {Jiang},
		\citenamefont {Hsieh}, \citenamefont {Jones}, \citenamefont {Majchrzak},
		\citenamefont {Sahoo}, \citenamefont {Watanabe}, \citenamefont {Taniguchi},
		\citenamefont {Miwa}, \citenamefont {Chen},\ and\ \citenamefont
		{Ulstrup}}]{Jiang2023}%
	\BibitemOpen
	\bibfield  {author} {\bibinfo {author} {\bibfnamefont {Z.}~\bibnamefont
			{Jiang}}, \bibinfo {author} {\bibfnamefont {K.}~\bibnamefont {Hsieh}},
		\bibinfo {author} {\bibfnamefont {A.~J.~H.}\ \bibnamefont {Jones}}, \bibinfo
		{author} {\bibfnamefont {P.}~\bibnamefont {Majchrzak}}, \bibinfo {author}
		{\bibfnamefont {C.}~\bibnamefont {Sahoo}}, \bibinfo {author} {\bibfnamefont
			{K.}~\bibnamefont {Watanabe}}, \bibinfo {author} {\bibfnamefont
			{T.}~\bibnamefont {Taniguchi}}, \bibinfo {author} {\bibfnamefont {J.~A.}\
			\bibnamefont {Miwa}}, \bibinfo {author} {\bibfnamefont {Y.~P.}\ \bibnamefont
			{Chen}},\ and\ \bibinfo {author} {\bibfnamefont {S.}~\bibnamefont
			{Ulstrup}},\ }\bibfield  {title} {\bibinfo {title} {Revealing flat bands and
			hybridization gaps in a twisted bilayer graphene device with {microARPES}},\
	}\href {https://doi.org/10.1088/2053-1583/acf775} {\bibfield  {journal}
		{\bibinfo  {journal} {2D Materials}\ }\textbf {\bibinfo {volume} {10}},\
		\bibinfo {pages} {045027} (\bibinfo {year} {2023})}\BibitemShut {NoStop}%
	\bibitem [{\citenamefont {Wang}\ \emph {et~al.}(2023)\citenamefont {Wang},
		\citenamefont {Chen}, \citenamefont {Lin}, \citenamefont {Hou}, \citenamefont
		{Lin},\ and\ \citenamefont {Chou}}]{Wang2023}%
	\BibitemOpen
	\bibfield  {author} {\bibinfo {author} {\bibfnamefont {W.-C.}\ \bibnamefont
			{Wang}}, \bibinfo {author} {\bibfnamefont {F.-W.}\ \bibnamefont {Chen}},
		\bibinfo {author} {\bibfnamefont {K.-S.}\ \bibnamefont {Lin}}, \bibinfo
		{author} {\bibfnamefont {J.~T.}\ \bibnamefont {Hou}}, \bibinfo {author}
		{\bibfnamefont {H.-C.}\ \bibnamefont {Lin}},\ and\ \bibinfo {author}
		{\bibfnamefont {M.-Y.}\ \bibnamefont {Chou}},\ }\bibfield  {title} {\bibinfo
		{title} {Origin of magic angles in twisted bilayer graphene: {The} magic
			ring}\ }\href {https://doi.org/10.48550/arXiv.2309.10026}
	{10.48550/arXiv.2309.10026} (\bibinfo {year} {2023})\BibitemShut {NoStop}%
	\bibitem [{\citenamefont {Bi}\ \emph {et~al.}(2019)\citenamefont {Bi},
		\citenamefont {Yuan},\ and\ \citenamefont {Fu}}]{Bi2019}%
	\BibitemOpen
	\bibfield  {author} {\bibinfo {author} {\bibfnamefont {Z.}~\bibnamefont
			{Bi}}, \bibinfo {author} {\bibfnamefont {N.~F.~Q.}\ \bibnamefont {Yuan}},\
		and\ \bibinfo {author} {\bibfnamefont {L.}~\bibnamefont {Fu}},\ }\bibfield
	{title} {\bibinfo {title} {Designing flat bands by strain},\ }\href
	{https://doi.org/10.1103/PhysRevB.100.035448} {\bibfield  {journal} {\bibinfo
			{journal} {Physical Review B}\ }\textbf {\bibinfo {volume} {100}},\ \bibinfo
		{pages} {035448} (\bibinfo {year} {2019})}\BibitemShut {NoStop}%
	\bibitem [{\citenamefont {Sinner}\ \emph {et~al.}(2023)\citenamefont {Sinner},
		\citenamefont {Pantaleón},\ and\ \citenamefont {Guinea}}]{Sinner2023}%
	\BibitemOpen
	\bibfield  {author} {\bibinfo {author} {\bibfnamefont {A.}~\bibnamefont
			{Sinner}}, \bibinfo {author} {\bibfnamefont {P.~A.}\ \bibnamefont
			{Pantaleón}},\ and\ \bibinfo {author} {\bibfnamefont {F.}~\bibnamefont
			{Guinea}},\ }\bibfield  {title} {\bibinfo {title} {Strain-{Induced}
			{Quasi}-{1D} {Channels} in {Twisted} {Moiré} {Lattices}},\
	}\href {https://doi.org/10.1103/PhysRevLett.131.166402} {\bibfield  {journal}
		{\bibinfo  {journal} {Physical Review Letters}\ }\textbf {\bibinfo {volume}
			{131}},\ \bibinfo {pages} {166402} (\bibinfo {year} {2023})}\BibitemShut
	{NoStop}%
	\bibitem [{\citenamefont {Kögl}\ \emph {et~al.}(2023)\citenamefont {Kögl},
		\citenamefont {Soubelet}, \citenamefont {Brotons-Gisbert}, \citenamefont
		{Stier}, \citenamefont {Gerardot},\ and\ \citenamefont {Finley}}]{Koegl2023}%
	\BibitemOpen
	\bibfield  {author} {\bibinfo {author} {\bibfnamefont {M.}~\bibnamefont
			{Kögl}}, \bibinfo {author} {\bibfnamefont {P.}~\bibnamefont {Soubelet}},
		\bibinfo {author} {\bibfnamefont {M.}~\bibnamefont {Brotons-Gisbert}},
		\bibinfo {author} {\bibfnamefont {A.~V.}\ \bibnamefont {Stier}}, \bibinfo
		{author} {\bibfnamefont {B.~D.}\ \bibnamefont {Gerardot}},\ and\ \bibinfo
		{author} {\bibfnamefont {J.~J.}\ \bibnamefont {Finley}},\ }\bibfield  {title}
	{\bibinfo {title} {Moiré straintronics: a universal platform for
			reconfigurable quantum materials},\ }\href
	{https://doi.org/10.1038/s41699-023-00382-4} {\bibfield  {journal} {\bibinfo
			{journal} {npj 2D Materials and Applications}\ }\textbf {\bibinfo {volume}
			{7}},\ \bibinfo {pages} {1} (\bibinfo {year} {2023})}\BibitemShut {NoStop}%
	\bibitem [{\citenamefont {Escudero}\ \emph {et~al.}(2024)\citenamefont
		{Escudero}, \citenamefont {Sinner}, \citenamefont {Zhan}, \citenamefont
		{Pantaleón},\ and\ \citenamefont {Guinea}}]{Escudero2024}%
	\BibitemOpen
	\bibfield  {author} {\bibinfo {author} {\bibfnamefont {F.}~\bibnamefont
			{Escudero}}, \bibinfo {author} {\bibfnamefont {A.}~\bibnamefont {Sinner}},
		\bibinfo {author} {\bibfnamefont {Z.}~\bibnamefont {Zhan}}, \bibinfo {author}
		{\bibfnamefont {P.~A.}\ \bibnamefont {Pantaleón}},\ and\ \bibinfo {author}
		{\bibfnamefont {F.}~\bibnamefont {Guinea}},\ }\bibfield  {title} {\bibinfo
		{title} {Designing moiré patterns by strain},\ }\href
	{https://doi.org/10.1103/PhysRevResearch.6.023203} {\bibfield  {journal} {\bibinfo
			{journal} {Physical Review Research}\ }\textbf {\bibinfo {volume} {6}},\ \bibinfo
		{pages} {023203} (\bibinfo {year} {2024})}\BibitemShut {NoStop}%
	\bibitem [{\citenamefont {Huder}\ \emph {et~al.}(2018)\citenamefont {Huder},
		\citenamefont {Artaud}, \citenamefont {Le~Quang}, \citenamefont
		{de~Laissardière}, \citenamefont {Jansen}, \citenamefont {Lapertot},
		\citenamefont {Chapelier},\ and\ \citenamefont {Renard}}]{Huder2018}%
	\BibitemOpen
	\bibfield  {author} {\bibinfo {author} {\bibfnamefont {L.}~\bibnamefont
			{Huder}}, \bibinfo {author} {\bibfnamefont {A.}~\bibnamefont {Artaud}},
		\bibinfo {author} {\bibfnamefont {T.}~\bibnamefont {Le~Quang}}, \bibinfo
		{author} {\bibfnamefont {G.~T.}\ \bibnamefont {de~Laissardière}}, \bibinfo
		{author} {\bibfnamefont {A.~G.}\ \bibnamefont {Jansen}}, \bibinfo {author}
		{\bibfnamefont {G.}~\bibnamefont {Lapertot}}, \bibinfo {author}
		{\bibfnamefont {C.}~\bibnamefont {Chapelier}},\ and\ \bibinfo {author}
		{\bibfnamefont {V.~T.}\ \bibnamefont {Renard}},\ }\bibfield  {title}
	{\bibinfo {title} {Electronic {Spectrum} of {Twisted} {Graphene} {Layers}
			under {Heterostrain}},\ }\href
	{https://doi.org/10.1103/PhysRevLett.120.156405} {\bibfield  {journal}
		{\bibinfo  {journal} {Physical Review Letters}\ }\textbf {\bibinfo {volume}
			{120}},\ \bibinfo {pages} {156405} (\bibinfo {year} {2018})}\BibitemShut
	{NoStop}%
\end{thebibliography}
\end{document}